\documentclass[twoside,11pt]{PhDthesisPSnPDF}

\usepackage{enumerate}
\usepackage{amsmath}
\usepackage{epstopdf}
\usepackage{slashed}
\usepackage{verbatim}
\usepackage{pdfpages}
\usepackage{lettrine}

\usepackage{tikz}
\usetikzlibrary{trees}
\usetikzlibrary{snakes}
\usetikzlibrary{decorations.pathmorphing}
\usetikzlibrary{decorations.markings}

\tikzset{
    photon/.style={decorate, decoration={snake}, draw=red},
    electron/.style={draw=blue, postaction={decorate},
        decoration={markings,mark=at position .55 with {\arrow[draw=blue]{>}}}},
    positron/.style={draw=blue, postaction={decorate},
        decoration={markings,mark=at position .55 with {\arrow[draw=blue]{<}}}},
    gluon/.style={decorate, draw=magenta,
        decoration={coil,amplitude=4pt, segment length=5pt}}, 
    higgs/.style={draw=black,dashed},
    slepton/.style={draw=blue,dashed, postaction={decorate},
        decoration={markings,mark=at position .55 with {\arrow[draw=blue]{>}}}}
}

\newcommand{\refer}[1]{(\ref{#1})}
\newcommand{\diff}{\mathrm{d}}
\newcommand{\comm}[2]{\left[ #1, #2 \right]}

\newcommand{\lineint}{\int\limits_{-\infty}^{\infty}\diff}

\newcommand{\Exp}[1]{\mathrm{e}^{#1}}
\newcommand{\imply}{\Rightarrow}
\newcommand{\oper}[1]{{\cal #1}}
\newcommand{\iunit}{\mathrm{i}}
\newcommand{\abs}[1]{\left|#1\right|}

\newcommand{\Tr}{\mathrm{Tr}}

\renewcommand{\vec}[1]{\mathbf{#1}}

\ifpdf  
    \pdfcatalog { /PageMode (/UseOutlines)
                  /OpenAction (fitbh)  }
\fi

\title{
Dynamick\'a realizace sc\'en\'a\v{r}e br\'anov\'ych sv\v{e}t\r{u} pomoc\'i topologick\'ych
soliton\r{u} \\ \vspace*{1ex}
Dynamical Realization of the Brane world Scenario Using Topological Solitons}

\ifpdf
  \author{\href{mailto:filip.blaschke@fpf.slu.cz}{Filip Blaschke}}
  \cityofbirth{Obor: Teoretick\'a fyzika a astrofyzika} 
  \collegeordept{\href{http://www.fpf.slu.cz}{Filozoficko-p\v{r}\'irodov\v{e}deck\'a 
  fakulta v Opav\v{e}}}
  \university{\href{http://www.slu.cz}{SLEZSK\'A UNIVERZITA V OPAV\v{E}}}

  \crest{\includegraphics[width=4cm]{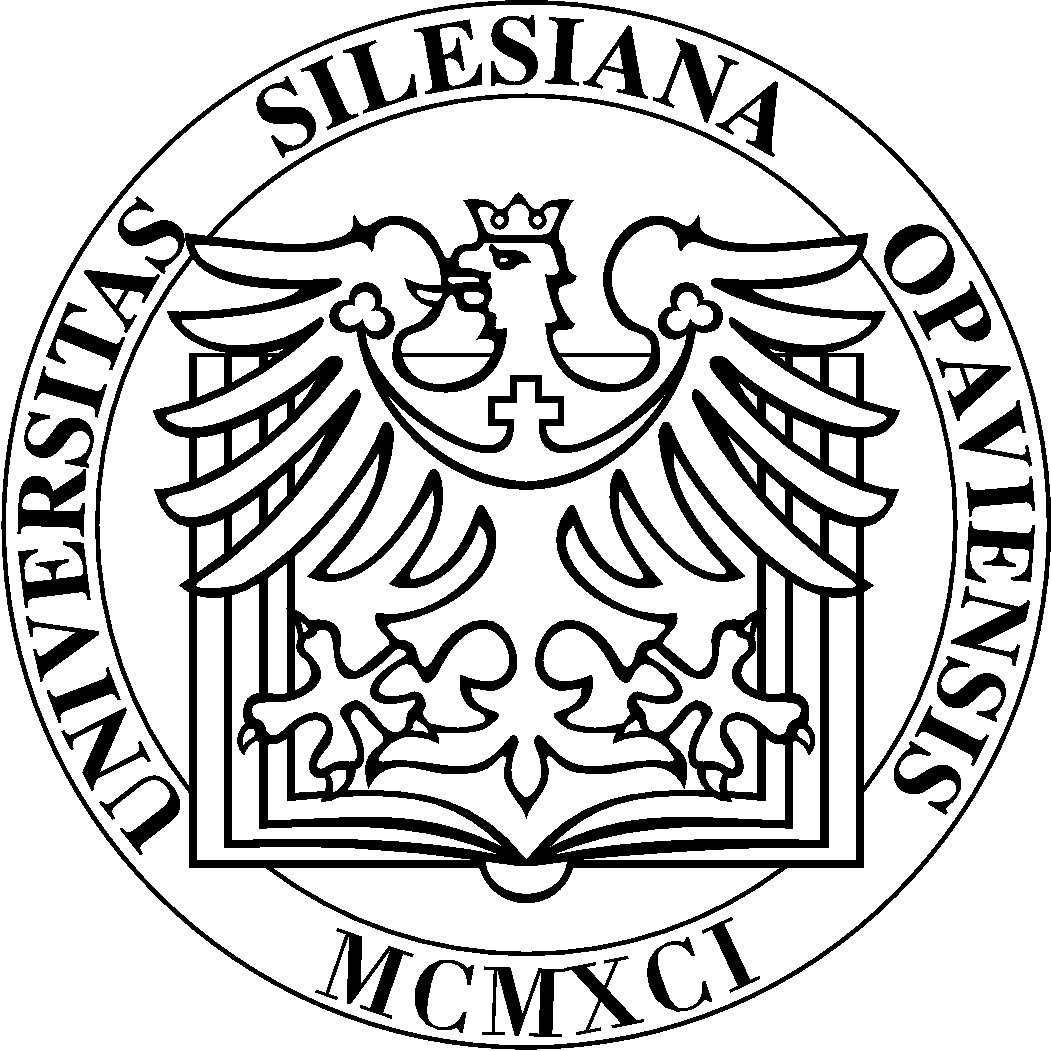}}
  
\else
  \author{Filip Blaschke}
  \cityofbirth{Obor: Teoretick\'a fyzika a astrofyzika}
  \collegeordept{Filozoficko-p\v{r}\'irodov\v{e}deck\'a fakulta v Opav\v{e}}
  \university{SLEZSK\'A UNIVERZITA V OPAV\v{E}}
  \crest{\includegraphics[width=4cm]{logo}}
\fi

\degree{ Diserta\v{c}n\'i pr\'ace}
\degreedate{Opava 2013 \hfill \v{S}kolitel diserta\v{c}n\'i pr\'ace:\\ \vspace{-3ex}
\phantom{a} \hfill Dr. Masato Arai, PhD. \\ \vspace{-3ex} \phantom{a} \hfill Prof. Ing. Peter Lichard, DrSc.}

\begin{document}


\renewcommand\baselinestretch{1.2}
\baselineskip=18pt plus1pt


\maketitle  









\begin{abstracts}
{\small

In this thesis we discuss how the brane world scenario can be realized dynamically within the 
field theoretical framework using topological solitons.  As a playground we consider a bosonic 
sector of a (4+1)-dimensional supersymmetric gauge theory, which naturally supports soliton of 
co-dimension one, a domain wall. We first discuss separate localization of matter fields and 
gauge fields on the world-volume of the domain wall and then we present two explicit five-dimensional
 models, 
where both matter fields and gauge fields are localized together with minimal interactions. 
We show that matter fields localize in the adjoint representation of the non-Abelian gauge group and we
calculate the effective interaction Lagrangian of these matter fields up to the second order in 
derivatives. We discuss similarities of our models with effective models describing pions in QCD 
and with D-branes from string theory.

\vspace{5mm}

\vspace{5mm}
\begin{center}
  \vspace*{1.0cm}
  {\Large \bfseries  Anotace}
  \vspace{5mm}
\end{center}
  
  V t\'eto diserta\v{c}n\'i pr\'aci se zab\'yv\'ame ot\'azkou dynamick\'e realizace sc\'en\'a\v{r}e
  br\'anov\'ych sv\v{e}t\r{u} pomoc\'i topologick\'ych soliton\r{u}. Nejprve diskutujeme
  lokalizaci hmotov\'ych a kalibra\v{c}n\'ich pol\'i na dom\'enov\'e st\v{e}n\v{e} odd\v{e}l\v{e}n\v{e}
  a potom uva\v{z}ujeme explictn\'i modely ve (4+1)-dimenz\'ich, ve kter\'ych se hmotov\'a
  a kalibra\v{c}n\'i pole lokalizuj\'i sou\v{c}asn\v{e} s minim\'aln\'i 
  interakc\'i.
  Pot\'e vypo\v{c}\'it\'ame efektivn\'i interak\v{c}n\'i Lagrangi\'an 
  hmotov\'ych pol\'i do \v{c}len\r{u} druh\'eho \v{r}\'adu v 
  derivac\'ich.
  Nakonec diskutujeme podobnosti na\v{s}ich model\r{u} s efektivn\'imi modely pion\r{u} v QCD a s
  D-br\'anami ve strunov\'e teorii.
  
\vspace{5mm}
{\bfseries Kl\'i\v{c}ov\'a slova: Sc\'en\'a\v{r} br\'anov\'ych sv\v{e}t\r{u}, 
topologick\'y soliton, supersymetrie.}
}
\end{abstracts}




\frontmatter



\begin{declaration}        

I herewith declare that I have produced this paper without the prohibited assistance of third parties and without making use of aids other than 
those specified; notions taken over directly or indirectly from other sources have been identified as such. This paper has not previously been 
presented in identical or similar form to any other domestic or foreign examination board.

The thesis work was conducted from Sebtember 2009 to Sebtember 2013 under the supervision of 
Masato Arai, PhD. and Prof. Ing. Peter Lichard, DrSc.

\vspace{10mm}

In Opava, June 2013 \hfill \dots\dots\dots\dots\dots\\
\phantom{a} \hfill Filip Blaschke

\newpage\null\addtocounter{page}{4}\thispagestyle{fancy}\newpage

\end{declaration}


\begin{acknowledgements}      

I would like to thank my supervisor, Masato Arai for his guidance, patience and support during those
stormy years of my PhD course. Thanks to his care I have been given an opportunity to explore
both the world of  theoretical physics and that of a theoretical physicist. For that I am forever gratefull.
The same goes to Peter Lichard, who has been my mentor since the very beginning of my study and who
actively encouraged me to expand my profesional horizons. Special thanks to Norisuke Sakai and 
Minoru Eto for their kind counsel and the opportunity of collaboration, which I hold as a singular
privilige.

I would have not chosen physics as my field have not been for support and encouragement from my family, especially my brothers.
And my wife, who has been (and always shall be) a constant source of inspiration.

\newpage\null\thispagestyle{fancy}\newpage
\end{acknowledgements}



\begin{dedication} 










  

The truth is usually just an excuse for a lack of imagination.

    -Garak, Star Trek: Deep Space Nine
\newpage\null\thispagestyle{fancy}\newpage
\end{dedication}

\begin{abstractslong}        

In this thesis we discuss how certain aspects of the so-called brane world scenario can be realized within the field-theoretical framework. In accordance with an old proposition, and for simplicity as well, we use a domain wall as a model of the brane. Thus, we consider the case of single, infinitely long extra dimension, along which domain wall varies. We show that, in the presence of the brane, certain modes of fluctuations of background fields are trapped inside the domain wall, representing particles localized on the brane. Then, according to specified rules, we derive an effective four-dimensional theory describing nonlinear interactions of these modes. The ultimate goal of our work is to localize a full spectrum of Standard model particles in this way. Here, we present an important step towards this goal, a model where various scalar particles (Nambu-Goldstone bosons) with non-trivial quantum charges are localized on the brane along side with non-Abelian gauge fields. We show that the corresponding effective theory shares many aspects with low-energy theory of pions. In addition, we explore the full structure of non-linear interactions between pions up to second order terms in derivatives.
We also show that our model naturally incorporates the Higgs mechanism which is, in contrast to the Standard Model, based purely on geometry. Lastly we address the issue of stability and typicality of our construction.
 
This thesis is organized as follows. In Ch.~\ref{ch:1} we make short historical overview of extra dimensions and their phenomenological implications for high energy physics. Ch.~\ref{ch:2} has an introductory purpose too, as we consider there a simple toy model realizing the brane world scenario.
Along the presentation we encounter several important concepts, which we utilize in the later chapters, such as a BPS soliton, zero modes and their localization and description by an effective Lagrangian. In Ch.~\ref{ch:3} we make a detour and discuss another key concept, supersymmetry. First we establish connection between BPS solitons with supersymmetry in (1+1)-dimensions as a natural extension of our toy model of Ch.~\ref{ch:2}. Then we discuss basic aspects of supersymmetric theories in four and subsequently in five space-time dimensions. As we will demonstrate, supersymmetric gauge theories in five-dimensions are very convenient laboratories to explore the brane world scenario based on domain walls.
The culmination of this discussion is represented by chapter \ref{ch:4}, where we address the crucial issue of localization of gauge fields. At the end of this chapter all tools will be prepared to make a next logical step of Ch.~\ref{ch:5}, with which we start to review original results. 

In Ch.~\ref{ch:5} we introduce simple, yet robust field-theoretical model of the brane world scenario. We demonstrate how the matter fields and gauge fields can be localized on a domain wall with nontrivial interactions. This chapter is based on the paper I listed below. 
In Ch.~\ref{ch:6} we present a similar model to that of Ch.~\ref{ch:5}, but with more economical charge assignments. This subtle change in turn ensure questionable stability of the gauge sector of the previous model and also results in new features in the effective Lagrangian. This chapter is based on the paper II listed below. Finally we conclude in Ch.~\ref{ch:7} by discussing possible extensions of our work.

The author of this thesis took an active part in this research and main aspects of his contribution were twofold. First, he found closed
formulas for low energy effective Lagrangians presented in Ch.~\ref{ch:5} and Ch.~\ref{ch:6}. And second,
he investigated the mechanism, which removed the potential instability of the gauge sector and which represents the main part of Ch.~\ref{ch:6}. 

For the reference we list below all
publications to which the author of this thesis contributed during his PhD study.

\begin{description}

\item{\bf Paper I:\\}
{\sc M.~{Arai}, F.~{Blaschke}, M.~{Eto}, and N.~{Sakai}}.
\newblock {\bf {Matter fields and non-Abelian gauge fields localized on
  walls}}.
\newblock {\em Progress of Theoretical and Experimental Physics}, {\bf
  2013}(1):010003, January 2013.

\item{\bf Paper II:\\}
{\sc M.~{Arai}, F.~{Blaschke}, M.~{Eto}, and N.~{Sakai}}.
\newblock {\bf {Stabilizing matter and gauge fields localized on walls}}.
\newblock {\em ArXiv e-prints}, March 2013.

\item{\bf Paper III:\\}
{\sc Masato Arai, Filip Blaschke, Minoru Eto, and Norisuke Sakai}.
\newblock {\bf Localization of matter fields and non-Abelian gauge fields on
  domain walls}.
\newblock {\em Journal of Physics: Conference Series}, {\bf 411}(1):012001,
  2013.

\item{\bf Paper IV:\\}
{\sc Masato Arai and Filip Blaschke}.
\newblock {\bf {Cotangent bundle over Hermitian symmetric space $E_7/E_6 \times
  U(1)$ from projective superspace}}.
\newblock {\em JHEP}, {\bf 1302}:045, 2013.
\end{description} 

\end{abstractslong}


\setcounter{secnumdepth}{3} 
\setcounter{tocdepth}{3}    
\tableofcontents            


\listoffigures	

\listoftables  




\markboth{\MakeUppercase{\nomname}}{\MakeUppercase{\nomname}}


\nomenclature{ADD}{Arkani-Hamed-Dimopoulos-Dvali}
\nomenclature{AdS}{Anti-de Sitter}
\nomenclature{BPS}{Bogomol'nyi-Prasad-Sommerfield}
\nomenclature{CS}{Chern-Simons}
\nomenclature{DBI}{Dirac-Born-Infeld}
\nomenclature{DoF}{Degrees of Freedom}
\nomenclature{EoM}{Equations of Motion}
\nomenclature{FI}{Fayet-Iliopulous}
\nomenclature{KK}{Kaluza-Klein}
\nomenclature{LED}{Large Extra Dimensions}
\nomenclature{NG}{Nambu-Goldstone}
\nomenclature{ODE}{Ordinary Differential Equation}
\nomenclature{QCD}{Quantum chromodynamics}
\nomenclature{SM}{Standard Model}
\nomenclature{SUSY}{Supersymmetry}
\nomenclature{VEV}{Vacuum Expectation Value}
\nomenclature{WZ}{Wess-Zumino}

\begin{small} 

\printnomenclature[1.5cm] 
\label{nom} 

\end{small}


\mainmatter

\renewcommand{\chaptername}{} 




\chapter{Introduction}\label{ch:1}

\ifpdf
    \graphicspath{{1_introduction/figures/PNG/}{1_introduction/figures/PDF/}{1_introduction/figures/}}
\else
    \graphicspath{{1_introduction/figures/EPS/}{1_introduction/figures/}}
\fi


The Standard
Model (SM), developed in early 1970’s, successfully describes three
of the four fundamental interactions (namely the strong, weak and electromagnetic
interactions) of elementary particles on energy scales accessible to particle
accelerators such as the Large Electron Positron Collider, the Large Hadron Collider
at CERN in Switzerland and the Tevatron at Fermilab in the United States.
The SM predicts phenomena observed in these particle accelerators quite well and
actually there is no inconsistency between the SM and current experimental
data. However, it is widely recognized that SM is not a complete theory by
a number of theoretical and phenomenological reasons. First of all, the SM does
not provide answers for questions such as why the spectrum of particle masses is
hierarchical or why the number of generations is three. It also does not explain the
relic abundance of dark matter in the universe, as none of the particles in the SM can
be a candidate for dark matter. Furthermore, the SM has a conceptual problem
called the gauge hierarchy problem, stating that there is a huge hierarchy (17
orders of magnitude) between the electroweak scale and the gravity (Planck) scale.
Finally, the SM does not describe gravity at all, as the correct quantum theory of gravity is still missing. 
The common expectation is that there is a more fundamental theory beyond the SM, in which
these problems are naturally resolved.

A very popular and one of the most promising ideas beyond the SM  is an intriguing notion of extra dimensions. The possibility, that there might be more dimensions of space than three, has become a part of serious science more than a hundred years ago and since that time it influenced physics in many direct and indirect ways. 
Consequently this subject, that one could call ``extra-dimensional physics'', is very vast and progressively more richer in content as we follow its development up to the present day. Therefore, given limited space of this chapter, our presentation of this topic is accordingly coarse-grained.
 In fact, from the plethora of different models/theories, based on extra dimensions, we will
  focus only on the four most influential ones. Each of these corresponds to a shift
   in thinking about the role of extra dimensions in high-energy physics and also marks the
    beginning of a fruitful and ongoing investigation.
These four milestones
reflect in the organization of this chapter. Its goal is to provide enough background 
material to properly set the remaining content of this thesis in the landscape of extra-dimensional 
scenarios.      

The beginning of extra dimensions, which we describe in the first section, started from attempts to unify different forces in nature. 
Even though first investigations in this direction were made by Nordstr\"om, we are going to introduce the more famous work of Kaluza and Klein \cite{Kaluza, Klein}. As we will see, in the Kaluza-Klein scenario the mathematical consequences of additional, fifth dimension result in remarkable unification of four-dimensional gravity and electromagnetism. However, the presence of the fifth dimension is considered here as an unwanted feature and, in order not to alter phenomenology, it is compactified to an unobservable size. We will also very shortly describe developments of Kaluza-Klein ideas within string theory.

In the second section we discuss a proposition of Rubakov and Shaposhnikov \cite{Rubakov, Rubakov2}, in which the role of an extra dimension is completely different. 
The central idea of their work, which later becomes identified as a \emph{brane world scenario}, is based on an assumption, that our Universe is a three-dimensional defect evolving in a multidimensional space, called a \emph{bulk}. All matters are trapped on this defect and, if the energy is not sufficiently high, it cannot propagate into other dimensions. Rubakov and Shaposhnikov in their famous paper \cite{Rubakov2} discussed a simple model, in which this scenario is realized. We will revisit this model in great detail in chapter \ref{ch:2}. In fact, the work we are going to present in this thesis, might be considered as an extension of this model towards to its logical conclusion.  

The canonical form of the brane world scenario, however, was fully realized later in the work 
of Arkadi-Hamed, Dimopoulos and Dvali \cite{Dvali}, which is described in the third section. 
Their model, known as the ADD model, rather than hide additional dimensions under the carpet, 
used the bulk as an instrument to explain phenomenological truths about the nature. 
In ADD scenario the extra dimensions are compactified, but 
in contrast to the Kaluza-Klein scenario, not to the Planck scale $M_P\sim 10^{19}$~GeV,
 but rather to experimentally more accessible Weak scale $M_{W}\sim 10^3$~GeV. As a consequence, 
 they are called \emph{large extra dimensions}, even though from human perspective they are still
  rather small (around micrometer or less). The reason for this is to account for apparent 
 weakness of gravity.
In the brane world scenario, gravitons are not, like ordinary matter, trapped on the brane, 
but they are free to roam in the bulk. This results in weakening of gravitation from a point 
of view of a brane-bound matter.
We also briefly describe different approach to the brane world scenario, which is the work
of Randall and Sundrum \cite{RS,RS2}. To localize particles on the brane, they use 
warped extra-dimension, compactified 
on a ${S^1/\mathbb{Z}_2}$ orbifold and a negative cosmological constant in the bulk.
In contrast to ADD scenario, the fundamental scale in models of Randall and Sundrum is the Planck scale.
However, the warping of the extra dimension causes exponential screening of the fundamental scale down to 
$M_W$. 

The dichotomy between gravity and other forces in the brane world scenario 
actually originated in string theory.
Indeed, the very word ``brane'' comes from string theory.
In 1995 Polchinski \cite{Polchinski} showed that string theory possesses a remarkable degree of freedom: a \emph{D-brane}.
The D-brane naturally realizes a brane world scenario, since open strings (matter and gauge particles) ends on the D-brane.
On the other hand, closed strings (gravitons) cannot be confined to the D-brane
and must propagate into extra dimensions.
Introduction of D-branes led to new ideas in phenomenology such as the above mentioned large
extra dimensions. 
Moreover, it is widely recognized that topological soliton
such as a domain wall, vortex and instantons can be seen as an effective low-energy
analog of a D-brane. This opens up the possibility to realize brane world scenario
dynamically and to directly investigate the spectrum of particles localized on the
brane in the field-theoretical framework. We discuss this issue in the four section, where we 
also set the goals of this thesis. 

\section{Kaluza-Klein theory}

In 1919 Theodor Kaluza came up with an elegant way how to put gravity and electromagnetism into one field-theoretical frame. 
Considering that at that time those were only known forces, there is no wonder that his theory raised serious attention in the scientific community.
Indeed, upon submitting his paper to Einstein himself, Kaluza had to wait two years (an unusually long time even by the standards of those 
slow-paced days) to receive a confirmatory response and recommendation for publication \cite{Kaluza}. 

In his work he noticed, that if one starts with
pure five-dimensional gravity one can reduce it to Einstein-Maxwell theory
in four dimensions. Curiously, the only \emph{ad hoc} assumption he had to make, was that no fields depends on the fifth direction (which
we from now on denote as $y$).
He discovered that this assumption alone naturally decomposes five-dimensional metric tensor into three four-dimensional fields of spin $2,1$ and $0$ 
and, upon plugging
the appropriate ansatz into the five-dimensional Einstein equations, one obtains separate sets of equations for each field,\footnote{Apart from
scalar field which is coupled with both gravitational and gauge kinetic terms.} which
are precisely the correct equations of motion for each spin. It worked like a miracle.

Let us briefly expose the mechanism of this miracle from a modern point of view. The spectrum of fields in effective four dimensional theory
is easily shown to be degrees of freedom (DoF) consistent. Indeed, we started with five-dimensional metric tensor $^5\! g_{MN}$ with 
$M,N = {0,1,2,3,y}$ which contains
$5(5+1)/2 = 15$ real DoF. With respect to $y$ we can split $^5\! g_{MN}$ into the four-dimensional metric tensor $\hat g_{\mu\nu}\equiv\, ^5\! g_{\mu\nu}$ with 10 
DoF and a
 vector field $\hat A_{\mu} \equiv\, ^{5}\! g_{\mu y}$ of altogether four DoF. The last degree of freedom is associated 
with the scalar field $\hat \phi \equiv\, ^{5}\! g_{yy}$, which sums up back to 15, as required. 

The spectrum itself, however, can be deduced from symmetry considerations. The independence of metric tensor $^{5}\! g_{MN}$ on extra-dimensional
coordinate $y$ is formulated by the so-called Cylinder condition $\partial_{y}\,^5\! g_{MN} = 0$, which is obviously not a covariant statement. 
However, there are two kinds of transformations which are compatible with this 
condition, namely
\begin{align}
\mbox{I type:} && x^{M} & \to f^{\mu}(x)\delta_{\mu}^{M}+y\delta_{y}^{M}\,, \\
\mbox{II type:} && x^{M} & \to x^{\mu}\delta_{\mu}^{M}+\bigl(y+\varepsilon(x)\bigr)\delta_{y}^{M}\,. 
\end{align}
The first one represents a general change of coordinates in four dimensions with no change to extra dimension (for simplicity we ignore 
here global transformations such as rescaling and constant shifts). Under this type of transformations four-dimensional fields 
$\hat g_{\mu\nu}$, 
$\hat A_{\mu}$ and $\hat \phi$ transform respectively as a rank-two tensor, vector and a scalar.
The second type can be respected as a gauge transformation. For infinitesimal shifts $\abs{\varepsilon(x)} \ll 1$ we get
\begin{align}
\hat g_{\mu\nu} & \to \hat g_{\mu\nu}-\partial_{\mu}\varepsilon \hat A_{\nu}-\partial_{\nu}\varepsilon \hat A_{\mu} \,, \\
\hat A_{\mu} & \to \hat A_{\mu} - \partial_{\mu}\varepsilon \hat \phi \,, \\
\hat \phi & \to \hat \phi\,.
\end{align}
This leads to introduction of the following quantities
\begin{equation}
g_{\mu\nu} \equiv \hat g_{\mu\nu}-\hat A_{\mu}\hat A_{\nu}/\hat \phi\,, \hspace{5mm} A_{\mu} \equiv \hat A_{\mu}/\hat \phi\,,
\hspace{5mm} \phi \equiv \hat \phi\,.
\end{equation}
As one can easily check, these have correct transformation properties under the gauge transformation and enable us to identify them as a 
four-dimensional metric tensor, $U(1)$ gauge field and a real scalar field.

At the time of Kaluza, 
the scalar field $\phi$ was an unwanted feature, which was forced out of the picture.
In the present time, however, it bears the name \emph{radion} (or \emph{graviscalar} or also \emph{dilaton}) and it has a significance to 
modifications of Einstein theory, supergravity and string theories.

The Cylinder condition of Kaluza $\partial_{y}\,^5\! g_{MN} = 0$ was a main source of criticism of his otherwise stunning
discovery. Another obvious objection was the manifest lack of the fifth direction one supposedly should be able to see, if one takes the 
Kaluza idea literally. 
Kaluza himself, however, did not believe in 
actual existence of this fifth dimension, but considered it as a useful mathematical construction. Such a view, however, was changed 
few years later when Swedish physicist Oskar Klein \cite{Klein} presented an idea, that the invisibility of the fifth dimension is due to the fact that it is \emph{compact}, more precisely a circle of extremely small radius.
This construction not only settled most of the qualms but also gave an unexpected bonus: a quantization of charge.

If the fifth coordinate is periodic, the Cylinder condition is not necessary. To obtain a desired four-dimensional effective theory, one decomposes 
all fields into a series of its Fourier modes:
\begin{align}
g_{\mu\nu}(x,y) & = \sum\limits_{k= -\infty}^{\infty}g_{\mu\nu}^{(k)}(x)\Exp{-\iunit k y/R}\,, \\
A_{\mu}(x,y) & = \sum\limits_{k= -\infty}^{\infty}A_{\mu}^{(k)}(x)\Exp{-\iunit k y/R}\,, \\
\phi(x,y) & = \sum\limits_{k= -\infty}^{\infty}\phi^{(k)}(x)\Exp{-\iunit k y/R}\,,
\end{align}    
where $R$ is a radius of the circle.

It can be seen most easily from five-dimensional Klein-Gordon equation
\begin{equation}
(\partial^2-\partial_y^2)\phi(x,y) = 0 \hspace{5mm}\imply \hspace{5mm} \sum\limits_{k=-\infty}^{\infty}(\partial^2-m_k^2)\phi^{(k)}(x) = 0
\end{equation}
that all modes with $k\not = 0$ are massive with the mass $m_k^2 = k^2/R^2$. Only zero modes $k = 0$ remains massless. 
These zero modes are then identified with the known four-dimensional fields our universe is filled with (i.e. gravitons and photons).
The presence of an extra dimension is only visible 
through the infinite Kaluza-Klein (KK) tower of
massive modes. In addition, if the radius $R$ is sufficiently small, even the lightest of massive modes will be too heavy to 
affect the physics at our energy scales and thus
the whole theory is effectively four-dimensional.

Let us also point out, that spectrum of massive modes consists only of massive gravitons and not of massive photons nor scalars. 
The reason is that while massless graviton has only two 
polarizations, massive one has five. Similarly, massive gauge field must have three polarizations. In order to accommodate these
extra degrees of freedom, 
a graviton in the process very alike the Higgs mechanism eats a scalar field  and a gauge field to form a proper massive mode with 
five physical degrees of freedom. 

The quantization of the charge is a direct consequence of a periodic nature of the extra dimension. The charge is identified with the momentum
along the $y$ direction, which is a conserved quantity. The compact nature of this dimension then forces the momentum to be quantised as multiples
of $\sqrt{8\pi G^2}/R$. Upon identification of this with elementary charge $e$ one can estimate the radius of the extra dimension to be
$R \approx 10^{-33}$~cm. In other words, the Klein-Gordon picture demands that characteristic length scale of extra dimension is close to the Planck
length. If this is correct, there is really little hope that we would be ever able to probe such tiny distances by our experimental devices directly.  

\subsection{Compactification and string theory}

Despite the lack of direct phenomenological consequences of the Planck length sized extra dimensions, the idea was pursued further, mainly to 
explore its indirect consequences.  
In the 1920's it was Schr\"odinger, Gordon and Fock, while in 
the 1930's it was Einstein \cite{Einstein}, Mandel \cite{Mandel} and Pauli \cite{Pauli}. Most notably in 1953, a year before the introduction
of non-Abelian gauge symmetries into the quantum field theory by Yang and Mills, Pauli discovered all its essential features in the work 
based on Kaluza-Klein ideas where he compactified a sphere rather then a circle \cite{Straumann}. However, since he could not avoid having a 
massless vector meson in the spectrum, which was not observed experimentally,
 he decided not to publish his results.\footnote{Today, we base our model of electro-weak interaction on almost the same 
gauge group as in Pauli's work. However, massless vector mesons (gauge bosons in modern terminology) are not present in the spectrum 
due to the Higgs mechanism, where the symmetry is spontaneously broken and these particles become massive.}
 
After this, there was a long pause before the compact extra dimensions were revived in the string theory framework.
It started with the work of Scherk and Schwarz in 1972 \cite{Scherk}. They proposed a superstring model based on a product 
space of our four-dimensional space-time with six-dimensional compact manifold. This observation
lead to the discovery of Candelas \cite{Candelas} that, if the compact manifold is the so-called Calabi-Yau manifold, in the low energy 
limit one recovers $E(8)\times E(8)$ gauge group which contains the Standard Model gauge symmetry and three generations of chiral fermions.
Actually, this discovery (published in 1985) triggered what is usually referred to as the superstring revolution.
Since the typical size of a Calabi-Yau manifold in superstring theory is the Planck length, this picture can be regarded as a 
culmination of the Kaluza-Klein ideas.

Systematic research of superstring-based models of the SM type then started in 1990's. With the turn of the century,
however, the D-brane concept \cite{Polchinski} dramatically increased plausibility of such models \cite{Cremades, Cremades2}.
In these models, intersecting D-branes in low energy limit naturally provides the right structure of gauge interactions and 
give the correct number of quark/lepton families such that they are anomaly-free. It also gives Dirac neutrino masses and stable proton
quite naturally.
This development based on superstring/D-brane engineering, although very promising, does leave zero room for experimental observations of 
extra dimensions, being all confined to the Planck size. 

\section{Rubakov \& Shaposhnikov model}\label{sec-RS}
 
In 1983 two Russian physicists V.~A.~Rubakov and M.~E.~Shaposnikov asked a simple question, which 
also makes the title of 
their famous paper: ``Do we live inside a domain wall?''
Their three-pages long text \cite{Rubakov2} marked the beginning of new thinking about 
extra dimensions, 
nowadays commonly referred to as the \emph{brane world scenario}.\footnote{
It turned out, as so many times before, that Rubakov and Shaposhnikov were not the first one to propose a brane world scenario.
In 1982 Japanese physicist K. Akama published a paper \cite{Akama}, where he discuss a possibility that our world is an object called a
\emph{vortex}, a soliton of co-dimension 2, immersed in a six dimensional flat space-time. He then argues, using rather 
involved path-integral methods, that small fluctuations of the vortex background induce an Einsteinian gravity on the vortex's world-volume.
Although the approach is very different, the idea exactly match what we now perceive as brane world scenario, with its huge 
phenomenological implications.
This paper, however, was not widely known and so its origin is consistently connected with a year older paper 
of Rubakov and Shaposhnikov.}

Their idea is based on the following observation. Assuming there are additional dimensions than three, the phenomenology dictates that these extra dimensions must be hidden. One possible way how to do this is a compactification as in the Kaluza-Klein theory. Another possibility, as Rubakov and Shaposhnikov argue, relies on dynamics. In their scenario all particles are trapped inside a potential well, sufficiently narrow in the direction of extra dimensions and flat in ordinary  space directions. This means that propagation in extra dimensions is energetically unfavored resulting in effective three-dimensional appearance of our Universe. Notice, however, that particles are not forbidden to climb out of the well, provided they have enough energy. Such an event would seem to violate energy and momentum conservations from a point of view of observes inside the well. 

The origin of the well-like potential can be purely dynamical.  
In their work \cite{Rubakov2} Rubakov and Shaposhnikov illustrate this fact in a toy model, where the potential is realized by a background solitonic solution: a domain wall. As we are going to give all the details in Ch.~\ref{ch:2}, let us here only touch the most important points. 

Consider a five-dimensional theory defined by an action
\begin{equation}\label{ch1:action}
S =  \int \diff^5 x\, \oper{L} = \int \diff^5 x\, \Bigl[\frac{1}{2}\partial_{M}\phi \partial^{M}\phi 
- V(\phi)\Bigr]\,,
\end{equation}
where $M=0,1,2,3,y$ and where
\begin{equation}
V(\phi) = \frac{\lambda}{4}\Bigl(v^2-\phi^2\Bigr)^2\,.
\end{equation}
Notice that this model has two degenerate vacua $\phi = \pm v$. A domain wall solution is given by
\begin{equation}
\phi_{K}(y) = v \tanh\Bigl(v\sqrt{\frac{\lambda}{2}}(y-y_0)\Bigr)\,.
\end{equation} 
The non-trivial topology of this solution can be seen from the fact, that $\phi_{K}(y)$ approaches different vacuum value $\pm v$ as $y \to \pm \infty$.
Since both vacua are true vacua, it costs no energy
for a field to be near them. Therefore almost all energy of a domain wall  
is stored near the point $y_0$,
where the transition between vacua takes place. 
Since most of the energy is localized at the wall's position, it does act as a barrier, 
separating the space into two pieces.  

Let us investigate small fluctuations around this solution.
First we denote $\phi(x^{\mu},y) = \phi_K(y)+\varepsilon(x^{\mu},y)$, where $\mu = 0,1,2,3$ and
then we plug this into corresponding equations of motion. Retaining only linear terms in $\varepsilon$ we have
\begin{equation}
\partial_5^2 \varepsilon - \lambda\varepsilon\Bigl(v^2-3\phi_{K}^2\Bigr) = 0\,.
\end{equation}
To find normal modes we employ the ansatz $\varepsilon = b(y)\Exp{-\iunit
\vec k\vec x+\iunit \omega t}$. Putting this back into the above equation we obtain a 
Schr\"odinger-like eigen-value problem for the ``wave function'' $b(y)$
\begin{equation}
-\partial_y^2 b(y)+U(y)b(y) = (\omega^2-\vec{k}^2)b(y)\,,
\end{equation} 
where $U(y)$ is a potential given as
\begin{equation}
U(y) = \lambda v^2\frac{2\sinh^2\Bigl(v\sqrt{\frac{\lambda}{2}}(y-y_0)\Bigr)-1}
{\cosh^2\Bigl(v\sqrt{\frac{\lambda}{2}}(y-y_0)\Bigr)}\,.
\end{equation}
This is the well-like potential Rubakov and Shaposhnikov had in mind (see Fig.~\ref{ch2:fig02}).
In Ch.~\ref{ch:2} we provide explicit formulas for the entire spectrum of eigenmodes of this potential. We will also argue that a few lowest modes are trapped inside the well. These are identified as particles localized on the brane.

The above analysis, however, considers only scalar fields. Rubakov and Shaposhnikov also showed that fermionic field can be localized as well.
One simply adds following terms to the model \refer{ch1:action}
\begin{equation}
\iunit \bar\psi \slashed \partial \psi + h\bar\psi \psi \phi\,,
\end{equation}
which describes a Dirac spinor coupled to a scalar field $\phi$ via Yukawa term with the coupling constant $h$. The equation of motion for fermionic fluctuations $\psi(x,y)$ is given as
\begin{equation}
\iunit \slashed \partial \psi(x,y) + h \phi_{K}(y)\psi(x,y) = 0\,,
\end{equation}  
with the solution
\begin{equation}
\psi(x,y) = \Exp{-h\int^y \diff y^{\prime}\phi_K(y^{\prime})}\psi^{(0)}(x)\,,
\end{equation}
where $\psi^{(0)}(x)$ is a (3+1)-dimensional left handed massless spinor. Moreover, the profile $\psi(x,y)$ along the extra dimension shows that $\psi^{(0)}(x)$ is, indeed, localized on the domain wall.

Obviously, the toy model of Rubakov and Shaposhnikov cannot be considered as a realistic realization of the brane world scenario, as there is only single scalar and single fermion localized on a domain wall. One can, however, continue to improve the methods how to simultaneously localize more fields and how to make them interact. Indeed, developing such a ``domain-wall engineering'' is the aim of this thesis. 
The true value of the work of Rubakov and Shaposhnikov  lies in the ``existence proof'' that the brane world scenario can be realized within the framework of (quantum) field theory. In other words, they dispelled the Kaluza-Klein paradigm and opened a new line of investigation of extra-dimensional physics. 
It took, however, another 15 years before people realized the full power of the brane world scenario with respect to the SM phenomenology.
We will describe this shift in thinking in the next section. 

\section{Brane world scenario}

One of the yet unsolved problems of the SM is the gauge hierarchy problem. Simply speaking, 
this problem points to the stupefying mismatch between what we think should be a fundamental scale
 of the SM and what it actually is. The fundamental scale of the SM should be the Planck scale. 
Recall that the Planck scale is a unit independent energy scale obtained as a combination of
 fundamental constants of physics, namely the gravitational constant, speed of light and the 
 Planck constant $M_P = \sqrt{\tfrac{\hbar c^5}{G}} \sim 10^{19}$~GeV. This is the energy at 
 which effects of quantum gravity cannot be neglected and where the SM breaks down. 
 This statement is supported from analysis of running of coupling constants of all forces, 
 which shows that they meet (roughly) at $M_P$. The fact that the natural size of Higgs 
 dimensionful parameter, which controls masses of W and Z bosons and fermion masses, is 
 around the Weak scale $M_W\sim 10^3$~GeV, 17 order of magnitudes lower than our expectation 
 is the crux of the gauge hierarchy problem.

There is, however, a deep assumption at work here. While non-gravitational forces has been
 experimentally probed reasonably close to the Weak scale, precision measurement of gravity
  is, at best, approaching 1~MeV or a micrometer distances \cite{Poli}. Can we really be 
  confident that our low-energy concepts of how gravity works holds  down to the weak scale 
  distances $1\ \mathrm{TeV} \sim 10^{-19}$~m or even down to the very Planck length 
  $10^{-35}$~m? It might be the case that the Planck scale is only a low-energy illusion and 
  that the true fundamental scale of gravity is much lower.

This possibility is realized within the brane world scenario quite naturally. 
There are two typical models of the brane world scenario. 
One of them is proposed by Arkani-Hamed, Dimopoulos and Dvali (ADD) \cite{ADD}.
The mechanism in this model is the following.
One can argue 
that gravity, connected with the geometry of space-time, cannot be confined to the brane 
and  must ``leak out'' into the bulk. This will lower the apparent strength of gravitational force from a point of view of brane-bound observers, setting the effective four-dimensional Planck scale $M_P^{(4)}$ much higher than what the actual $(4+n)$-dimensional Planck scale $M_P^{(4+n)}$ is. 
This ``diluting effect'' of $n$ compact extra dimensions can be easily quantified. Let us assume that typical size of extra dimensions is $R$. Then, for distances $r\ll R$, the effective four-dimensional gravitational force may be written either in terms of effective Planck scale $M_P^{(4)}$ or in terms of actual Planck scale $M_P^{(4+n)}$ as  
\begin{equation}
F(r) \sim \frac{m_1 m_2}{(M_P^{(4)})^2}\frac{1}{r^2} = \frac{m_1m_2}{(M_P^{(4+n)})^{2+n}R^n}\frac{1}{r^2}\,.
\end{equation}
This is giving us the relation
\begin{equation}
(M_P^{(4)})^2 \sim (M_P^{(4+n)})^{2+n}R^n\,.
\end{equation}
If we set $M_P^{(4)} \sim 10^{19}$~GeV, then the radius of extra dimensions must be of the order
\begin{equation}
R\sim 10^{\tfrac{30}{n}-19}\,\mathrm{m}\times \biggl(\frac{1\ \mathrm{TeV}}{M_P^{(4+n)}}\biggr)^{1+\tfrac{2}{n}}\,.
\end{equation}  

In the ADD model,
most interesting case of $M_P^{(4+n)} = M_W$ is discussed, which simplifies the above formula to
\begin{equation}
R\sim 10^{\tfrac{30}{n}-19}\,\mathrm{m}\,.
\end{equation}
For $n=1$ we obtain $R\sim 10^{11}$~m implying deviations from Newtonian gravity over Solar system distances, which is clearly unacceptable. For $n=2$, however, we have $R\sim 10^{-4}$~m, which is experimentally accessible by today's instruments, promising an exiting new physics just behind the current experimental boundaries. Compared to Kaluza-Klein models, where $R$ is typically around the Planck length, we see that in ADD scenario, $R$ is quite large. For that reason, such a medium scale extra dimensions are called \emph{large extra dimensions} (LED), even though from a human point of view, they are still quite small.   

In the work \cite{ADD} the authors presented a particular
realization of $n=2$ scenario, by considering a six-dimensional theory with our Universe localized at the core of a vortex. We are not going to describe the details of this model, since the phenomenological consequences of LED idea are not sensitive to them. In fact, the same trio of authors later discussed the phenomenology  in the paper \cite{ADD2} and,  together with Antoniadis \cite{AADD}, they outlined a way of embedding the ADD model in superstring theory.

The ADD scenario, which can be seen as peculiar marriage of Kaluza-Klein ideas with the concept of brane-worlds, sparked a paradigm shift in thinking about the extra dimensions, known as the LED paradigm (a term, which we borrow from \cite{Shifman}). After ADD many other studies followed, giving rise to a fruitful new direction in the high-energy physics, which continues to the present day. The lasting popularity of LED-based theories is a result of a wish to have new physics at the Weak scale. Moreover, LED paradigm opens up possibilities to explain some of the properties of SM in a new way. Let us, as an example, mention the work of Dvali and Shifman \cite{Dvali}, where a hierarchy of fermion masses is explained without invoking broken family symmetries. Instead, Dvali and Shifman consider a case, where each generation lives on a separate brane with gauge fields propagating freely in the bulk. The Higgs fields are localized on another brane with an exponentially decaying profile in the bulk. In this way the closest brane has much larger overlap than the brane further away and even larger than the furthest one. Since the overlap of profiles in the bulk determines the interaction strength of fermions with Higgs field, this arrangement leads to a large hierarchy between masses in each generations, even though there is no hierarchy of distances between the branes.

Many old problems were revisited within the LED approach. Let us, again, pick up a few examples.
The issue of proton stability, for one, must have been completely rethought. The reason is that with strong gravity at $M_W$ scale, any global symmetry, such as baryon number, is violated by virtual black hole and wormhole production, leading to a rapid proton decay in LED-based theories. Many clever ideas how to circumvent this catastrophe were invented. In the original ADD model, the authors resolved this by invoking some new physics above $M_W$ scale, which 
suppresses these dangerous processes. More concrete possibility is based on an idea to turn the baryon number conservation into a discrete gauge symmetry \cite{Krauss}, which is protected by the gauge principle. The phenomenological implications of this method was shown to be compatible with experimental
data (see \cite{Nath} for a review). More geometrical approach to stabilize the proton was put forward by Arkani-Hamed and Schmaltz \cite{Arkani-Hamed}.
Their idea relies again on exponentially decaying profiles of localized particles on a brane. They considered a case of a thick wall with  quarks and leptons localized on different ends. Because of this, interaction terms between quarks and leptons, which are proportional to the overlap of their profiles, are exponentially suppressed, leading to the experimentally save lifetime of the proton.

The conventional approach to the issue of neutrino masses is based on the so-called \emph{seesaw} mechanism, which relies on the enormity of the fundamental scale of the SM.   This is an example of another problem, conventionally regarded as ``solved'', which requires a  reconsideration within LED paradigm.
Its possible solution was actually proposed by the authors of ADD scenario themselves \cite{ADDM}. Their idea was to introduce a right-handed neutrino, freely propagating in the bulk. The neutrino mass term is proportional to the overlap of the left-handed neutrino, confined to the brane, and the right handed neutrino (see \cite{Shifman} for detailed formula). Since the right-handed neutrino propagates in the entire bulk, its wave function is suppressed by a factor $V_n^{-1/2}$, where $V_n$ is the volume of extra dimensional space. This effect places the neutrino mass in the right ballpark. Thus, from a point of view of LED-based theories, light neutrinos and the weakness of gravity are both consequence of the same phenomenon: large extra dimensions.

The other model was proposed by Randall and Sundrum \cite{RS,RS2}, which is called a warped 
extra-dimensional model.
This is a five-dimensional model, where one extra-dimension is compactified 
on a ${S^1/\mathbb{Z}_2}$ orbifold and a negative cosmological constant is introduced in the bulk. 
Two D3-branes are placed at fixed points of the orbifold $\phi=0$ and $\phi=\pi$ 
($\phi$ is an angle of $S^1$) with opposite brane tensions. 
A brane at $\phi=0$ with a positive tension is called the hidden brane and the other one at 
$\phi=\pi$ 
with a negative tension is called the visible brane on which the SM fields localize. 
Solving the Einstein equation of this system, the five-dimensional bulk geometry is found to be
 a slice 
of anti-de Sitter (AdS) space, 
\begin{equation}
 \diff s^2=\Exp{-2 \kappa r_c |\phi|}\, \eta_{\mu\nu}\, \diff x^\mu \diff x^\nu - r_c^2 \diff\phi^2\,,
 \hspace{5mm} \eta_{\mu\nu}={\rm diag}(1,-1,-1,-1)\,,
 \label{geometry}
\end{equation}
where $\kappa$ is the AdS curvature in five dimensions and $r_c$ is a compactification radius. 
This background geometry allows us to take the Planck scale as a fundamental scale. 
Indeed, in the effective four-dimensional description an effective mass scale on the visible brane 
is warped down to $\Lambda_\pi = \bar{M}_{P}^{(5)}e^{-\pi\kappa r_c}$ due to the effect of the warped geometry, where
$\bar{M}_{P}^{(5)}$ is the reduced Planck scale in the five-dimensional space-time.
Therefore, with a mild parameter tuning, $\kappa r_c\simeq 12$, we can realize
 $\Lambda_\pi = {\cal O}$(1 TeV) and obtain 
a natural solution to the gauge hierarchy problem.

Phenomenological aspects of the warped extra-dimensional model have been vigorously 
investigated, as was performed in the ADD model.
Apart from that, the warped extra-dimensional model has an interesting connection to the
 AdS/CFT correspondence 
\cite{Maldacena,Witten:1998qj,Gubser:1998bc}.
This correspondence tells us that four-dimensional theories with strong interactions is
 related to weak-interacting five-dimensional 
supergravity theories. In other words, one can understand the properties of five-dimensional 
fields as those of four-dimensional composite 
state \cite{ArkaniHamed:2000ds}.
This has opened up new directions for tackling questions in particle physics such as the 
flavor problem, grand unification
and the origin of the electro-weak symmetry breaking or supersymmetry breaking.

\section{Purpose of this thesis}

The brane world scenario explained briefly above has a lot of interesting theoretical and
 phenomenological features, but there is a big assumption 
that the SM fields are localized on the three-dimensional hyper-surface, a brane. As mentioned
 above, such a situation is naturally realized in the string
theory. One of the promising candidates of the fundamental theory is the superstring theory, where 
the number of space-time dimensions is ten. 
Its low-energy effective theory is the so-called supergravity that is described as the 
ten-dimensional field theory. If the SM is realized as a low energy 
effective theory of the supergravity/the string theory, it is necessary to make a gap of the 
dimensions between the SM and the supergravity. One of
ways is to compactify the space-time dimensions, but it is usually performed by hand and 
therefore it is not dynamical.
The other way is to use a topological soliton, where fields naturally localize around that, 
as explained in subsection \ref{sec-RS}.

However, there is an obstacle that the mechanism of Rubakov and Shaposhnikov works only for 
bosons and fermions.
In order to realize the SM model on a topological solitons, it is necessary to localize not 
only bosons and fermions, but also massless gauge fields
associated with the SM gauge groups $SU(3)\times SU(2)\times U(1)$. Naive application of 
the mechanism by Rubakov and Shaposhnikov to
bulk gauge fields tells us that gauge fields localize around a topological solution, but 
inevitably become massive, leading to the violation of the 
four-dimensional Coulomb's law (details will be discussed in Ch.~\ref{ch:4}). It has been a 
long-standing problem in construction of a brane world 
scenario in the field theory framework. However, recently, a very simple mechanism to localize 
massless gauge fields has been proposed by 
Ohta and Sakai \cite{Otha}. 
The key ingredient of this mechanism is a position-dependent gauge coupling that is a coefficient 
of the gauge kinetic term. Consider the
five-dimensional space-time. If the position-dependent gauge coupling has a profile such that  it 
goes zero at infinities of extra dimension and has
a peak around a topological soliton, it can be shown that the massless gauge field localizes and that
 the four-dimensional Coulomb's law holds.
It is a great step towards a construction of a realistic brane world scenario. 

For a realistic model building, we further need to proceed along this line. In this thesis, 
we will propose two five-dimensional models where gauge fields 
and charged matters localize on a domain wall solution. The first model, introduced in 
Ch.~\ref{ch:5}, successfully 
makes matters and gauge fields localize, but it has an
instability that the position-dependent gauge coupling can be negative. The second model of Ch.~\ref{ch:6}
improves this shortcomings. We will also derive the low energy effective actions on a domain wall, 
based on a technique called the moduli approximation \cite{Manton} (we discuss it in Ch.~\ref{ch:4}). 
Resultant actions are similar
 to the 
chiral Lagrangian being the effective theory of QCD.


\chapter{Brane world scenario and solitons} \label{ch:2} 


The key ingredient  how to realize the brane world scenario in the field-theoretical framework is a topological soliton.
In order to illustrate this fact, in this chapter we discuss in detail a simple model containing a 3-brane. This is a (4+1)-dimensional theory
with single scalar field and a quartic potential with degenerate vacua. Non-trivial vacuum structure allows us to construct a 
solitonic solution, varying only in one dimension, which we identify as a brane. 
Then, studying  small fluctuations on the background of this solution,  we discover that in the 
direction of extra-dimensional coordinate the effective potential has a well-like shape, centered at the position of the brane. 
Inevitably, some modes of the fluctuations will be trapped
in this well and we speak of them as of particles localized on the brane.
The culmination of our discussion will be the construction of the effective Lagrangian describing dynamics of these trapped modes on 
(3+1)-dimensional world-volume of the brane. 

\section{A brane-like solution}

Let us consider a theory defined by an action
\begin{equation}\label{ch2:action}
S =  \int \diff^5 x\, \oper{L} = \int \diff^5 x\, \Bigl[\frac{1}{2}\partial_{M}\phi \partial^{M}\phi 
- V(\phi)\Bigr]\,,
\end{equation}
where $\phi$ is a scalar field and $M=0,1,2,3,y$ is a coordinate index of (4+1)-dimensional
spacetime with mostly negative signature (+,-,-,-,-). 
In such theories, the question of existence and properties of solitonic solutions is determined by 
the topological structure of the so-called vacuum manifold.
The vacuum manifold is simply a set containing all homogeneous (i.e. space-time independent) 
solutions to a equation of motion (EoM) with zero energy $E=0$,\footnote{We will always assume that the potential has at least single global minimum 
with energy $E=0$.} where:
\begin{equation}\label{ch2:energy}
E =   \int \diff^4 x\, \oper{E} =  \int \diff^4 x\, \Bigl[\frac{1}{2}\dot\phi^2+\frac{1}{2}
\vec\nabla\phi^2 + V(\phi)\Bigr]\,.
\end{equation}
These solutions are called vacua of the theory. Since they are space-time independent, they 
correspond to the global minima of the potential.
If the potential has a single minimum, such as the potential of the $\phi^4$ theory
\begin{equation}
V(\phi) = \frac{1}{2}m^2\phi^2+\frac{\lambda}{4!}\phi^4\,,
\end{equation}
there is only a single solution to EoM with $E=0$ (as one can always arrange things, by adding 
physically unimportant constant to the Lagrangian, so
that this condition is fulfilled), which is $\phi = 0$. In this case the vacuum manifold
 $\oper{V}_1$ contains just a single point and it is, therefore, topologically trivial
  $\oper{V}_1\sim \{0\}$.
In consequence, no topological solitons exist.

However, in the case of a potential 
\begin{equation}\label{ch2:hat}
V(\phi) = \frac{\lambda}{4}\Bigl(v^2-\phi^2\Bigr)^2\,,
\end{equation}
there are two  vacua $\phi = \pm v$, giving topologically nontrivial vacuum manifold $\oper{V}_2$.  
It may seem strange to think about a set of two points as a manifold, but geometrically speaking, $\oper{V}_2$ is equivalent
 to a zero-dimensional circle $S^0$ of a radius $v$. 
At this point, it is hardly necessary to plunge ourselves into technical details of the homotopy theory 
to see, that set of two points is (topologically) 
different than a
set containing just one. This topological non-triviality is usually a strong signal that solitonic 
solutions may exist.\footnote{It is not a priory guaranteed, however,
whether solitonic solutions are physical.
For example, a topologically non-trivial structure of the so-called $O(3)$ non-linear sigma model gives rise to 
a solitonic solution called
a \emph{lump}. Upon inspection of its dynamics, it turns out that this solution has a scaling 
instability, meaning that it can evolve into a constant zero or 
a delta peak in a finite time. This makes it unphysical (see \cite{Manton} for details).} And,
 indeed, in this case there exists a well-known solitonic solution, namely
\begin{equation}\label{ch2:kink}
\phi_{K}(y) = v \tanh\Bigl(v\sqrt{\frac{\lambda}{2}}(y-y_0)\Bigr)\,,
\end{equation} 
What makes this solution topological? In general, any physically acceptable solution must approach 
a vacuum value at a spatial infinity sufficiently fast,
otherwise it would have an infinite energy. A non-topological solution  approaches 
the same vacuum everywhere, which makes it 
diffeomorphic to it. In other words, it can be transformed to a constant by smooth deformations only.
On the other hand, topological solution takes different values in vacuum manifold
 as we go along spatial infinity and it is, by 
definition, impossible to deform it to a constant (vacuum).  

\begin{figure}
\begin{center}
\includegraphics[width=0.7\textwidth]{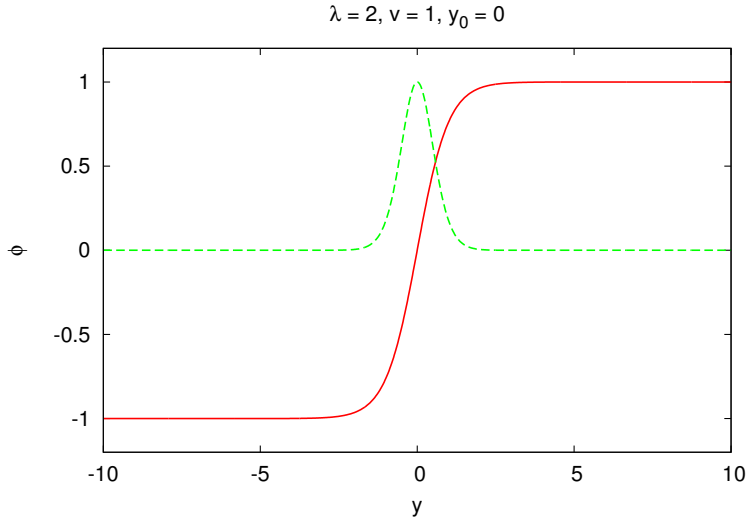}
\caption[Profiles of the kink solution and its energy density.]{\small Profiles of the kink solution \refer{ch2:kink} (solid red line) and of its energy
density (green dashed line).}
\label{ch2:fig01}
\end{center}
\end{figure}

Indeed, as $y\to -\infty$ function $\phi_{K}(y)$ tends to $-v$ and to $v$ in the opposite infinity 
$y\to \infty$. Therefore, as far as we concentrate only
on the spatial infinity in the $y$-direction, the solution $\phi_{K}(y)$ is topological. 
The solution \refer{ch2:kink} is known as \emph{kink} or, if it is embedded in higher-dimensional
 model like in our case, a \emph{domain wall}.
The reason to call $\phi_K$ a ``kink'' its clearly identifiable from the Fig.~\ref{ch2:fig01} (red solid line), 
which shows a rapid transition around a point $y_0$ and otherwise 
nearly constant profile. The reason to call it a (domain) wall
is connected with the fact, that in other coordinates $\phi_{K}(y)$ is uniform. Moreover,
 if we plot its energy density (Fig.~\ref{ch2:fig01} green dashed line) we
observe that it is clearly localized around $y_0$, where the transition takes place. 
This promotes the intuition that a kink is indeed a wall. It creates a barrier, which for
 example, affects incoming radiation, etc.

Another important fact about $\phi_{K}$ is its stability. Once a soliton exists, it is
 very hard to destroy, for example by
some external forces. In fact, within the formal boundaries of our model, it is impossible. 
One can quantify this by introducing a conserved charge,
carried by solitons, called a topological charge:
\begin{equation}
Q= \frac{1}{2v}\lineint y\, \partial_y \phi = \frac{\phi(\infty)-\phi(-\infty)}{2v}\,,      
\end{equation}
where the normalization has been chosen so that $Q(\phi_K) = 1$. Due to the $\mathbb{Z}_2$ 
symmetry $\phi\to -\phi$ of our theory, there also 
exists an \emph{anti}-kink solution $\phi_{\bar K} = -\phi_{K}$ with the corresponding charge 
$Q(\phi_{\bar K}) = -1$. This solution behaves in many ways
as an anti-particle to a kink. For example, a gentle collision of kink and anti-kink results in their 
 annihilation into radiation.\footnote{This is not, however, the only possibility. In general, the outcome of a collision between kink and anti-kink depends very sensitively on their initial velocity (in the center-of-mass frame). Numerical investigations of
kink and anti-kink collisions in (1+1)-dimensions revealed delicate pattern of resonances, which are formed after the initial annihilation, and which then decay back into kink-anti-kink pair. See i.e. \cite{Goodman} for details.}   

One can easily construct associated (two-dimensional) conserved current 
$j^{\mu} = \varepsilon^{\mu\nu}\partial_{\nu}\phi$, where $\mu,\nu \in \{0,y\}$ and 
$\varepsilon^{\mu\nu}$ is a completely antisymmetric symbol with $\varepsilon^{0y}=1$. 
The  curious feature of this current is that unlike the Noether current, which is connected with
some continuous symmetry of the theory, $j^{\mu}$
is conserved identically, without a need of EoM:
\begin{equation}
\partial_{\mu}j^{\mu} = \varepsilon^{\mu\nu}\partial_{\mu}\partial_{\nu}\phi = 0\,.
\end{equation}

Given its wall-like character and its stability, ensured by topological properties, a kink
 provides an excellent field-theoretical
model of a brane.

\section{BPS argument}

The solution \refer{ch2:kink} has strictly speaking infinite energy with respect to all dimensions,
 but if restricted only to $y$-coordinate, we obtain a finite number, 
called a tension
\begin{equation}
T_{kink} = \lineint y\, \oper{E}(\phi_K) = \frac{2\sqrt{2\lambda}}{3}v^3\,.
\end{equation}
It turns out, that $T_{kink}$ is a minimal amount of tension for \emph{any} 
static solution of EoM with topological charge $Q=1$. 
Let us prove this assertion. In doing so we will employ a famous trick due to Bogoml'nyi,
 which will also lead us to the solution to the question we
have been avoiding so far:
How one finds a soliton solution such as $\phi_K$?

Let us write down a tension of a static field $\phi(y)$
\begin{equation}
T = \lineint y\, \Bigl[\frac{1}{2}{\phi^{\prime}}^2+V(\phi)\Bigr]\,.
\end{equation} 
Here $\phi^{\prime}$ denotes $y$-derivative. We can rewrite this expression as follows
\begin{align}
 T & = \lineint y\, \frac{1}{2}\Bigl[\phi^{\prime}-\sqrt{2V(\phi)}\Bigr]^2+\lineint y\, 
 \sqrt{2V(\phi)}\phi^{\prime}\,, \label{ch2:bb1} \\
& \Downarrow \nonumber \\
T & \geq  \lineint y\, \sqrt{2V(\phi)}\phi^{\prime}\,. \label{ch2:bb2}
\end{align}
Since the first term in Eq.~\refer{ch2:bb1} is always positive, one can create a lower bound on the 
tension $T$, called after its inventor \emph{the Bogomol'nyi bound}, which 
is Eq.~\refer{ch2:bb2}. Furthermore, if the potential $V(\phi)$ can be recast as 
\begin{equation}\label{ch2:supot}
V(\phi) = \frac{1}{2}\Bigl(\frac{\diff \oper{W}}{\diff \phi}\Bigr)^2\,,
\end{equation}
this bound is simply
\begin{equation}  \label{ch2:bb3}
T \geq \oper{W}(\phi_{\infty})-\oper{W}(\phi_{-\infty})\,,
\end{equation}
where $\phi_{\infty}$ and $\phi_{-\infty}$ are values of $\phi$ at respective infinities.

The topological nature of this bound is clear. If the field $\phi$ interpolates between same
 vacuum values we have $\phi_{\infty}=\phi_{-\infty}$ and 
$T\geq 0$. For topological solutions, however, there is a minimal value of a tension they 
must have. In the case of the potential \refer{ch2:hat} one
finds, up to an irrelevant integration constant
\begin{equation}\label{ch2:supp}
\oper{W}(\phi) = \frac{\sqrt{\lambda}}{\sqrt{2}}\phi \Bigl(v^2-\frac{1}{3}\phi^2\Bigr)\,.
\end{equation}
Using the fact that a topological solution (up to $\mathbb{Z}_2$ transformation) 
$\phi_{\pm \infty} = \pm v$, we obtain for the bound \refer{ch2:bb3}
\begin{equation}
T \geq \frac{2\sqrt{2\lambda}}{3}v^3 = T_{kink}\,.
\end{equation}
This completes the proof that any static solution of EoM with $Q=1$ (or equally as well as $Q=-1$)
must have the tension at least $T_{kink}$. It also provides us with a
tool how to construct $\phi_K$, since from Eq. \refer{ch2:bb1} we see that the bound will be 
saturated ($T=T_{kink}$) if the following first order ordinary differential equation (ODE) is fulfilled:
\begin{align}\label{ch2:bps}
\phi^{\prime} -\sqrt{2V(\phi)} &= 0\,, \\
& \Downarrow \nonumber \\ \label{ch2:bps2}
\phi^{\prime}-\frac{\sqrt{\lambda}}{2}(v^2-\phi^2) & = 0\,. 
\end{align} 
Eq.~\refer{ch2:bps} is called \emph{Bogomol'nyi-Prasad-Sommerfield} (BPS) equation. One can check
 that, by taking its derivative, BPS equation is 
fully compatible with EoM,  meaning that any solution to BPS equation is also a solution to EoM.
 But there is also an indirect way how to see this. By saturating the bound we have clearly found a field configuration in the minimum of the tension
 functional $T[\phi]$. However, since for static fields a minimum of a tension must also be a 
 minimum of an action, solution to BPS equation must also be a solution to EoM.

It is a straightforward exercise to check that integration of Eq.~\refer{ch2:bps2} leads to the solution
 $\phi_{k}$ of Eq.~\refer{ch2:kink}. Solutions of BPS equation are called BPS solitons. We will have more to say on BPS
solitons in the chapter \ref{ch:3}, where we establish important connection between BPS solitons and supersymmetry.

\section{Trapped modes}

In this subsection we will show that the domain wall solution $\phi_K$ can be regarded as a field-theoretical model of a brane. This can be established by investigating a spectrum of fluctuations of $\phi_K$ and showing that few lowest-lying modes are trapped in a well-like potential along the extra dimension centered around the position of the wall. Thus, these trapped modes can freely propagate only in the four-dimensional world-volume of the wall, functionally playing a role of particles living on the brane.
 
\begin{figure}
\begin{center}
\includegraphics[width=0.7\textwidth]{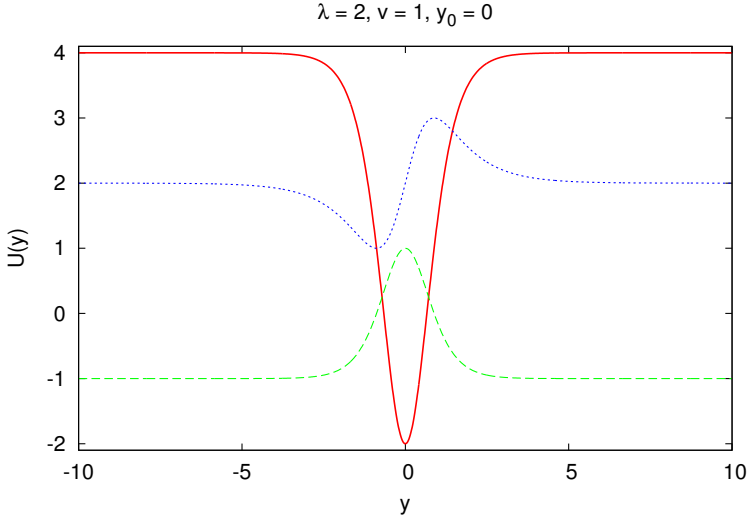}
\caption[Profile of the potential together with unnormalized profiles of a zero mode and a massive mode.]{\small Profile of the potential $U(y)$ (red solid line) together with unnormalized profiles
of the zero mode $2b_0(y)-1$ (green dashed line) and the massive mode $2b_1(y)+2$ (blue dotted line).}
\label{ch2:fig02}
\end{center}
\end{figure}

Let us denote $\phi(x^{\mu},y) = \phi_K(y)+\varepsilon(x^{\mu},y)$, where $\mu = 0,1,2,3$ and
$\abs{\varepsilon(x^{\mu},y)}\ll v$. 
Upon pluging this into EoM and
retaining only linear terms in $\varepsilon$ we get
\begin{equation}
\partial_5^2 \varepsilon - \lambda\varepsilon\Bigl(v^2-3\phi_{K}^2\Bigr) = 0\,.
\end{equation}
To find normal modes we employ an ansatz $\varepsilon(x^{\mu},y) = b(y)\Exp{-\iunit
\vec k\vec x+\iunit \omega t}$ and plug it back into the previous equation to obtain a 
Schr\"odinger-like eigenvalue problem for a ``wave function'' $b(y)$
\begin{equation}\label{ch2:stabil}
-\partial_y^2 b(y)+U(y)b(y) = (\omega^2-\vec{k}^2)b(y)\,,
\end{equation} 
where $U(y)$ is a potential given as
\begin{equation}
U(y) = \lambda v^2\frac{2\sinh^2\Bigl(v\sqrt{\frac{\lambda}{2}}(y-y_0)\Bigr)-1}
{\cosh^2\Bigl(v\sqrt{\frac{\lambda}{2}}(y-y_0)\Bigr)}\,.
\end{equation}
One can easily check that the so-called \emph{zero mode}\footnote{We adjusted overall constant 
factor so that $b_0$ is properly normalized: 
\begin{equation*}
\lineint y\, b_0^2(y) = 1\,.
\end{equation*}}
\begin{equation}\label{ch2:zeromode}
b_0(y) \sim \frac{\diff}{\diff y}\phi_{K}(y) = 3^{1/4}\frac{\sqrt{m_1}}{2\cosh^2\Bigl(v\sqrt{\frac{\lambda}{2}}(y-y_0)\Bigr)}\,,
\end{equation}
is a solution to the eigenvalue-problem \refer{ch2:stabil} with $\omega_0^2 = \vec{k}^2$ and
$m_1^2 = 3\lambda v^2/2$. Here ``zero mode''
simply means zero mass. Indeed, this solution corresponds to a freely propagating wave in 
(3+1)-dimensions with a usual dispersion relation; the factor $b_0(y)$ determines
spreading of this wave into the extra dimension. From the Fig.~\ref{ch2:fig02} (green 
dashed line) we can clearly see, that the 
zero mode is localized very closely to the domain-wall's position.

The existence of zero modes is actually dictated by symmetry. Background solution
$\phi_K$, although invariant under translations in four dimensions, spontaneously
breaks translation invariance in the $y$-direction. According to the Goldstone theorem, for
every spontaneously broken symmetry generator, there must be a massless particle in the spectrum.
 Hence a particle associated with the zero mode
is often called a Nambu-Goldostone (NG) boson.

We can easily see that our zero mode \refer{ch2:zeromode} is indeed connected with the breaking of 
the translational invariance in the
$y$-direction. The quantity, 
parameterizing this breaking is, of course, the position of the wall $y_0$. An infinitesimal shift 
$y_0 \to y_0-\delta y_0$ gives
\begin{equation}
\phi_K(y_0-\delta y_0) = \phi_K(y_0)+b_0(y)\delta y_0 + O\Bigl((\delta y_0)^2\Bigr)\,.
\end{equation}
Therefore, since
$\phi_K$ is a solution to EoM for all values of $y_0$, the zero mode \refer{ch2:zeromode} must be a
solution to \refer{ch2:stabil}.

Another solution to the eigenvalue-problem \refer{ch2:stabil} is a \emph{massive mode}
\begin{equation}\label{ch2:massivemode}
b_1(y) =3^{1/4}\sqrt{\frac{m_1}{2}}\frac{\sinh\Bigl(v\sqrt{\frac{\lambda}{2}}(y-y_0)\Bigr)}
{\cosh^2\Bigl(v\sqrt{\frac{\lambda}{2}}(y-y_0)\Bigr)}\,,
\end{equation}
with $\omega_1^2 = \vec{k}^2+\frac{3}{2}\lambda v^2$, which corresponds to a particle with mass 
$m_1^2 = 3\lambda v^2/2$. 
From Fig.~\refer{ch2:fig02} (blue dotted line) we see that, although somewhat wider than in the 
case of the zero mode, profile of the massive mode is 
definitely localized around the wall's position.

Being massive a massive mode is hard to excite. If one is interested only in low-energy processes 
it is a common practice to discard all massive
modes from the picture, in the same spirit as one neglects the existence of massive Kaluza-Klein (KK)
 modes in the KK theory. Unlike the KK scenario, however,
in our case there is not an infinite tower of massive excitations. The massive mode $b_1(y)$ is 
the only one in the eigenvalue-problem
\refer{ch2:stabil} (as the number of bound states depends very non-trivially on the form of the 
potential $U(y)$).

When the frequency reaches a threshold $\omega_{\infty}^2 = 2\lambda v^2$ the spectrum becomes 
continuous and 
corresponding particles are no longer localized around the position of the wall. 
Surprisingly the ``wave function'' of such an unbound state can be found
explicitly:\footnote{The wave function $b_q(y)$, however, is not a normalizable state and in
 practice one should consider some appropriate wave packet of
these states.}
\begin{equation}
b_q(y) = \Exp{\iunit q y}\Bigl(3\tanh^2y-1-q^2-3\iunit q \tanh y\Bigr)\,,
\end{equation}
The corresponding energy is   $\omega_{q} = \bigl(2+q^2/2\bigr)\lambda v^2$.
This type of excitation is  a ``radiation'', which 
can escape from the brane into the bulk or other way around. 
From the point of view of a brane-bound
observer an excitation of these particles corresponds to processes with missing energy and momentum.

\section{Effective theory on a domain-wall}

Investigation of small fluctuations around the domain wall solution revealed two modes
 trapped on its world-volume. We can capture the effective 
(3+1)-dimensional dynamics of
these modes (interpreted as particles living on the brane) by constructing an effective 
Lagrangian. 
Let us again employ the following ansatz 
\begin{equation}
\phi(x^{\mu},y) = \phi_K(y) +\varepsilon(x^{\mu},y)\,.
\end{equation}  
Let us decompose the general fluctuation field $ \varepsilon(x^{\mu},y)$ into the trapped 
normal modes:
\begin{equation}\label{ch2:decomp}
 \varepsilon(x^{\mu},y) = b_0(y) u_0(x^{\mu})+b_1(y) u_1(x^{\mu})\,,
\end{equation}
with $b_0(y)$ and $b_1(y)$ given in Eqs.~\refer{ch2:zeromode} and \refer{ch2:massivemode}.
Now we plug this into the Lagrangian and integrate the result over the extra-dimensional coordinate $y$
\begin{equation}
\oper{L}_{\mathrm{eff}} = \lineint y\, \Bigl[\frac{1}{2}\partial_{\mu}\phi\partial^{\mu}\phi 
-\frac{1}{2}(\partial_y \phi)^2-\frac{\lambda}{4}(v^2-\phi^2)^2\Bigr] = \oper{L}_{\mathrm{eff}}^{(0)}+\oper{L}_{\mathrm{eff}}^{(2)}+O(\varepsilon^3)\,,
\end{equation}
where
\begin{align}
\oper{L}_{\mathrm{eff}}^{(0)}  & = \lineint y\, \Bigl[-\frac{1}{2}(\partial_y \phi_K)^2-
V(\phi_K)\Bigr]\,, \\
\oper{L}_{\mathrm{eff}}^{(2)} &= \lineint y\,\Bigl[\frac{1}{2}\partial_{\mu}\varepsilon
\partial^{\mu}\varepsilon -\frac{1}{2}(\partial_y \varepsilon)^2-\frac{1}{2}V^{\prime\prime}
(\phi_K)\varepsilon^2\Bigr]\,. 
\end{align} 
Notice that first order contributions from the kinetic term and the potential term exactly 
cancel each other, as it should be since the background solution $\phi_K$ lies in the minimum
 of the action.

The first term $\oper{L}_{\mathrm{eff}}^{(0)}$ can be easily evaluated as
\begin{equation}
\oper{L}_{\mathrm{eff}}^{(0)} = -\frac{2\sqrt{2\lambda}}{3}v^3 = -T_{kink}\,,
\end{equation}
which is (classically) just unimportant constant equivalent to minus the tension of the kink.
In the second term, using the decomposition \refer{ch2:decomp}, equation of motion for the 
normal modes and the fact that they are orthonormal
\begin{equation}
\lineint y\, b_i(y)b_j(y) = \delta_{i,j}\,, \hspace{5mm} i,j = 0,1\,,
\end{equation}
we obtain
\begin{equation}\label{ch2:eff}
\oper{L}_{\mathrm{eff}}^{(2)} = \frac{1}{2}\partial_{\mu}u_0\partial^{\mu}u_0+\frac{1}{2}\partial_{\mu}u_1\partial^{\mu}u_1-\frac{1}{2}m_1^2u_1^2\,.
\end{equation}
Thus, we found that at the lowest order of approximation, the effective Lagrangian
 contains two noninteracting scalar fields, one of which is massive, with mass
$m_1^2 = \tfrac{3}{2}\lambda v^2$.  This result is in accordance with our interpretation of trapped 
modes as particles living on a world-volume of a domain wall. 
Up to this point, however, we ignored continuous modes. Such approximation is correct if the 
total energy of a massive mode does not exceed $2\lambda v^2$. If so, radiation would carry 
the excess away from the brane. Notice, however, that similar discussion does not apply to a zero
mode, which can have an arbitrary velocity (lower than speed of light). This follows from the fact, 
that a wall with an exited zero mode can be interpreted as having a non-zero speed in the $y$-direction.
Due to the Lorentz invariance, this should be physically equivalent to a static wall. Therefore, no radiation
should occur.

We could
 go on and calculate remaining higher order contributions, namely $\oper{L}_{\mathrm{eff}}^{(3)}$
  and $\oper{L}_{\mathrm{eff}}^{(4)}$. These turn out to be
\begin{align}
\oper{L}_{\mathrm{eff}}^{(3)} & =-\frac{1}{3!}\lineint y\, V^{\prime\prime\prime}
(\phi_K)\varepsilon^3 =  -\frac{\lambda\pi 3 3^{1/4}\sqrt{m_1}}{64\sqrt{2}} \bigl(3u_0^2u_1+2u_1^3\bigr)\,, \\
\oper{L}_{\mathrm{eff}}^{(4)} & = -\frac{1}{4!}\lineint y\, V^{\prime\prime\prime\prime}
(\phi_K)\varepsilon^4 = -\frac{\lambda 3 \sqrt{3}m_1}{70} \Bigl(u_0^4+2u_0^2u_1^2+
\frac{1}{2}u_1^4\Bigr)\,.
\end{align} 
No higher corrections exists. Whether we should take these corrections seriously, however,
is a highly non-trivial question. 

Apart from the similar issue with continuum modes, there are deeper
reasons to be cautious. In Eq.~\refer{ch2:decomp} we have artificially restricted a
space of all possible field configurations to a very small subspace. 
But it is not clear whether such a subspace contains all typical field configurations, which 
emerges if we turn on the full dynamics. And indeed, since we ignored continuous modes, we know that for 
sufficiently high energy, it does not. On the other hand, 
if there is a reason to believe that the field stays close to \refer{ch2:decomp} 
in the configuration space, interaction terms in $\oper{L}_{\mathrm{eff}}^{(3)}$ and 
$\oper{L}_{\mathrm{eff}}^{(4)}$ should be reliable.
This has to be checked a posteriori, however, for example by numerically solving the
 exact EoM in an energy range of interest.
If the energy of fluctuations is small, the intuition supports the view, that $u_0$ and $u_1$ should 
be free fields, so the leading effective Lagrangian $\oper{L}_{\mathrm{eff}}^{(2)}$ seems 
to be most reliable.

In conclusion,  for low energy of fluctuations the effective Lagrangian \refer{ch2:eff} is our best bet. Moreover, in the case at hand, we can reasonably 
assume that the mass of massive mode $m_1$ is very high and, therefore, we can ignore the field $u_1$ altogether. All we are left with, then, is a free zero mode
fluctuation propagating along the wall.

As we will see, in a more complicated theories it is very laborious to derive an effective Lagrangian in the direct-fashion way we employed here. Moreover, 
due to reasons similar to  those we have just discussed, it is not guaranteed that such a 
direct approach yields correct higher order corrections, as the question 
of beyond leading terms is notoriously difficult. For this reason in this entire text we will 
concentrate on leading order terms only. In fact, we will
be satisfied with an effective description of dynamics of zero modes. Fortunately, there is a very
 beautiful and straightforward way, how to obtain it.  As we will rely on this method, 
 called the \emph{moduli approximation} invented by Manton \cite{Manton}, quite heavily in the
 subsequent discussions, we will briefly explain it below.

Let us first stress once more, that existence of zero modes in the spectrum is closely connected
 with the symmetry breaking by the background solution.
Geometrically, zero modes corresponds to flat directions in the configuration space, 
meaning that deformation of the field in these directions costs no energy. Therefore, an associated
 mode of fluctuation (particle) is massless.
Interestingly, what Manton found is that a low-energy dynamics of zero modes can be 
understood in geometrical terms too.  
It turns out, that leading order effective Lagrangian is generally given as
 a sum of kinetic terms of zero modes, i.e.
\begin{equation}
\oper{L}_{\mathrm{eff}} = \frac{1}{2}g_{ij}(u_0)\partial_{\mu}u_0^{i}\partial^{\mu}u_0^j\,,
\end{equation} 
where $i,j$ runs from all types of zero modes $u_0^i$ and where $g_{ij}(u_0)$ is a metric tensor.
In other words, low energy dynamics of zero modes is equivalent to geodetic motion in a curved
 space, called a \emph{moduli space}.
It is even possible to write down an explicit formula for the metric $g_{ij}$. Geometrically, 
if the starting theory has a global symmetry group $G$ and the background
solution break it down to some subgroup $H$, the zero modes should take values in the coset 
space $G/H$. If $H$ is a normal subgroup, then
the coset $G/H$ itself is a group and its continuous part corresponds to a manifold. Since
breaking of discrete symmetries does not lead to  
NG particles, it is only this continuous part of $G/H$ which we identify with a 
moduli space. The metric $g_{ij}$ should be then a metric
of a continuous part of $G/H$.

The formula itself can be written down quite easily, using the following observation. Since the 
background solution breaks a part of the global symmetry 
of the theory, it must contains a number of parameters (such as $y_0$ in \refer{ch2:kink}) called \emph{moduli}, which describe the breaking.
Let us denote all moduli in our generic example as  $y_i$, where $i = 1,\ldots , \mathrm{dim}(G/H)$. The metric is then given as
(up to an overall factor)
\begin{equation}
g_{ij} \sim \lineint y\, \abs{\frac{\partial \phi_{b}}{\partial y_i}\frac{\partial \phi_{b}}{\partial y_j}}\,,
\end{equation} 
where $\phi_b$ is a background solution.
Going back to the case of a domain wall solution \refer{ch2:kink} we have a single molulus $y_0$ and a metric is just number
\begin{equation}
g_{11} \sim \lineint y\, \abs{\frac{\partial \phi_{K}}{\partial y_0}\frac{\partial \phi_{K}}{\partial y_0}} = T_{kink}\,,
\end{equation}
which can be absorbed into the normalization of the field $u_0$.

Putting geometry aside, Manton's method relies on a intuitively obvious picture, that low-energy 
fluctuations of the background solution are deformations of its moduli parameters. 
In other worlds, it is possible to derive effective Lagrangian of zero modes just by promoting
moduli to fields $y_i \to y_i(x^{\mu})$ on the soliton's world-volume.
Let us demonstrate this in our toy model.

As the first step we promote the only modulus in the background solution \refer{ch2:kink} to a four-dimensional field $y_0 \to y_0(x^{\mu})$. 
Now we plug this promoted 
solution back to the Lagrangian and integrate the result over extra-dimensional coordinate $y$. For the moduli approximation, however, it is sufficient just
to consider kinetic term, while the potential term is dropped:
\begin{align}
\oper{L}_{\mathrm{eff}} & = \lineint y\, \frac{1}{2}\partial_{\mu}\phi_{K}(y,y_0(x))\partial^{\mu}\phi_K(y,y_0(x)) \nonumber \\
& = \lineint y\, \frac{\lambda v^4}{4}
\frac{1}{\cosh^4\bigl(v\sqrt{\tfrac{\lambda}{2}}(y-y_0)\bigr)}\partial_{\mu}y_0\partial^{\mu}y_0 = \frac{1}{2}T_{kink}\partial_{\mu}y_0\partial^{\mu}y_0  \,.
\end{align} 
If we denote $u_0 = T_{kink}^{1/2}y_0$ we  obtain
\begin{equation}
\oper{L}_{\mathrm{eff}} =  \frac{1}{2}\partial_{\mu}u_0\partial^{\mu}u_0  \,,
\end{equation}
which is just a Lagrangian of a single massless scalar field, in accordance with our previous result.
In our toy model, the effective Lagrangian contains single zero mode and as we see, 
the metric is trivial, making the moduli space just $\mathbb{R}$. This indeed corresponds to a
coset $G/H$, where $G = \mathbb{R}$ is a translational invariance with respect to the 
$y$-coordinate, which 
is broken down to the identity $H =\mathbf{1}$.\footnote{
One may be tempted to identify $G$ with $O(4,1)$ the group of space-time symmetries in (4+1)-dimensions and $H$ with $O(3,1)$ the Lorentz group in 
(3+1)-dimensions, which is preserved by the domain wall. Since the number of generators of $O(n-1,1)$ is $\tfrac{n(n-1)}{2}$, it follows that the dimension of the coset is $\mathrm{dim}(O(4,1)/O(3,1)) = 4$. Indeed, the domain wall breaks translational symmetry in the direction perpendicular to the wall and three rotational symmetries around the three axes tangent to the wall. Based on this naive counting, there should be four zero modes, not just one. The reason why that is not the case can be intuitively understood by realization that small fluctuation on the domain wall can be equally well respected as a local translation or a local rotation. Therefore, zero modes connected with the breaking of the rotational invariance are not independent of the translational zero mode and there exist really only one independent mode of fluctuation. More informations about this issue can be found in \cite{Low}.   
}

In chapters \ref{ch:4}, \ref{ch:5} and \ref{ch:6}, where models we consider becomes progressively more complicated, the moduli approximation method of obtaining effective Lagrangian will be one of the keys for the realization of brane world scenario.


\chapter{Some aspects of supersymmetry} \label{ch:3}






Since BPS solitons obey first order
rather than second order differential equations, namely the BPS equations, they can be often
 expressed in a closed form. This 
fact greatly reduces the labor needed for derivation of an effective Lagrangian. In particular, with a closed formula at hand, we
can employ the moduli approximation introduced in the previous chapter to obtain the low-energy effective Lagrangian almost effortlessly. 
Therefore, it seems practical to prefer BPS solitons over their non-BPS analogues as field theoretical models of a brane.

However, as we want to localize particles other than scalar fields, presumably we need to deal with much more complicated theories than the toy model of 
Ch.~\ref{ch:2}. But how can we guarantee that solitonic solutions in those theories are going to be BPS solitons?
More precisely, the question we need to answer is: what is the proper matter content of a five-dimensional theory to allow for a BPS domain wall
rather just \emph{a} domain wall?

The key to answering that question is both simple and surprising and it forms the content of this chapter.
It is a well-known fact, that if the underlying theory is topologically non-trivial and supersymmetric (SUSY),  BPS solitons exist.
Therefore, before we can discuss more advanced model building of Ch.~\ref{ch:4}, Ch.~\ref{ch:5} and Ch.~\ref{ch:6}, we need to introduce few concepts from SUSY and explain their connection with BPS solitons.

Let us first, however, clarify several things. At the most basic level, SUSY can be regarded as an
invariance with respect to mixing of bosonic and fermionic degrees of freedom.
The toy model of Ch.~\ref{ch:2} is seemingly not supersymmetric, since it contains only bosons.
But we can view it
as a bosonic part of a  supersymmetric theory. As far as classical solutions are concerned 
(i.e. no macroscopic fermionic fields) we can ignore the
other part completely. Therefore, our toy model can be enlarged to SUSY theory 
 without changing any result of the previous chapter. 
As we will see repeatedly later on, our models are chosen in such a way that they can
 be enlarged, by adding appropriate fermionic and bosonic fields, to be
supersymmetric. This property allows us to construct BPS solitons, as we will explain below. 

Also, the fact that we use BPS solitons may seem as a simplifying assumption about our 
model (in gauge theories, for example, this corresponds to a special relations between 
gauge coupling and interaction couplings of charged matter fields), but we will see that, 
at least at the lowest order
of approximation, this will not affect the result.
Therefore, even though we will rely on SUSY concepts at the motivational level,
our final model need not be supersymmetric in order to arrive at the same low-energy effective Lagrangian.
 
Let us now proceed with a brief introduction of supersymmetry. The first objective of
this chapter is to clarify the connection of SUSY with BPS solitons.  
This can be most easily/pedagogically done in (1+1)-dimensions. We devote the first section to that. 
The second and ultimate goal of this chapter is to describe general properties of SUSY theories in (4+1)-dimensions. This is described in the second section.
It will turn out, that several objects, namely Fayet-Iliopulous (FI) terms and the so-called prepotential, which arise in five-dimensional SUSY gauge theories, are necessary to localize gauge fields on the domain wall. We will address these issues in chapter \ref{ch:4}.      

\section{BPS solitons and supersymmetry}

The main goal of this section is to demonstrate that  generic SUSY theories in (1+1)-dimensions naturally contain BPS domain wall solution. In fact, we will see that their bosonic part is essentially the same as the toy model of Ch.~\ref{ch:2}.\footnote{In this section we employ the notation
and conventions of  \cite{Shifmanch2}.} 

\subsection{Superspace in (1+1)-dimensions}


There are two ways how to represent SUSY. Either we use ordinary fields, in which case we end up 
having different \emph{multiplets} of bosonic and fermionic fields, all of which fall into the same
irreducible representation of SUSY, or we use \emph{superfields}, which are objects transforming naturally
under SUSY transformation. The advantage of the former approach is its clear interpretation, since it
involves only known objects, namely fields on space-time. 
On the other hand this representation is SUSY non-manifest and it heavily relies on 
the representation theory, producing rather cumbersome expressions which are equally cumbersome to manipulate with.
If the theory is formulated in terms of superfields, SUSY is manifest and resulting formulas 
are elegant and concise. The cost, however, is the need to introduce new objects, namely 
superfields, which are fields on the so-called \emph{superspace}.

A superspace is a combination of ordinary space-time dimensions and additional Grassmann dimensions.
Thus a superspace in (1+1)-dimensions has not only the usual commuting coordinates $x^\mu \in (x^0,x^1)$,
but also two real anti-commuting coordinates $\theta_{\alpha} \in (\theta_1,\theta_2)$, where $\alpha$ is a spinor index, with the
property
\begin{equation}
\{\theta_{\alpha},\theta_{\beta}\} = \delta_{\alpha \beta}\,, \hspace{5mm} \theta_{\alpha}^{\dagger}
= \theta_{\alpha}\,,
\end{equation}
where $\{A,B\} = AB+BA$. Given their anticommuting nature, the inner product between Grassmann
coordinates cannot be symmetrical, i.e. $\theta^2 = \theta_\alpha \theta_\alpha = 0$. In other 
words, the metric tensor must be antisymmetric as well:
\begin{equation}
\bar\theta \theta = \bar \theta_{\alpha}\theta_{\alpha} = \theta_{\beta}\gamma_{\beta\alpha}^{0}
\theta_{\alpha} = -2\iunit \theta_1\theta_2\,,
\end{equation}
where we introduced Dirac conjugation $\bar \theta = \theta^{\dagger}\gamma^0$.
With this convention Grassmann cordinates $\theta_{\alpha}$ are Majorana spinors, if the 
gamma matrices are in the Majorana representation $\gamma^0 = \sigma_2$ and 
$\gamma^1 = \iunit \sigma_3$, with Pauli matrices given as
\begin{equation}\label{ch2:pauli}
\sigma_1 = \begin{pmatrix}
0 & 1 \\
1 & 0 \end{pmatrix}\,, \hspace{5mm}
\sigma_2 = \begin{pmatrix}
0 & -\iunit \\
\iunit & 0 \end{pmatrix}\,, \hspace{5mm}
\sigma_3 = \begin{pmatrix}
1 & 0 \\
0 & -1 \end{pmatrix}\,.
\end{equation}
It is easy to check that for two different spinors $\psi$ and $\phi$ it holds:
\begin{equation}
\bar\psi\phi = \bar\phi\psi\,, \hspace{5mm}\bar\psi\gamma^{\mu}\phi = -\bar\phi\gamma^{\mu}\psi\,.
\end{equation}
As a consequence we have $\bar\theta\gamma^{\mu}\theta=0$, which is true only in two dimensions.

Since in  superspace we treat both commuting and anti-commuting coordinates equally, we must 
also invent differentiation and integration with respect to Grassmann coordinates. 
Derivatives can be introduced quite naturally:
\begin{align}\label{ch2:der1}
\partial_{\alpha} & \equiv \frac{\partial}{\partial \theta_{\alpha}}\,, &
\partial_{\alpha}\theta_{\beta} & =\delta_{\alpha\beta}\,,  &
\partial_{\alpha}\bar\theta_{\beta} & = \gamma_{\alpha\beta}^0\,, \\\label{ch2:der2}
\bar\partial_{\alpha} & \equiv \frac{\partial}{\partial \bar\theta_{\alpha}} = \gamma_{\alpha\beta}^0\frac{\partial}{\partial\theta_{\beta}}\,, &
\bar\partial_{\alpha}\bar\theta_{\beta} & =\delta_{\alpha\beta}\,,  &
\bar\partial_{\alpha}\theta_{\beta} &= -\gamma_{\alpha\beta}^0\,.
\end{align}
Integration, on the other hand, is defined as
\begin{align}\label{ch2:int1}
\int\diff \theta_{\alpha} \cdot 1 & = 0\,, & \int\diff\theta_{\alpha}\theta_{\beta} &= \delta_{\alpha\beta}\,,
 & \int\diff\theta_{\alpha}\bar\theta_{\beta} &= \gamma_{\alpha\beta}^0\,, \\\label{ch2:int2}
\int \diff \bar\theta_{\alpha} \cdot 1 & = 0\,, 
& \int\diff\bar\theta_{\alpha}\bar\theta_{\beta} &= \delta_{\alpha\beta}\,, 
& \int\diff\bar\theta_{\alpha}\theta_{\beta} &= -\gamma_{\alpha\beta}^0\,,
\end{align}
which may not be everyone's first guess. 
In particular, why is $\int\diff \theta_{\alpha} \cdot 1 = 0$ and not $\int\diff \theta_{\alpha} \cdot 1 = \theta_{\alpha}$?
The reason is that we think of $\int\diff\theta$ as a direct analog to $\lineint x$ rather than to an anti-derivative.  
One of characteristic properties of a line integral in ordinary space is that a
constant shift of an integration variable does not change the result. If we demand that the analog
holds for $\theta$ integration, we obtain exactly the properties \refer{ch2:int1}-\refer{ch2:int2}.
Moreover, comparing 
\refer{ch2:int1}-\refer{ch2:int2} with \refer{ch2:der1}-\refer{ch2:der2} we see that
fomally $\int \diff \theta_{\alpha} \equiv \partial_{\alpha}$, from which 
\(\int\diff \theta_{\alpha} \cdot 1 = \partial_{\alpha}\cdot 1 = 0\) follows immediately. 

Let us also define a volume integral as 
\begin{equation}
\int\diff^2\theta = \int \diff\bar\theta_\alpha \diff\theta_\alpha = 2\iunit \int \diff \theta_2\diff \theta_1 = -\bar\partial\partial 
= 2\iunit\partial_2\partial_1\,.
\end{equation}
And in particular
\begin{equation}\label{ch2:integ}
\int\diff^2\theta\, \bar\theta\theta = -\bar\partial\partial\, \bar\theta\theta = 4\,.
\end{equation}

\subsection{SUSY generators and superfields}

Let us look at symmetry transformations of superspace coordinates which leaves the volume measure 
$\diff^2 x\,\diff^2\theta$ invariant. These obviously include Lorentz transformations and constant 
shifts of space-time coordinates $x^{\mu}$ and in analogy, transformations that 
mix together anti-commuting coordinates $\theta_\alpha$. Both of these are not particularly 
important for the following discussion.
A physically interesting symmetry mixes both type of coordinates together. Let us
introduce a SUSY transformation $\delta_{\xi}$ in the following way
\begin{equation}\label{ch2:deltas}
\delta_{\xi}\theta_{\alpha} = \xi_{\alpha}\,, \hspace{5mm} \delta_{\xi}x^{\mu} = -\iunit \bar\theta
\gamma^{\mu}\xi\,,
\end{equation} 
where $\xi$ is a constant Majorana spinor parametrizing the SUSY transformation. 
From Eq.~\refer{ch2:deltas} we can extract the formulas for fermionic operators $Q_{\alpha}$,
which generate the SUSY transformation $\delta_{\xi}=\bar\xi_{\alpha} Q_{\alpha} = 
\bar Q_{\alpha}\xi_{\alpha}$
\begin{equation}
Q_{\alpha} = \bar\partial_\alpha+\iunit\gamma_{\alpha\beta}^{\mu}\theta_{\beta}\partial_{\mu}\,,
\hspace{5mm}
\bar Q_{\alpha} = -\partial_\alpha-\iunit\bar\theta_{\beta}\gamma_{\beta\alpha}^{\mu}\partial_{\mu}\,.
\end{equation}
It can be easily checked that these generators satisfy an algebra
\begin{equation}
\{Q_{\alpha},\bar Q_{\beta}\} = -2\iunit \bigl(\slashed \partial\bigr)_{\alpha\beta}\,,
\end{equation}
where  the Feynman slash notation is expanded as
$\slashed \partial = \gamma^{\mu}\partial_{\mu}$. All other anti-commutators are zero.
Moreover,
one can check that adding $Q_1,Q_2$ to the list of generators of Poincar\'{e} group of symmetries of (1+1)-dimensional
space-time results in a well-behaving \emph{graded} Lie algebra (or simply \emph{superalgebra}), which 
is closed under repetitive use of (anti)commutators between all its elements. In other words, SUSY generators extends 
the usual space-time algebra in a non-trivial way (i.e. not as a direct product).

Back in 1967 a famous no-go theorem of Coleman and Mandula \cite{Coleman} stated that exactly this kind of mixing of 
internal and space-time symmetries is impossible. To be more precise, the Coleman-Mandula theorem claims that any realistic theory (non-trivial $S$-matrix) 
with a mass gap in (3+1)-dimensions can only have symmetry Lie algebra which is always a direct product of the Poincar\'e group and some 
internal group. As it stands, the theorem is indeed correct. However, it turned out that assumptions of the theorem are too restrictive.
In 1971 Golfand and Likhtman \cite{Golfand} explicitly constructed a non-trivial extension of the Poincar\'e algebra by adding to it a set of fermionic operators (forbidden in Coleman-Mandula analysis), which
marked the beginning of SUSY. Four years after this breakthrough the search of possible extensions of the Poincar\'e algebra culminated in 
 the so-called Haag-Lopusanski-Sohnius theorem \cite{Haag}, showing that under fairly general conditions SUSY is the only extension beyond the four-dimensional Poincar\'e algebra.   

After this historical sketch let us now proceed to introduce another vital concept of SUSY: a \emph{superfield}.
As far as we are interested in supersymmetric theories in (1+1)-dimensions, there is only one
type of superfield $\Phi\equiv \Phi(x,\theta)$, which is simply an arbitrary function of superspace coordinates.
In more dimensions, where superspace description is in general redundant, there are more types of 
superfields.

Given their anti-commuting nature, any function of Grassmann coordinates $\theta_{\alpha}$ can be 
Taylor-expanded into a finite series of terms. Thus, a superfield can be always decomposed into 
 \emph{component fields} as
\begin{equation}\label{ch2:components}
\Phi(x,\theta) = \phi(x) +\bar\theta\psi(x)+\frac{1}{2}\bar\theta \theta F(x)\,,
\end{equation}
where $\phi(x),F(x)$ are scalar fields and $\psi(x)$ is a Majorana spinor field.
These components are various projections of a superfield $\Phi$
\begin{equation}
\phi = \Phi\Big|_{\theta = 0}\,, \hspace{5mm}
\psi_{\alpha} = \bar\partial_{\alpha}\Phi\Big|_{\theta = 0}\,, \hspace{5mm}
F = -\frac{1}{2}\bar\partial\partial\,\Phi\Big|_{\theta = 0}
\end{equation}
and they are independent of each other.
Under SUSY transformation, a (scalar) superfield only changes by change of its arguments:
\begin{equation}
\delta_{\xi}\Phi = \Phi(x+\delta_{\xi}x,\theta+\delta_{\xi}\theta)-\Phi(x,\theta) =
\bar\xi \psi +\bar\theta (F-\iunit\slashed\partial \phi)\xi +\frac{\iunit}{2}\bar\theta\theta \,
\partial_{\mu}\bar\psi\gamma^{\mu} \xi\,.
\end{equation}
Finally, matching up the powers of $\theta$ we can express the SUSY transformation in terms of
components as
\begin{align}
\delta_{\xi} \phi &= \bar\xi \psi\,, \\
\delta_{\xi}\psi & = -\iunit\slashed\partial \phi \xi +F\xi\,, \\
\delta_{\xi}F & = \iunit \partial_{\mu}\bar\psi\gamma^{\mu} \xi\,.
\end{align}

\subsection{SUSY theories in (1+1)-dimensions}

Notice that the $\bar\theta \theta$ component of a superfield 
transforms into a total derivative: $\delta_\xi F = \iunit \partial_{\mu}(\bar\psi \gamma^\mu \xi)$.
This observation allows us to construct supersymmetric theories in a
very simple fashion, since its immediate consequence is that integration of the $F$-term 
over space-time is a SUSY scalar. Indeed,
\begin{equation}
\delta_{\xi}\int\diff^2 x\, F = \iunit \int\diff^2 x \partial_{\mu} (\bar\psi \gamma^\mu \xi) = 0\,,
\end{equation}
where the last equality follows from the Stokes theorem, provided that $F$ vanishes at the boundary. 
Using \refer{ch2:integ} and \refer{ch2:components} we can reformulate this observation in 
a SUSY invariant way as
\begin{equation}
\int \diff^2x\, F = \frac{1}{2}\int \diff^2x\,\diff^2\theta\, \Phi = \mbox{SUSY scalar.}
\end{equation}
This result is rather pleasing. It tells us that if we integrate a superfield over 
the whole superspace we obtain a SUSY invariant quantity (which is by construction also a Lorentz invariant).
Let us stress, however, that the above relation holds for \emph{any} superfield.
As a consequence, the way how to write down an action for SUSY invariant theory in (1+1)-dimensions
is simply
\begin{equation}
S = \int\diff^2 x\, \oper{L} = \frac{1}{2}\int\diff^2 x\,\diff^2\theta\, \oper{L}_S\,,
\end{equation}
where $\oper{L}_S$ is a superfield. Factor $\tfrac{1}{2}$ is added for convenience.

Let us in the following reserve the symbol $\Phi$ to represent a superfield,
whose components are fundamental degrees of freedom
of our theory, namely $\phi, \psi$ and $F$. We can use $\Phi$ to construct other superfields. 
In particular, any function of $\Phi$, say $f(\Phi)$, is also a superfield.
We would also like to construct terms, which on the component level contain quantities such as 
$\partial_{\mu}\phi$ and $\partial_{\mu}\psi$ and are superfields at the superspace level.
It turns out, that 
individual derivatives $\partial_\mu \Phi$ and $\partial_{\alpha}\Phi$ do not 
transform in a desired way under SUSY transformation. However, if we combine them into a 
covariant derivative
\begin{equation}
D_{\alpha} = \bar\partial_{\alpha} -\iunit (\gamma^{\mu}\theta)_{\alpha}\partial_{\mu}\,,
\hspace{5mm}
\bar D_{\alpha} = -\partial_{\alpha}+\iunit(\bar\theta\gamma^{\mu})_{\alpha}\partial_{\mu}\,,
\end{equation}    
the associated quantities $D_{\alpha}\Phi = \gamma_{\beta\alpha}^{0}\bar D_{\beta}\Phi$ are 
superfields.
Therefore any Lagrangian density of the form $\oper{L}_S(\Phi,D_{\alpha}\Phi)$, with properly 
contracted Lorentz and spinor indices, is
a viable candidate for supersymmetric theory in (1+1)-dimensions.\footnote{Arguments coming
from dimensional analysis and locality further reduce this general form into 
\begin{equation*}
\oper{L}_S = \frac{1}{4}\oper{K}(\Phi)\bar D_{\alpha}\Phi D_{\alpha}\Phi + \oper{W}(\Phi)\,,
\end{equation*}
with both $\oper{K}$ and $\oper{W}$ being arbitrary functions of $\Phi$.} In addition, if we demand that 
component fields are to have canonical kinetic terms, we end up with a general expression
\begin{equation}\label{ch2:lagr1}
\oper{L}_S = \frac{1}{4}\bar D_{\alpha}\Phi D_{\alpha}\Phi +\oper{W}(\Phi)\,,
\end{equation} 
where the so-called \emph{superpotential} $\oper{W}(\Phi)$ is an arbitrary function of $\Phi$.

We can now Taylor-expand the supersymmetric Lagrangian density $\oper{L}_S$ of Eq.~\refer{ch2:lagr1} with respect
to $\theta$ coordinates and perform the integration $\int \diff^2\theta$, which picks out
the $\bar\theta \theta$ component of $\oper{L}_S$. The result is  
\begin{equation}
\oper{L} = \frac{1}{2}\partial_{\mu}\phi\partial^{\mu}\phi+
\frac{\iunit}{2}\bar\psi\slashed\partial\psi+\frac{1}{2}F^2
+\frac{\diff \oper{W}(\phi)}{\diff\phi}F-\frac{1}{2}\frac{\diff^2 \oper{W}(\phi)}{\diff \phi^2}\bar\psi \psi\,.
\end{equation}
Since the $F$ component is not dynamical it can be eliminated through its EoM
\begin{equation}
\frac{\partial \oper{L}}{\partial F} = 0\,, \hspace{5mm}\imply\hspace{5mm} F = -\frac{\diff \oper{W}(\Phi)}{\diff \Phi}\,,
\end{equation} 
which leads to the final form
\begin{equation}\label{ch2:susylagr}
\oper{L} = \frac{1}{2}\partial_{\mu}\phi\partial^{\mu}\phi+
\frac{\iunit}{2}\bar\psi\slashed\partial\psi-\frac{1}{2}\Bigl(\frac{\diff \oper{W}(\Phi)}{\diff\Phi}\Bigr)^2
-\frac{1}{2}\frac{\diff^2 \oper{W}(\phi)}{\diff \phi^2}\bar\psi \psi\,.
\end{equation}

The bosonic part of the Lagrangian \refer{ch2:susylagr} ($\psi = 0$) is in fact the one, which we used as our toy
model in the previous chapter, with the superpotential given in Eq.~\refer{ch2:supp}. Thus, we confirmed the claim we made at the beginning 
of this section. Let us now settle the remaining issue, namely of the interplay between SUSY and BPS solitons. We will discover that a BPS solution has a unique property of preserving a part of the SUSY. A property which will not only provide an alternative way of constructing BPS solitons, but also automatically localize fermions.

\subsection{BPS solitons and breaking of SUSY}

Let us discuss what kind of solutions the theory \refer{ch2:lagr1} can have, with respect to their behavior under SUSY transformation.
The equations of motion can be written down either in the superfield formalism as
\begin{equation}
\frac{1}{2}\bar D D\,\Phi - \frac{\diff \oper{W}}{\diff\Phi} = 0\,,
\end{equation}  
or by using components
\begin{align}
\partial^2 \phi -\frac{\diff^2 \oper{W}}{\diff \phi^2}F+\frac{1}{2}\frac{\diff^3\oper{W}}{\diff\phi^3}\bar \psi \psi & = 0\,, \\ \label{ch2:eompsi}
\iunit \slashed \partial \psi - \frac{\diff^2 \oper{W}}{\diff \phi^2}\psi & = 0\,, \\
F + \frac{\diff \oper{W}}{\diff \phi} & = 0\,.
\end{align} 
Both ways are, of course, equivalent, which can be checked by $\theta$ expansion of the superfield EoM, with the help of the identity
\begin{equation}
\bar D D = \bar \partial \partial +\iunit\bigl(\bar \theta \gamma^\mu \bar\partial +\partial \gamma^\mu \theta\bigr)\partial_{\mu}+\bar\theta\theta \partial^2\,.
\end{equation}

A generic solution $\Phi$ of the above system of partial differential equations breaks SUSY, meaning that 
\begin{equation}
Q_{1}\Phi \not = 0\,, \hspace{5mm} Q_2\Phi \not = 0\,.
\end{equation}
Therefore, if we took $\Phi$ as a background solution, the effective theory of its fluctuations will not be supersymmetric (moreover, as the $\psi_\alpha$ component of $\Phi$ is generically non-zero, even the Lorentz invariance will be broken). But is this true for all solutions? Certainly not. It can be easily checked that one type of solutions, let us denote them as $\Phi^0$, are SUSY invariant. This means that apart being solutions to EoM, they also satisfy conditions  $Q_1\Phi^0 = Q_2\Phi^0 = 0$. Solving for those conditions we discover that $\Phi^0 = \phi^0$, where $\phi^0$ are vacua of the theory $\tfrac{\diff \oper{W}(\phi^0)}{\diff \phi} = 0$.

Since we have two SUSY generators, there exists a third possibility, that some solution $\Phi^{1/2}$ breaks only one generator, for instance
the first one $Q_1\Phi^{1/2} \not = 0$, while the second one is preserved $Q_2\Phi^{1/2} = 0$.  Expansion of the last condition to components yields
\begin{align}
Q_2\Phi^{1/2} & = \bigl(\bar\partial_2+\iunit \gamma_{2\alpha}^{\mu}\theta_\alpha\partial_{\mu}\bigr)\bigl(\phi^{1/2}+\bar\theta\psi^{1/2}+
\frac{1}{2}\bar\theta\theta F^{1/2}\bigr) \nonumber \\
& = \psi_2^{1/2}+\theta_2 F^{1/2}+\iunit \gamma_{2\alpha}^{\mu}\theta_{\alpha}\partial_{\mu}\phi^{1/2}-\frac{\iunit}{2}\bar{\theta}\theta
\gamma_{2\alpha}^{\mu}\partial_{\mu}\psi_{\alpha}^{1/2}\,.
\end{align}
Setting each order of $\theta_{\alpha}$ to zero, we obtain
\begin{align}\label{ch2:cond1}
\partial_t\phi^{1/2} &= 0\,, & F^{1/2}+\partial_x\phi^{1/2} & = 0\,, \\ \label{ch2:cond2}
\partial_t\psi_1^{1/2} & = 0\,, & \psi_2^{1/2} & = 0\,. 
\end{align}
We have to check whether these conditions are compatible with EoM. In particular we have $F^{1/2} = -\tfrac{\diff \oper{W}}{\diff \phi}$, which together
with the first and second condition gives
\begin{equation}
\partial_x\phi^{1/2}(x) = \frac{\diff \oper{W}}{\diff \phi}\Big|_{\phi = \phi^{1/2}}\,.
\end{equation}
This is nothing else but the BPS equation \refer{ch2:bps}! We can see this directly by recalling Eq.~\refer{ch2:supot}.
For the $\psi_\alpha^{1/2}$ the 1/2 SUSY preserving conditions \refer{ch2:cond2} demand that $\psi_2^{1/2} = 0$ and that $\psi_1^{1/2}\equiv \psi_1^{1/2}(x)$ is an arbitrary function of $x \equiv x^1$. The exact $x$ dependence of $\psi_1^{1/2}$ can be easily obtained from EoM \refer{ch2:eompsi} and it is uniquely determined through the superpotential by $\phi^{1/2}(x)$ up to an integration constant:
\begin{equation}
\psi_1^{1/2} = \Exp{-\int^x \frac{\diff^2\oper{W}}{\diff\phi^2}}\eta\,,
\end{equation} 
where $\eta$ is a constant Grassmann number. Thus, we can always consistently set $\psi = 0$, which we will always do for classical solutions of EoM. 
If we would allow for nonzero $\psi^{1/2}$ as a part of the background solution, resulting effective theory will not be Lorentz invariant. Let us also point out
that setting $\psi^{1/2}$ to zero allows us to recast superfield solution $\Phi^{1/2}$ into a particularly nice form
\begin{equation}\label{ch2:niceform}
\Phi^{1/2} = \phi_{K}\bigl(x-\tfrac{1}{2}\bar\theta\theta\bigr)\,,
\end{equation} 
where $\phi_{K}$ is a kink solution of Eq.~\refer{ch2:kink}. 
If we demand that the first generator is preserved $Q_1 \Phi^{1/2} = 0$, 
we simply obtain the anti-kink solution and the corresponding superfield is $\Phi^{1/2} = -\phi_K\bigl(x+\tfrac{1}{2}\bar\theta\theta\bigr)$.

In the background of the BPS solution $\Phi^{1/2}$ of Eq.~\refer{ch2:niceform} a half of the SUSY generators is spontaneously broken.
Therefore we expect that in the effective theory there will be a corresponding Goldstone fermion or \emph{goldstino}. This new particle is a superpartner
to a bosonic zero mode, which arises from breaking of the translational symmetry.
This observation can be easily verified by direct calculations. If we write down linearized EoM for fermionic fluctuations $\delta \psi_\alpha = \Exp{\iunit \omega t}c_{\alpha}(x)$, with use of Eqs.~\refer{ch2:kink} and \refer{ch2:supp} and setting $\lambda = 2$, $v=1$ for convenience, we obtain 
\begin{align}
\partial_x c_1 -\iunit \omega c_2 - 2\tanh(x)c_1 & = 0\,, \\
\partial_x c_2 -\iunit\omega c_1+2\tanh(x)c_2 & = 0\,.
\end{align}
This system has a non-trivial solution only if $\omega =0$, for $c_1, c_2$ are real. Moreover, from general conditions \refer{ch2:cond2} we have $c_1=0$ and solving for $c_2$ yields
\begin{equation}
c_2 = \frac{1}{\cosh^2(x)}\eta\,,
\end{equation} 
where $\eta$ is a constant Grassmann number. The profile of $c_2(x)$ is exactly the same as the profile of a bosonic zero mode \refer{ch2:zeromode}. Thus, we have found a localized fermionic fluctuations, which is the sought goldstino.  
Let us stress, that similar analysis can be applied to (4+1)-dimensional case in the same fashion as we did for the toy model of Ch.~\ref{ch:2}. The same goldstino localizes on (3+1)-dimensional brane, together with previously found scalar zero mode. The resulting effective action, however, will not be supersymmetric in (3+1)-dimensional sense, but only in (0+1)-dimensional sense. This is so, because SUSY of the parent theory is a rather special one, as it completely ignores three spatial dimension. Indeed, in the next section, we will set up a  theory with a genuine (4+1)-dimensional SUSY, which will give us much richer spectrum of particles in the 
effective theory. 
 
In conclusion, we have arrived at very important fact: BPS solutions partially preserve supersymmetry. This characteristic property of BPS solitons is universal and it is not, for instance, restricted to 
(1+1)-dimensions. It also gives us an alternative way how to find solitonic solutions. The Bogomol'nyi trick we introduced in the previous chapter is not 
so straight-forward to implement, as it involves some guessing work and it may become cumbersome as the number of field in the model increases. On the other hand, partial SUSY preserving conditions are just a set of first order differential equations, from which one can read off corresponding BPS equations  upon inspection.

It is also worth mentioning that on the quantum level topological charges become central charges of the superalgebra (see e.g. \cite{Witten} for details).
 
\section{Supersymmetry in higher dimensions}\label{susy45d}

In order to make our discussion of localization of gauge fields of Ch.~\ref{ch:4} 
unimpeded by technical details, it is convenient to develop basic understanding of SUSY theories in higher space-time dimensions. In this section, we are going to present basic facts about supersymmetry with four and subsequently eight supercharges. Our exposition will be focused on the model building aspects, namely on the matter
content of generic $N=1$ SUSY theories in (3+1)-dimensions and (4+1)-dimensions. This will help us to cement our notation used in the rest of this text and also to expose vital field-theoretical ingredients, required for localization of gauge fields (see section \ref{sec:otha} in Ch.~\ref{ch:4}). A more detailed discussion of supersymmetry in four and five dimensions can be found in \cite{Wess}, \cite{Seiberg}, \cite{Hassan}, \cite{Bilal}, \cite{Cortes}, to name just a few.

\subsection{A SUSY invariant theories in (3+1)-dimensions}

A superspace realization of $N=1$ SUSY in (3+1)-dimensions is obtained by adding four Grassmann coordinates $\theta^{\alpha}, \bar\theta_{\dot\alpha}$ to the space-time coordinates $x^{\mu}$, with $\mu=0,1,2,3$. Here $\alpha = 1,2$ and $\dot \alpha = \dot 1, \dot 2$ are both spinor indices, but while first one is reserved for spinors in $(\tfrac{1}{2},0)$ representation of the Lorentz group, the second one is used for spinors in  $(0,\tfrac{1}{2})$ representation. The dotting  notation makes explicit that indices from different representations should never be contracted together. This also means that although $\theta^{\alpha}$ and $\bar\theta_{\dot\alpha}$ look formally conjugated to each other, they are in fact independent objects. 
Raising and lowering of spinor indices  is performed by $\varepsilon$ symbols as $\theta^{\alpha} = \varepsilon^{\alpha\beta}\theta_{\beta}$ and $\bar\theta^{\dot\alpha}=\varepsilon^{\dot\alpha\dot\beta}\bar\theta_{\dot\beta}$, where conventions are set in such a way that  
\begin{equation}
\theta^2 = \theta^{\alpha}\theta_{\alpha} = -2\theta^1\theta^2\,, \hspace{5mm} \bar\theta^2 = \bar\theta_{\dot\alpha}\bar\theta^{\dot\alpha} = 2\bar\theta_{\dot 1}\bar\theta_{\dot 2}\,,
\end{equation}
holds.  Following identities are frequently used
\begin{equation}
\theta^{\alpha}\theta^{\beta} = \frac{1}{2}\varepsilon^{\alpha\beta}\theta^2\,,\hspace{5mm} \bar\theta_{\dot\alpha}\bar\theta_{\dot\beta} = 
-\frac{1}{2}\varepsilon_{\dot\alpha\dot\beta}\,\bar\theta^2\,, \hspace{5mm} \theta\sigma^{\mu}\bar\theta \theta\sigma^{\nu}\bar\theta = 
\frac{1}{2}g^{\mu\nu}\theta^2\bar\theta^2\,,
\end{equation}
where  
\begin{equation}
 \theta\sigma^{\mu}\bar\theta =  \theta^{\alpha}\sigma_{\alpha\dot\alpha}^{\mu}\bar\theta^{\dot\alpha} =
- \bar\theta_{\dot\alpha}\bar\sigma^{\mu\,\dot\alpha\alpha}\theta_{\alpha} =  - \bar\theta\bar\sigma^{\mu}\theta\,,
\end{equation}
with $\sigma^{\mu} = (\mathbf{1},\sigma_i)$ and $\bar\sigma^{\mu} = \varepsilon^{T}\sigma^{\mu}\varepsilon = (\mathbf{1},-\sigma_i)$ being four-vector extension of the Pauli matrices given in Eq.~\refer{ch2:pauli}. The sign convention for the metric tensor is $ (+,-,-,-)$. 

Under supersymmetry transformation $\delta_{\xi,\,\bar\xi}$, superspace coordinates $(x^{\mu},\theta^\alpha,\bar\theta_{\dot\alpha})$ undergo a translation
\begin{align}
\delta_{\xi,\,\bar\xi}\,x^{\mu} & = \iunit \theta\sigma^{\mu}\bar\xi +\iunit \bar\theta\bar\sigma^{\mu}\xi\,, \\
\delta_{\xi,\,\bar\xi}\,\theta^{\alpha} & = \xi^{\alpha}\,, \\
\delta_{\xi,\,\bar\xi}\,\bar\theta_{\dot\alpha} &= \bar\xi_{\dot\alpha}\,,
\end{align}
where $\xi$, $\bar\xi$ are independent Weyl spinors parameterizing SUSY transformation. This transformation can be represented as an operator on the superspace $\delta_{\xi,\,\bar\xi} = \xi^{\alpha}Q_{\alpha}+\bar\xi_{\dot\alpha}\bar Q^{\dot\alpha}$, where the supercharges $Q_{\alpha}$ and $\bar Q_{\dot\alpha}$ are given as
\begin{equation}
Q_{\alpha} = \frac{\partial}{\partial \theta^{\alpha}}-\iunit \sigma_{\alpha\dot\alpha}^{\mu}\bar\theta^{\dot\alpha}\partial_{\mu}\,,\hspace{5mm}
\bar Q_{\dot\alpha} = -\frac{\partial}{\partial\bar\theta^{\dot\alpha}}+\iunit \theta^{\alpha}\sigma_{\alpha\dot\alpha}^{\mu}\partial_{\mu}\,.
\end{equation}
Correspondingly, the covariant derivatives are
\begin{equation}\label{ch2:covar}
D_{\alpha} = \frac{\partial}{\partial \theta^{\alpha}}+\iunit \sigma_{\alpha\dot\alpha}^{\mu}\bar\theta^{\dot\alpha}\partial_{\mu}\,,\hspace{5mm}
\bar D_{\dot\alpha} = -\frac{\partial}{\partial\bar\theta^{\dot\alpha}}-\iunit \theta^{\alpha}\sigma_{\alpha\dot\alpha}^{\mu}\partial_{\mu}\,.
\end{equation}

\subsubsection{Chiral superfields}

A superfield is in general an arbitrary function of superspace coordinates $f(x^{\mu},\theta^{\alpha},\bar\theta_{\dot\alpha})$. However, we are at liberty to impose covariant constrains on a superfield. It is known that a chiral superfield, denoted as $\Phi$, defined by the constraint
\begin{equation}
\bar D_{\dot \alpha}\Phi = 0\,,
\end{equation}
forms an irreducible representation.
The above constraint can be easily solved, if one realizes that there are two primitive superfields, namely $y^{\mu}\equiv 
x^{\mu}+\iunit \theta\sigma^{\mu}\bar\theta$ and $\theta^{\alpha}$, which are chiral superfields
\begin{equation}
\bar D_{\dot\alpha} y^{\mu} = \bar D_{\dot\alpha} \bigl(x^{\mu}+\iunit \theta\sigma^{\mu}\bar\theta\bigr) = 0\,, \hspace{5mm}
\bar D_{\dot\alpha} \theta^{\alpha} = 0\,.
\end{equation} 
An arbitrary function of these primitive chiral superfields $\Phi(y,\theta)$ is also a chiral superfield. We can expand $\Phi(y,\theta)$ in terms of components as
\begin{align}
\Phi(y,\theta) & = \phi(y)+\sqrt{2}\theta \psi(y) + \theta^2 F(y)
\end{align}
and further expansion of Grassmann variables inside $y$ yields
\begin{align}
\Phi(x,\theta, \bar\theta) & = \phi(x)+\iunit \theta\sigma^{\mu}\bar\theta \partial_{\mu}\phi(x)-\frac{1}{4}\theta^2\bar\theta^2 \partial^2 \phi(x) \nonumber \\ \label{ch3:chiralsup}
& + \sqrt{2}\theta\psi+\frac{\iunit}{\sqrt{2}}\theta^2 \bar\theta\bar\sigma^{\mu}\partial_{\mu}\psi(x)+\theta^2F(x)\,.
\end{align}
We see that a chiral multiplet corresponding to this chiral superfield consists of a complex scalar $\phi$, Weyl spinor $\psi$ and an auxiliary field $F$.   
Similarly, an anti-chiral superfield $\Phi^{\dagger}$ is defined by the constraint $D_{\alpha}\Phi^{\dagger} = 0$
\begin{align}
\Phi^{\dagger}(y^{*},\bar\theta) & = \phi^{\dagger}(y^{*})+\sqrt{2}\bar\theta \bar\psi(y^{*}) + \bar\theta^2 F^{\dagger}(y^{*})\,. \nonumber \\
\Phi^{\dagger}(x,\theta,\bar\theta) & = \phi^{\dagger}(x)-\iunit \theta\sigma^{\mu}\bar\theta \partial_{\mu}\phi^{\dagger}(x)-\frac{1}{4}\theta^2\bar\theta^2 \partial^2 \phi^{\dagger}(x) \nonumber \\
& + \sqrt{2}\bar\theta\bar\psi+\frac{\iunit}{\sqrt{2}}\bar\theta^2 \theta\sigma^{\mu}\partial_{\mu}\bar\psi(x)+\bar\theta^2F^{\dagger}(x)\,.
\end{align}
 A generic SUSY invariant theory constructed purely out of chiral and anti-chiral superfield can be written as
\begin{equation}\label{ch2:slagr}
\oper{L} = \int \diff^2\bar\theta\,\diff^2\theta\,\Phi^{\dagger}\Phi + \int\diff^2\theta\, \oper{W}(\Phi)+\int\diff^2\bar\theta\, 
\oper{W}^{\dagger}(\Phi^{\dagger})\,,
\end{equation}
where $\oper{W}(\Phi)$ is a superpotential.
The reasoning is as follows. First, any function of a chiral superfield is obviously 
also a chiral superfield. Since the $\theta^2$ component of a chiral superfield (or $\bar\theta^2$ component of an anti-chiral superfield) transforms into a total
derivative under SUSY transformation, the superpotential part of \refer{ch2:slagr} is indeed SUSY invariant. Second, a product of chiral and anti-chiral superfield $\Phi^{\dagger}\Phi$ is a general superfield. Consequently its $\theta^2\bar\theta^2$ component transforms into a total derivative. Moreover, at the component level (up to a surface term)
\begin{equation}
\Phi^{\dagger}\Phi\Big|_{\theta^2\bar\theta^2} = \partial_{\mu}\phi^{\dagger}\partial^{\mu}\phi +\iunit \bar\psi \bar\sigma^{\mu}\partial_{\mu}\psi
+F^{\dagger}F\,,
\end{equation}
which gives canonical kinetic terms for the fields $\phi$ and $\psi$. 

Let us note that $\Phi^{\dagger}\Phi$ is not the only possibility for acceptable SUSY invariant kinetic term. In fact, considering more chiral fields for a moment $\Phi_{i}$, with $i=1,\ldots N$, the most general structure is given by 
\begin{equation}
\int \diff^2\bar\theta\,\diff^2\theta\, \oper{K}(\Phi,\Phi^{\dagger})\,,
\end{equation} 
where $\oper{K}(\Phi,\Phi^{\dagger})$ is the so-called K\"ahler potential.  The reason why it is called a ``potential'' is that at the component level the
second derivatives of $\oper{K}(\Phi,\Phi^{\dagger})$
\begin{equation}
\oper{K}(\Phi,\Phi^{\dagger})\Big|_{\theta^2\bar\theta^2} = \frac{\partial^2 \oper{K}}{\partial \phi_i\partial\phi_j^{\dagger}}\partial_{\mu}\phi_i\partial^{\mu}\phi_j^{\dagger} +\ldots
\end{equation} 
 second derivatives of $\oper{K}(\phi,\phi^{\dagger})$ determines the metric tensor in a target space of $\phi_i$ fields. Moreover, as for any potential, there is certain amount of arbitrariness in definition of $\oper{K}(\Phi,\Phi^{\dagger})$. 
It is easy to see that adding any superfield of the form $f(\Phi)+f^{\dagger}(\Phi^{\dagger})$ to the K\"ahler potential  $\oper{K}(\Phi,\Phi^{\dagger})$ gives exactly the same action.\footnote{The reason is that
a superspace integral of the type
\begin{equation}
\int\diff^4 x\, \diff^2\bar\theta\, \diff^2\theta\, f(\Phi)
\end{equation} vanishes, since one can first make a change of  variables $x^{\mu}\to y^{\mu}$ and then use the fact that $f(\Phi)$ has no $\bar\theta^2$ term. A similar argument can be applied to $f^{\dagger}(\Phi^{\dagger})$.} An invariance under such a transformation (incidentally called a K\"ahler transformation) hints to the underlying geometrical structure of $N=1$ SUSY theories in (3+1)-dimensions, that a target space of bosonic fields $\phi_i$ must be a special type of a complex manifold called a K\"ahler manifold \cite{Zumino}.

\subsubsection{Vector superfields}

A vector superfield $V$ satisfies a reality constraint $V = V^{\dagger}$. Let us first consider the Abelian case. If we make the replacement
\begin{equation}\label{ch2:gauge}
V \to V + \iunit(\Lambda-\Lambda^{\dagger})\,,
\end{equation}  
where $\Lambda$ $(\Lambda^{\dagger})$ is a chiral (anti-chiral) superfield, we obtain another vector superfield. It can be shown that this transformation coincides for a particular component of $V$ with the usual gauge transformation. Therefore $V$ is considered as a SUSY extension of a gauge field and \refer{ch2:gauge} as a SUSY generalization of a gauge transformation. In the same way as we fix the gauge for a gauge field, we can use \refer{ch2:gauge} to reduce the number of components of $V$. The usual choice is the so-called Wess-Zumino (WZ) gauge \cite{Wess}, in which $V$ is decomposed as
\begin{equation}
V\Big|_{WZ} = -\theta \sigma^{\mu}\bar\theta A_{\mu}+\iunit \theta^2 \bar\theta\bar\lambda -\iunit \bar\theta^2\theta\lambda +\frac{1}{2}\theta^2\bar\theta^2 D\,, \label{ch3:vecsup}
\end{equation}
where $A_{\mu}$ is a $U(1)$ gauge field, $\lambda$ is a Weyl spinor (gaugino) and $D$ is an auxiliary field. Let us point out that the WZ gauge does not fix all the gauge freedom and the usual gauge transformations $A_\mu\to A_\mu +\partial_\mu\alpha$ are still allowed.
The SUSY generalization of a $U(1)$ field strength is defined as
\begin{equation}
W_{\alpha} = -\frac{1}{4}\bar D^2 D_{\alpha} V\,, \hspace{5mm} \bar W_{\dot\alpha} = -\frac{1}{4} D^2 \bar D_{\dot\alpha} V\,.
\end{equation} 
Using the identity 
\begin{equation}
\{D_\alpha,\bar D_{\dot\alpha}\} = -2\iunit \sigma_{\alpha\dot\alpha}^{\mu}\partial_{\mu}
\end{equation}
it can be easily verified that field strength $W_{\alpha}$ (and correspondingly $\bar W_{\dot \alpha}$) is gauge invariant and chiral (anti-chiral) superfield. Thus, in terms of components in the WZ gauge (with  $y^{\mu}\equiv x^{\mu}+\iunit \theta\sigma^{\mu}\bar\theta$)
\begin{equation}
W_{\alpha} =-\iunit \lambda_\alpha(y)+\theta_\alpha D(y)-\frac{\iunit}{2}(\sigma^\mu\bar\sigma^\nu\theta)_\alpha F_{\mu\nu}(y)+ \theta^2 \bigr(\sigma^\mu\partial_\mu \bar\lambda(y)\bigl)_\alpha\,,
\end{equation}  
where $F_{\mu\nu} = \partial_\mu A_\nu-\partial_\nu A_\mu$ is an Abelian field strength.

Since $W_\alpha$ is a chiral superfield, $W^{\alpha}W_{\alpha}$ is also a chiral superfield and its $\theta^2$ component can be used as a SUSY invariant term. In particular we have
\begin{equation}
\frac{1}{4g^2}\Bigl(\int\diff^2\theta\, W^\alpha W_\alpha +\int\diff^2\bar\theta\, \bar W_{\dot\alpha}\bar W^{\dot\alpha}\Bigr)  = -\frac{1}{4g^2}F_{\mu\nu}F^{\mu\nu} +\frac{1}{2g^2}D^2-\frac{\iunit}{g^2}\lambda \sigma^{\mu}\partial_{\mu}\bar\lambda\,.
\end{equation}
Also, let us note that inclusion of the so-called $\theta$-term (not to confuse with Grassmann coordinate!) can be achieved in a very compact way   
\begin{equation}
\frac{1}{8\pi}\mathrm{Im}\Bigl(\tau \int\diff^2\theta\,W^\alpha W_\alpha\Bigr)  = 
-\frac{1}{4g^2}F_{\mu\nu}F^{\mu\nu}+\frac{\theta}{32\pi^2}\tilde F^{\mu\nu}F_{\mu\nu}
+\frac{1}{2g^2}D^2-\frac{\iunit}{g^2}\lambda \sigma^{\mu}\partial_{\mu}\bar\lambda\,,
\end{equation} 
where $\tau = \theta/\pi+4\pi\iunit/g^2$ and $\tilde F^{\mu\nu} = \varepsilon^{\mu\nu\rho\sigma}F_{\rho\sigma}$.

Let us now write down SUSY Lagrangian with local $U(1)$ symmetry and interacting matter fields belonging to the chiral multiplet. We first need to modify  \refer{ch2:slagr} to be invariant under local $U(1)$ symmetry. 
Let us note that the Lagrangian \refer{ch2:slagr} is already invariant under a global $U(1)$ symmetry, which acts on the superfields as $\Phi \to \Exp{\iunit \Lambda}\Phi$ and $\Phi^{\dagger}\to\Exp{-\iunit \Lambda}\Phi^{\dagger}$, where $\Lambda \in \mathbb{R}$.\footnote{It can be shown that requiring global $U(1)$ invariance and renormalizibility severely restricts possible forms of a superpotential $\oper{W}(\Phi)$. If one allows for multiple chiral superfields $\Phi_i$, the superpotential can be at most a cubic function \cite{Wess}. In the case of a single chiral superfield, however, the superpotential is zero.} It turns out that to make this symmetry local, it is necessary to promote $\Lambda$ to a full chiral superfield and replace the kinetic term of \refer{ch2:slagr} with
\begin{equation}\label{ch2:kt}
\Phi^{\dagger}\Exp{-V}\Phi\,.
\end{equation}
Therefore, a simplest supersymmetric $U(1)$ gauge theory with a single chiral multiplet is defined:
\begin{equation}\label{ch2:slagr2}
\oper{L} = \frac{1}{4g^2}\Bigl(\int\diff^2\theta\, W^\alpha W_\alpha +c.c.\Bigr)+
 \int \diff^2\bar\theta\,\diff^2\theta\,\Phi^{\dagger}\Exp{-V}\Phi\,. 
\end{equation}
The generalization of the above Lagrangian to contain more chiral superfields is straight-forward. 
 
Similar considerations are applied in the non-Abelian case. The vector superfield $V=V^a T^a$ takes values in the adjoint representation of the gauge group and transforms under a gauge transformation as
\begin{equation}
\Exp{-V}\to \Exp{\iunit\Lambda^{\dagger}}\Exp{-V}\Exp{-\iunit \Lambda}\,.
\end{equation}  
The normalization of the generators $T^a$ is chosen as    $\Tr\bigl[T^aT^b\bigr]=\tfrac{1}{2}\delta^{ab}$.
The chiral superfields are in the fundamental representation of the gauge group and transforms as
\begin{equation}
\Phi \to \Exp{\iunit\Lambda}\Phi\,, \hspace{5mm} \Phi^{\dagger}\to \Phi^{\dagger}\Exp{-\iunit\Lambda^{\dagger}}\,.
\end{equation}
 Notice that the kinetic term \refer{ch2:kt} is still invariant under this transformation, if the ordering of superfields is exactly as denoted.
The non-Abelian field strength is defined as
\begin{equation}\label{ch2:fsna}
W_\alpha = \frac{1}{4}\bar D^2\Exp{V}D_\alpha\Exp{-V}\,,
\end{equation}
which is still a chiral superfield, but it transforms as an adjoint representation
\begin{equation}
W_\alpha \to \Exp{\iunit\Lambda}W_{\alpha}\Exp{-\iunit\Lambda}\,.
\end{equation}
The Lagrangian \refer{ch2:slagr2} can be readily modified to non-Abelian case just by including trace in the kinetic term for the vector superfield. 
Notice, however, that now $\Phi$ represents a column vector of chiral superfields.
\begin{equation}
\oper{L} = \frac{1}{2g^2}\Big(\int\diff^2\theta\, \Tr\bigl[W^\alpha W_\alpha\bigr] +c.c.\Bigr)+
 \int \diff^2\bar\theta\,\diff^2\theta\,\Phi^{\dagger}\Exp{-V}\Phi\,. 
\end{equation}
In terms of components we obtain
\begin{align}
\oper{L} & = -\frac{1}{2g^2}\Tr\bigl[F_{\mu\nu}F^{\mu\nu}\bigr] -2\frac{\iunit}{g^2}\Tr\bigl[\lambda \sigma^\mu\oper{D}_\mu \bar\lambda\bigr]
+\frac{1}{g^2}\Tr\bigl[D^2\bigr] \nonumber \\
& + \oper{D}_{\mu}\phi_i^{\dagger}\oper{D}^{\mu}\phi_i-\iunit \bar\psi_i\bar\sigma^{\mu}\oper{D}_\mu\psi_i \nonumber \\
&-\Tr\bigl[\phi\phi^{\dagger}D +\iunit\sqrt{2}\psi\phi^{\dagger}\lambda-\iunit\sqrt{2}\phi\bar\psi\bar\lambda-FF^{\dagger}\bigr] \nonumber \\
& + \frac{\partial \oper{W}}{\partial \phi_i}F_i+ \frac{\partial \bar{\oper{W}}}{\partial \phi_i^{\dagger}}F_i^{\dagger}
-\frac{1}{2}\frac{\partial^2\oper{W}}{\partial\phi_i\partial\phi_j}\psi_i\psi_j
-\frac{1}{2}\frac{\partial^2\bar{\oper{W}}}{\partial\phi_i^{\dagger}\partial\phi_j^{\dagger}}\bar\psi_i\bar\psi_j\,,
\end{align}
where index $i$ labels the fundamental representation of the gauge group and $\oper{D}_{\mu}$ are covariant derivatives. 
One can easily include $\theta$ term in the above Lagrangian by replacing
\begin{equation}
\frac{1}{2g^2}\Big(\int\diff^2\theta\, \Tr\bigl[W^\alpha W_\alpha\bigr] +c.c.\Bigr) \to
\frac{1}{4\pi}\mathrm{Im}\Bigl(\tau\,\Tr\int\diff^2\theta\, W^{\alpha}W_\alpha\Bigr)\,.
\end{equation}
 The auxiliary fields $F_i$ and $D^a$ can be eliminated through EoM. The resulting
scalar potential is given as
\begin{equation}
V(\phi) = \sum_{i}\Bigl|\frac{\partial\oper{W}}{\partial \phi_i}\Bigr|-\frac{g^2}{2}\bigl(\phi^\dagger T^a\phi\bigr)^2\,.
\end{equation}

\subsubsection{Fayet-Iliopoulos term}

The $D$-term of a vector superfield has some unique properties. It is gauge covariant and being $\theta^2\bar\theta^2$ component of a superfield, it transforms into a total derivative under
SUSY transformation. Thus, one can always consider the so-called Fayet-Iliopoulos (FI) term
\begin{equation}
c\int\diff\bar\theta^2\diff\theta^2\, V = c\, D
\end{equation}
as a valid term of a SUSY invariant theory. Notice that the non-Abelian analog would be $c\,\Tr[D]$. This is non-zero only for those components of $D$, which corresponds to a $U(1)$ subgroup the gauge group. Thus, there can be as many different FI terms as the number of different $U(1)$ gauge groups in the theory.  

Fayet and Iliopoulos have shown \cite{Fayet}, that presence of such terms in (3+1)-dimensional SUSY gauge theories leads to the spontaneous breakdown of supersymmetry. In the Abelian gauge theories in the five-dimensional space-time dimensions, however, 
FI terms do not break SUSY and can be freely included into the theory. This turns out to be important for the construction of domain walls, since constants appearing in front of FI terms, incidentally called FI constants, are related to vacuum expectation values of Higgs fields.

\subsection{A SUSY invariant theories in (4+1)-dimensions}

Our discussion of SUSY has been, so far, based on the superspace formalism, which is unarguably very 
powerful in (3+1)-dimensions or less. Interestingly, realization of the superspace idea in higher 
dimensions turns out to be rather non-trivial. To be precise, the practical usefulness of maintaining 
manifest SUSY via the superspace
formalism becomes questionable, when the number of supercharges is larger than four. For most of these 
situations, it is just more efficient to work directly with components. However, there are interesting 
alternatives in the case of eight supercharges, like the harmonic superspace approach \cite{Galperin}
or closely related projective superspace approach \cite{Rocek, Unge} (see \cite{Kuzenko} for a review). A more conservative possibility is to 
keep SUSY manifest only partially and use the language of $N=1$ superfields in four dimensions, which 
we developed in the previous subsection. This has been achieved in the work \cite{ArkaniHamed},
where five-dimensional $N=1$ SUSY gauge theories were described in terms of four-dimensional superfields,
but only for the price of sacrificing manifest Lorentz invariance.

In this section, we will adopt similar, but less direct approach.
Our goal is to write down (a bosonic part of) a generic $N=1$ SUSY non-Abelian gauge theory
with minimally interacting matters in (4+1)-dimensions. It will be sufficient for our purposes
to obtain the result in terms of component fields. However, instead of just citing the result, we will
motivate it by drawing an analogy between $N=1$ SUSY in five dimensions and the so-called extended $N=2$
SUSY in four dimensions.
Formally, there are many similarities
between these types of models. Both posses eight supercharges and both have
nearly identical structure of SUSY multiplets. We will use this similarity to our advantage and 
approach the problem in the following way.
First, we establish how to write down SUSY invariant Lagrangians for $N=2$ SUSY in (3+1)-dimensions,
through $N=1$ superfield formalism. 
In this part we rely heavily on the work \cite{Hassan}. After that 
we write down these Lagrangians in terms of componens and comment about how to elevate 
them to five dimensions.
 
There are two kinds of multiplets in $N=2$ four-dimensional SUSY.
The $N=2$ chiral multiplet, also called a \emph{hypermultiplet}, consists of a $N=1$ chiral multiplet $(H,\psi, F) \ni Q$ and anti-chiral multiplet $(\tilde H^{\dagger},\bar{\tilde\psi},\tilde{F}^{\dagger})
\ni \tilde Q^{\dagger}$, where $Q,\tilde Q^{\dagger}$ denotes corresponding superfields.
On the other hand, the $N=2$ vector multiplet is made of a chiral multiplet $(\Sigma,\zeta, Y) \ni \Phi$ and a $N=1$ vector multiplet $(W_{\mu},\lambda, D) \ni V$. 
Let us now develop an $N=1$ SUSY manifest way, how to describe these multiplets 
and how to construct supersymmetric $SU(N_c)$ gauge theory with $N_f$ minimally coupled hypermultiplets in the 
fundamental representation. In order to achieve it, let us briefly study 
both kinds multiplets and their Lagrangians separately. 

\subsubsection{Vector multiplet}

In addition to the space-time coordinates, the $N=2$ superspace is made of eight Grassmann coordinates $\theta^\alpha,\bar\theta_{\dot\alpha},\tilde\theta^{\beta},\bar{\tilde\theta}_{\dot \beta}$. A generic superfield $f(x,\theta,\bar\theta,\tilde\theta,\bar{\tilde\theta})$ contains many redundant degrees of freedom and one has to impose two or more constraints on it, in order to obtain irreducible representations.
It turns out (see i.e. \cite{Hassan}, \cite{Bilal}) that an $N=2$ vector multiplet is associated with the generalization of $N=1$ chiral superfield, or in other words, a superfield $\Psi$ constrained by conditions
\begin{equation}
\bar D_{\dot \alpha} \Psi = 0\,, \hspace{5mm} \bar{\tilde D}_{\dot\alpha}\Psi = 0\,. 
\end{equation}  
The covariant derivatives $\tilde D$ are the same as in Eq.~\refer{ch2:covar} with $\theta$ replaced by $\tilde\theta$ everywhere. 
Both constrains can be solved in analogy to the $N=1$ case, by having $\Psi$ depend only on primitive chiral superfields  $(\tilde y,\theta,\tilde\theta)$, where $\tilde y^{\mu} = y^{\mu}+\iunit\tilde\theta\sigma^{\mu}\bar{\tilde\theta} = x^{\mu}+\iunit\theta\sigma^{\mu}\bar\theta
+\iunit\tilde\theta\sigma^{\mu}\bar{\tilde\theta}$. This leads to the decomposition
\begin{equation}
\Psi(\tilde y,\theta,\tilde\theta) = \Psi^{(1)}(\tilde y,\theta)+\sqrt{2}\tilde\theta^\alpha \Psi^{(2)}_\alpha(\tilde y,\theta)+\tilde\theta^2 \Psi^{(3)}(\tilde y,\theta)\,.
\end{equation}
After additional constrains are imposed in order to remove certain unphysical degrees of freedom, $\Psi$ 
can be expressed in terms of $N=1$ superfields as \cite{Bilal}
\begin{equation}
\Psi(\tilde y,\theta,\tilde\theta) = \Phi(\tilde y,\theta)+\sqrt{2}\tilde\theta^\alpha W_\alpha(\tilde y,\theta)+\tilde\theta^2 G(\tilde y,\theta)\,,
\end{equation}  
where $W_\alpha$ is given as in Eq.~\refer{ch2:fsna} and
\begin{equation}
G(\tilde y,\theta) = \int\diff^2\bar\theta\,\bigl[\Phi(\tilde y-\iunit\theta\sigma\bar\theta, \theta, \bar\theta)\bigr]^{\dagger}\Exp{-V(\tilde y-\iunit\theta\sigma\bar\theta, \theta, \bar\theta)}\,.
\end{equation}
The $\diff^2\bar\theta$ integration is meant at fixed $\tilde y$ with $\Phi(x, \theta, \bar\theta)$ given in Eq.~\refer{ch3:chiralsup} and 
$V(x,\theta,\bar\theta)$ in Eq.~\refer{ch3:vecsup}.
We see that $N=2$ chiral superfield (also called a vector superfield) really contains (if in a rather complicated way) only two $N=1$ superfields, that is the chiral superfield $\Phi$ and the vector superfield $V$, as claimed in the beginning of this subsection.

The most general $N=2$ Lagrangian containing $SU(N_c)$ vector multiplet, can be written as
\begin{equation}
\oper{L} = \frac{1}{4\pi}\mathrm{Im}\biggl(\Tr\int\diff^2\theta\,\diff^2\tilde\theta\, \oper{F}(\Psi)\biggr)\,,
\end{equation}
where $\oper{F}(\Psi)$ is the so-called \emph{prepotential}. In four dimensions the prepotential 
can be for renormalizable theories at most quadratic function of $\Phi$. If we integrate out 
$\tilde \theta$ variables, we obtain the expansion
\begin{equation}
\oper{L} = \frac{1}{4\pi}\mathrm{Im}\biggl(\frac{1}{2}\int\diff^2\theta\,\frac{\partial^2\oper{F}}{\partial\Phi^a\partial\Phi^b}W^{a\,\alpha}W_{\alpha}^b+
\int\diff^2\theta\diff^2\bar\theta\Bigl(\Phi^{\dagger}\Exp{-V}\Bigr)^a\frac{\partial \oper{F}}{\partial\Phi^a}\biggr)\,,
\end{equation} 
where $a=1,\ldots ,N_c^2-1$ labels generators of $SU(N_c)$ group.
Let us write down the bosonic part of this Lagrangian. For that we choose the prepotential in the form
\begin{equation}\label{ch3:prepot}
\oper{F}(\Phi) = \frac{2\pi \iunit}{g^2}\Phi^{a}\Phi^a\,.
\end{equation}
The result is
\begin{equation}
\oper{L} = \frac{1}{g^2}\Tr\biggl(-\frac{1}{2}F_{\mu\nu}F^{\mu\nu}
+D_\mu\Sigma^{\dagger} D^\mu\Sigma+\frac{1}{2}D^2-\Sigma^{\dagger}\comm{D}{\Sigma}+F^{\dagger}F\biggr)\,,
\end{equation}
where the trace applies for the gauge index.
To turn this expression into the bosonic part of a five-dimensional SUSY gauge theory, we have to elevate partial derivatives and gauge fields into their five-dimensional counterparts and also consider only real adjoint scalar fields $\Sigma^{\dagger\, a} = \Sigma^a$, since one DoF is reserved for one of components
of the five-dimensional gauge field. Notice that this eliminates the fourth term in the above Lagrangian (by virtue of properties of the trace). We can now combine auxiliary fields $F^a$ and $D^a$ into $SU(2)_R$ triplet, which we call $Y_\alpha^{a}$, where 
$\alpha = 1,2,3$. $SU(2)_R$ is an internal symmetry of superspace, which mixes tilded and untilded quantities together. By taking the $SU(2)_R$ frame, so that only the third component $\alpha = 3$ is nonzero, we can write $Y_3^{a}\equiv Y^a$ and $Y_1^a = Y_2^a = 0$. Thus, in five dimensions we arrive at generic $N=1$ SUSY Lagrangian for a vector multiplet
\begin{equation}
\oper{L}_{SYM} = \oper{F}^{ab}(\Phi)\biggl(-\frac{1}{4}F_{MN}^{a}F^{b\, MN}+\frac{1}{2}D_M\Sigma^a D^M\Sigma^b+\frac{1}{2}Y^a Y^b\biggr)\,,
\end{equation}
where $\oper{F}^{ab}(\Phi) = \partial^{2}\oper{F}/\partial\Phi^a\partial \Phi^b$.
But the situation in five dimensions is even more complicated than that. It follows from general principles that $\oper{F}$ can be cubic function of chiral superfields \cite{Seiberg} and careful analysis of SUSY transformation of $\oper{F}(\Phi)$ reveals that also five-dimensional Chern-Simon (CS) term, which is proportional to third derivatives of the prepotential, might be included \cite{Cortes}
\begin{align}
\oper{L}_{CS} & =  \oper{F}^{abc}(\Phi)\biggl[-\frac{1}{24}\varepsilon^{MNPQR}W_M^a\Bigl(F_{NP}^b F_{QR}^c+\frac{1}{2}\comm{W_N}{W_P}^bF_{QR}^c \nonumber \\
& +\frac{1}{16}\comm{W_N}{W_P}^b\comm{W_Q}{W_R}^c\Bigr)\biggr]\,,
\end{align}
where $\oper{F}^{abc}(\Phi) = \partial^{3}\oper{F}/\partial\Phi^a\partial \Phi^b\partial\Phi^c$.
Lastly, as we discussed in the previous subsection, if the gauge group of the theory includes some $U(1)$'s it is always possible to add corresponding FI terms
to $\oper{L}_{SYM}$.

\subsubsection{Hypermultiplet}

As matter fields and gauge fields transform under different representation of the gauge group, they must belong to different multiplets.
As discussed at the beginning of this subsection, hypermultiplet can be understood as a pair of $N=1$ chiral multiplets $Q$, $\tilde Q$.
In order to write down $N=2$ SUSY Lagrangian with $N_f$ minimally coupled hypermultiplets, we have to understand how kinetic terms and potential terms of both chiral superfields must be combined to ensure $N=2$ SUSY invariance. 
It turns out that this is relatively easy. The kinetic term \refer{ch2:kt} for minimally coupled chiral superfield is unchanged, apart from ensuring that normalization of both $Q$ and $\tilde Q$ terms is the same. The superpotential, however, is restricted in $N=2$ SUSY. The reason is that $Q$ and $\tilde Q$ transform as a doublet under $SU(2)_R$ symmetry, which is an internal symmetry of the superspace. The term such as $\oper{W}(Q)$ would obviously break it. Furthermore, the interaction between $Q$ and $\tilde Q$ are restricted only to quadratic (mass) term. Because in renormalizable $N=2$ SUSY only possible interactions are gauge interactions, the coupling between the hypermultiplet and the vector multiplets enters the Lagrangian in form of a cubic interaction, which is the same for all flavors \cite{Hassan}. In conclusion, the Lagrangian for minimally coupled hypermultiplets, can be described in the $N=1$ language as follows\footnote{Notice that $SU(2)_R$ indices are suppressed.}
\begin{equation}
\oper{L} = \int\diff^2\theta\diff^2\bar\theta\,\biggl(Q_i^{\dagger}\Exp{-V}Q_i+\tilde Q_i\Exp{V}\tilde Q_i^{\dagger}\biggr)
+\int\diff^2\theta\,\Bigl(\sqrt{2}\tilde Q_i\Phi Q_i+m_i\tilde Q_i Q_i\Bigr) + c.c.\,,
\end{equation}
where $i=1,\ldots , N_f$. 

To obtain the bosonic part of the five-dimensional SUSY Lagrangian for minimally interacting matter fields we again elevate the bosonic part of the above Lagrangian into five dimensions by considering real adjoint scalars. The result is
\begin{align}
\oper{L}_{matter}  = \Tr \bigl[D_M H_{\lambda} D^M H_{\lambda}^{\dagger}\bigr]  & + \Tr\Bigl[H_{\lambda}^{\dagger}Y_\alpha (\sigma_\alpha)_{\lambda \kappa}H_{\kappa} \nonumber \\
&- (\Sigma H_{\lambda}-H_{\lambda} M)(\Sigma H_{\lambda}-H_{\lambda} M)^{\dagger}\Bigr] \,,
\end{align}  
where $\sigma_\alpha$ are the Pauli matrices \refer{ch2:pauli} and $\lambda, \kappa = 1,2$ label fundamental representation of $SU(2)_R$.
Here, the Higgs fields $H_{\lambda}$ are assembled into $N_c \times N_f$ matrices for convenience.
Thus, combining all our results, the $N=1$ SUSY $SU(N_c) $ gauge theory with $N_f$ hypermultiplets in (4+1) dimensions is defined by the Lagrangian
\begin{equation}\label{ch2:susyg}
\oper{L} = \oper{L}_{SYM}+\oper{L}_{CS}+\oper{L}_{matter}\,.
\end{equation} 

\subsubsection{A supersymmetric gauge theory for domain walls}

We started our discussion of supersymmetry out of interest what kind of theories supports BPS domain walls as their classical solutions.
Let us, therefore, conclude it by a brief account, what conditions must be imposed on a generic model such as \refer{ch2:susyg} in order to ensure existence of a domain wall(s) solution. As we discussed in Ch.~\ref{ch:2}, both existence and type of a topological soliton is determined by a structure of the vacuum manifold. Domain walls are bound to appear, when there are discrete vacua. Supersymmetry, on the other hand, ensures that at least some of the domain wall solutions are BPS domain walls.
Therefore, we only need to clarify what conditions lead to discrete vacua.
 
There are only two of them. First, the gauge group must contain at least
one $U(1)$ as a subgroup, which allows for non-zero FI term. And second, masses of hypermultiplets must not all be zero and they must be (at least partially \cite{Eto2}) non-degenerate.
Existence of FI term and non-degenerate masses guarantee that some vacua will be discrete. For example, in the $U(N_c)$ non-Abelian gauge theory coupled to $N_f$ flavors $N_f > N_c$, with non-degenerate masses and the FI terms, it has been shown be that the number of disconnected vacua is $\tfrac{N_f!}{(N_f-N_c)!N_c!}$ \cite{Arai}. 

To illustrate this, let us write down the bosonic part of $U(N_c)$ gauge theory coupled to $N_f$ hypermultiplets with the FI term and a simple prepotential,
\begin{equation}
\oper{F}(\Phi) = \frac{4\pi \iunit}{g^2}\Tr\bigl(\Phi^2\bigr)\,.
\end{equation}
The normalization is chosen such that $\Phi= \tfrac{1}{\sqrt{2N_c}}\Phi^0 \mathbf{1}_{N_c} +\Phi^{\hat a}T^{\hat a}$, where $\hat a$ runs from $1,\ldots , N_c^2-1$ and $\Tr \bigl[T^{\hat a},T^{\hat b}\bigr] = \tfrac{1}{2}\delta^{\hat a\hat b}$. The values of $U(1)$ and $SU(N_c)$ gauge couplings are taken to be the same for simplicity.
The Lagrangian is given as 
\begin{align}
\oper{L} & = -\frac{1}{4g^2}F_{MN}^{a}F^{a\, MN}+\frac{1}{2g^2}D_M\Sigma^a D^M\Sigma^a+\frac{1}{2g^2}Y_{\alpha}^{a} Y_{\alpha}^{a} - c_\alpha^a Y_\alpha^a \nonumber \\ \label{ch2:slagr3}
&  +\Tr \Bigl[D_M H_{\lambda} D^M H_{\lambda}^{\dagger}+H_{\lambda}^{\dagger}Y_\alpha (\sigma_\alpha)_{\lambda \kappa}H_{\kappa} - (\Sigma H_{\lambda}-H_{\lambda} M)(\Sigma H_{\lambda}-H_{\lambda} M)^{\dagger}\Bigr] \,.
\end{align} 
Here index  $a=0,1,\ldots N_c^2-1$ labels generators of both $U(1)$ and $SU(N_c)$ subgroups.
We take the mass matrix to be totally degenerate  $M = \mathrm{diag}(m_1,\ldots ,m_{N_f})$. 
 
After using $SU(2)_R$ rotation to fix FI parameters as $c_{\alpha} = (0,0,c)$, with $c>0$, we eliminate vector multiplet auxiliary fields $Y_{\alpha}$, obtaining the $D$-term potential in the form
\begin{equation}
V_D = \frac{1}{4}g^2\Tr\Bigl[(c\mathbf{1}_{N_c}-H_1H_1^{\dagger}-H_2H_2^{\dagger})^2+4H_2^{\dagger}H_2H_1^{\dagger}H_1\Bigr]\,.
\end{equation} 
The complete potential is given as a sum of $V_D$ and the $F$-term potential $V_F$ 
\begin{equation}
V = V_D+V_F\,, \hspace{5mm} V_F =  (\Sigma H_{\lambda}-H_{\lambda} M)(\Sigma H_{\lambda}-H_{\lambda} M)^{\dagger}\,.
\end{equation} 
Notice that if the mass matrix $M \sim \mathbf{1}_{N_f}$, the above model possesses $SU(N_f)$ global flavor symmetry of the Higgs fields $H_\lambda \to H_\lambda U_{N_f}$. 
Since we are considering the non-degenerate masses, this symmetry is explicitly broken down to a maximal torus $U(1)^{N_f-1}$.\footnote{We subtracted overall $U(1)$ which is gauged and therefore not a genuine global symmetry.}

To find vacua of this model amounts to finding configurations of the Higgs fields $H_\lambda$ and the adjoint scalar field $\Sigma$ for which $V=0$.
The second term of $V_D$ vanishes if we set one part of the Higgs doublet to be zero. Let us take $H_2 = 0$ and $ H_1 \equiv H$.\footnote{
Tha fact that $H_2$ does not contribute to domain walls can be proven in general, see the appendix in \cite{Isozumi}. } The potential now reads 
\begin{equation}
V = \frac{1}{4}g^2\Tr\Bigl[(c\mathbf{1}_{N_c}-HH^{\dagger})^2\Bigr]+\Tr\Bigl[(\Sigma H-M H)(\Sigma H -M H)^{\dagger}\Bigr].
\end{equation} 
Let us for clarity consider the simplest nontrivial case $N_c = 2$ and $N_f = 3$. The first part of the potential is given in components as 
\begin{align}
c\mathbf{1}_{N_c}- HH^{\dagger} & = c\begin{pmatrix} 1 & 0 \\ 0 & 1 \end{pmatrix} - 
\begin{pmatrix}
H_{00} & H_{01} & H_{02} \\
H_{10} & H_{11} & H_{12} \end{pmatrix}
\begin{pmatrix}
H_{00}^{*} & H_{10}^{*} \\
H_{01}^{*} & H_{11}^{*} \\
H_{02}^{*} & H_{12}^{*}
\end{pmatrix}
 \nonumber \\ & = -
\begin{pmatrix}
\abs{H_{00}}^2+\abs{H_{01}}^2+\abs{H_{02}}^2-c & H_{00}H_{10}^{*}+H_{01}H_{11}^{*}+H_{02}H_{12}^{*} \\
H_{00}^{*}H_{10}+H_{01}^{*}H_{11}+H_{02}^{*}H_{12} & \abs{H_{10}}^2+\abs{H_{11}}^2+\abs{H_{12}}^2-c
\end{pmatrix}\nonumber 
\end{align}
This will be zero if we choose for every row index of $H$, labeled by $r$, a color index $A_r$, such that $H_{rA_r} = \sqrt{c}$. All other components in $r$-th row of $H$ are set to zero. The phase of $H_{rA_r}$ can be always eliminated by using flavor symmetry. Thus, in order to specify a vacuum, we pair up every flavor index $A$ with a particular color index $r$. The list of all flavors $\{A_1\, \ldots\, A_{N_c}\}$ for each color represents a vacuum. In our example we can write down six possibilities $\{0,1\}, \{0,2\}, \{1,0\}, \{1,2\}, \{2,0\}, \{2,1\}$. However, the ordering of colors in the list does not really matter as the global gauge symmetry can always interchange any two colors. Thus, there are only three physically distinguishable combinations, say  $\{0,1\}, \{0,2\}, \{1,2\}$. Moreover, for each vacuum the adjoint scalar is uniquely given as $\Sigma = \mathrm{diag}(m_{A_1},m_{A_2},m_{A_3})$. This is the color-flavor locking. We can easily generalize this argument to a general case to prove the claim, that the number of discrete vacua are $\tfrac{N_f!}{(N_f-N_c)!N_c!}$.



\chapter{Localization of Abelian gauge fields on the domain wall} \label{ch:4} 






In chapter \ref{ch:2} we have seen how scalar particles are localized on the domain wall within a simple toy model. Although, we did not address the question of localization of fermions,
it is worth stressing, that a level of sophistication required to localize fermions is comparable with the one for bosons. For example, as we saw in Ch.~\ref{ch:3}, if we use supersymmetric theory in (4+1)-dimensions and BPS domain wall as a background solution, fermionic zero modes are naturally localized together with bosonic ones as a manifestation of unbroken supersymmetry in the (3+1)-dimensional effective theory. But even if the parent theory is not supersymmetric, there are
natural mechanisms how to localize fermions, for example by considering appropriate Yukawa coupling terms between fermionic and scalar fields \cite{Rubakov}.

In this chapter, we will focus on the remaining ingredient of any realistic model of elementary particles: a gauge field. To be more precise, our goal is to introduce a mechanism, which localizes massless gauge fields on a domain wall.
As we will see, this mechanism will be an essential ingredient for more realistic model building of effective theories, which we are going to undertake in the next chapter. But before we plunge ourselves into technical details, it is necessary to put this mechanism in the right context, by discussing main developments and setbacks of the gauge fields localization issue. 
 
Compared with scalar and fermionic fields, whose zero modes localize quite naturally, localization of gauge fields, turn out to be a difficult problem, which remained unsolved for a very long time. If we set its beginning with the original proposal to use domain wall as a model for a brane by Rubakov and Shaposhnikov \cite{Rubakov}, we count 30 years. 
During that time, there has been many proposals how to solve it (see \cite{Otha} for a brief description of this development), but they either turn out to be inapplicable in all cases or very hard to implement.   
The solution, we are going to present, has been developed quite recently \cite{Otha} and as we will show, do not suffer from any of these drawbacks. Calculations required for its implementation are actually quite simple and in principle, this mechanism can be used to localize Abelian and non-Abelian gauge fields on a domain wall in any dimensions, although its use is best justified in 
supersymmetric (4+1)-dimensional theories with eight supercharges.  However, it is neither unique nor complete solution of the localization problem, as the discussion still continues. For example, it is not yet clear whether the localization mechanism is not destroyed by quantum corrections or whether it can be induced by them. We are not going to address these issues in this text and we will employ the mechanism in the form it has been presented in \cite{Otha}.  

We will describe specifics of the gauge field localization problem in several steps. 
In the first section of this chapter we are going to attempt to localize an Abelian gauge field in a rather (with benefit of hindsight) naive way. We will investigate an effective action of a simple Abelian-Higgs model in $(4+1)$-dimensions, which supports a BPS domain wall solution. We will show that this model has two 
degenerate vacua, both of which spontaneously breaks $U(1)$ gauge symmetry. Thus, we will consider a so-called \emph{Higgs} phase (also a \emph{superconducting} phase) in the bulk, where gauge particles are massive. However, it is well known that at the core of the domain wall, the local $U(1)$
symmetry is restored and thus it should support a massless gauge field, propagating along its (3+1)-dimensional world volume. 
This expectation, however, turns out to be incorrect.
When we identify the physical reason for this setback, it  will be easy to motivate a proposal made by
Dvali and Shifman \cite{Dvali2} to overcome this problem. We will describe the so-called Dvali-Shifman mechanism in the second section. This will lead directly to the the localization mechanism introduced by Ohta and Sakai \cite{Otha}, which is the content of the third and last section.

\section{Abelian-Higgs model}\label{abelhiggs} 

Let us consider a simple model with local $U(1)$ symmetry in (4+1)-dimensions defined by the Lagrangian density
\begin{align}
\oper{L} &= - \frac{1}{4g^2}F_{MN}F^{MN} 
+ \frac{1}{2g^2}\partial_M\sigma\partial^M\sigma 
+ D_M H D^M H^{\dagger} - V\,, \label{ch2:abelian} \\
V &= \frac{g^2}{2}\left(|H|^2 - v^2\right)^2 
+ \left|\sigma H-HM\right|^2\,,
\end{align}
where $\sigma$ is a neutral scalar field and $H = (H_L, H_R)$ is a row vector of charged Higgs fields. The covariant derivatives are given as
\begin{equation}
D_M H = \partial_M H +\iunit A_M H\,
\end{equation}
and  $A_M$ is a $U(1)$ gauge field of the field strength
\begin{equation}
F_{MN} = \partial_MA_N-\partial_N A_M\,.
\end{equation}
The mass matrix $M$ is chosen in such a way to support a domain wall
\begin{equation}
M = \begin{pmatrix}
m & 0 \\
0 & -m \end{pmatrix}\,,
\end{equation} 
with $m > 0$. The vacuum manifold consists of two degenerate vacua, with vacuum expectation values (VEVs) of various fields given as
\begin{equation}
(\sigma, H_L, H_R) = (m, v, 0)\,,\, (-m,0,v)\,.
\end{equation}
Apart from the local $U(1)$ gauge symmetry, there also exists a global $U(1)_A$ flavor symmetry,\footnote{
Phase rotation of $H_{L}$ and $H_{R}$ in the same direction is gauged and the remaining global symmetry is in the opposite direction. Thus the notation 
$U(1)_{A}$. } which acts on the Higgs fields as $H_L \to \Exp{\iunit \alpha}H_L, H_R \to \Exp{-\iunit \alpha}H_R$. Both of these symmetries are spontaneously broken in both vacua. 

Notice that the same gauge coupling $g$ appears not only in front of the kinetic terms for the gauge fields and the neutral scalar field $\sigma$, but also in the quartic interaction term for the Higgs fields. This feature is motivated by supersymmetry. It can be shown, that the Lagrangian \refer{ch2:abelian} can be 
embedded into a supersymmetric model with eight supercharges by adding appropriate bosonic and fermionic fields, which however, do not contribute to a 
domain wall solution. As shown in the previous chapter, this property allows us to construct BPS domain wall solution, which will simplify much of the discussion below. Let us stress, however, that one may repeat all calculations with more generic coupling constant to arrive at essentially same conclusions.

To find a domain wall solution, let $y$ be the coordinate orthogonal to the domain wall and let all fields depend only on this coordinate. 
Also let $A_{\mu} = 0$, where $\mu = 0,1,2,3$ labels coordinates of (3+1)-dimensional world-volume of the domain wall. 
With this setup, the energy density $\oper{E} = -\oper{L}$ can be rewritten as 
\begin{align}
\oper{E} & = \frac{1}{2g^2}\bigl(\partial_y \sigma\bigr)^2 +\bigl|D_y H\bigr|^2+\frac{g^2}{2}\Bigl(\abs{H}^2-v^2\Bigr)^2
+ \bigl|\sigma H-HM\bigr|^2 \nonumber \\ \label{ch2:bog}
& = \frac{1}{2g^2}\Bigl(\partial_y \sigma +g^2 \bigl(\abs{H}^2-v^2\bigr)\Bigr)^2+\bigl|D_y H+\sigma H-HM\bigr|^2 \nonumber \\
& +v^2\partial_y\sigma -\partial_y\Bigl(H(\sigma \mathbf{1}-M)H^{\dagger}\Bigr)\,.
\end{align}
Since the second line in \refer{ch2:bog} is sum of squares, we see that $\oper{E}$ is bounded from below
\begin{equation}\label{ch2:bound}
\oper{E}  \geq  v^2\partial_y\sigma -\partial_y\Bigl(H(\sigma \mathbf{1}-M)H^{\dagger}\Bigr)\,.
\end{equation}
This is the Bogomol'nyi bound we have encountered in chapter \ref{ch:2}.
Solution, that saturates the Bogomol'nyi bound, obeys a set of first order differential equations, called BPS equations
\begin{align} \label{ch2:vec}
\partial_y \sigma +g^2 \bigl(\abs{H}^2-v^2\bigr) & = 0\,, \\ \label{ch2:hyper}
D_y H+\sigma H-HM & = 0\,.
\end{align}

The BPS equations, which arise in the bosonic sector of supersymmetric gauge theories with eight supercharges, are most easily solved using the so-called \emph{moduli matrix method} 
(see \cite{Eto} for a review).
In this method utilities the fact, that BPS equations are divisible into a vector multiplet part (in our case Eq.~\refer{ch2:vec}) and a hypermultiplet
part, Eq.~\refer{ch2:hyper}. This nomenclature corresponds to types of multiplets in SUSY theories with eight supercharges. The hypermultiplet part can be always solved by introduction of a ``Wilson line''
\begin{equation}
S(y) = \mathbf{P}\,\Exp{\int\diff y\, (\sigma+\iunit A_y) }\,,
\end{equation}   
where $\mathbf{P}$ denotes a path ordering.
One can easily verify that decomposition
\begin{equation}\label{ch2:modulmatrix}
H = v S^{-1} H_0\Exp{M y}\,,
\end{equation}
solves the hypermultiplet BPS equation \refer{ch2:hyper} identically for any values of $H_0$. In the case at hand, $H_0$ is a complex vector. But in models with larger gauge group, this object is a matrix. Since it contains free parameters (also called moduli) of the solution, it is called a moduli matrix.
The remaining equation \refer{ch2:vec} can be recast as the so-called \emph{master equation} for the gauge invariant 
quantity  $\Omega = \abs{S}^2$
\begin{equation}\label{ch2:master}
\frac{1}{2 g^2}\partial_y\bigl(\Omega^{-1}\partial_y\Omega\bigr) = v^2\bigl(1-\Omega^{-1}\Omega_0\bigr)\,,
\end{equation}
where $\Omega_0 =H_0\Exp{2M y}H_0^{\dagger}$.
Once the value of $H_0$ is fixed, solving the master equation amounts to evaluation of $\Omega$, from which we can deduce (up to a phase) $S$
and in turn $H$ and $\sigma$. Notice that, since the imaginary part of $S$ never enters the equation \refer{ch2:master}, we cannot determine the value of $A_y$. 
This ambiguity corresponds to our freedom to choose a gauge. In particular, we can always choose $A_y = 0$. Since we want the domain wall solution as a background solution, we will always use this gauge.

\begin{figure}[ht]
\begin{center}
\begin{tabular}{cc}
\includegraphics[height=4cm]{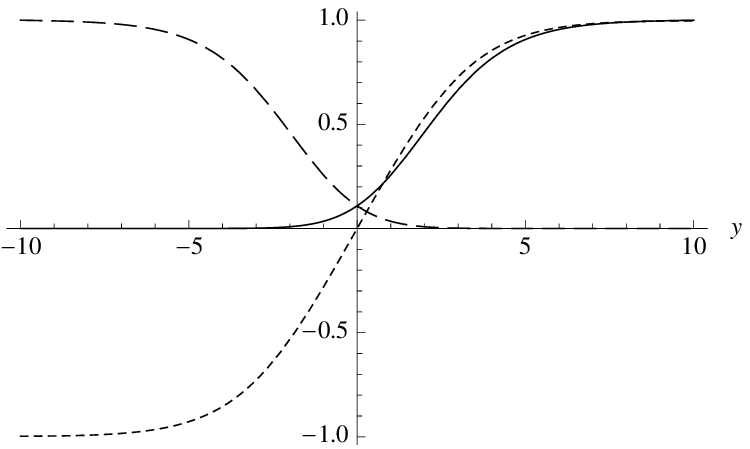} 
\includegraphics[height=4cm]{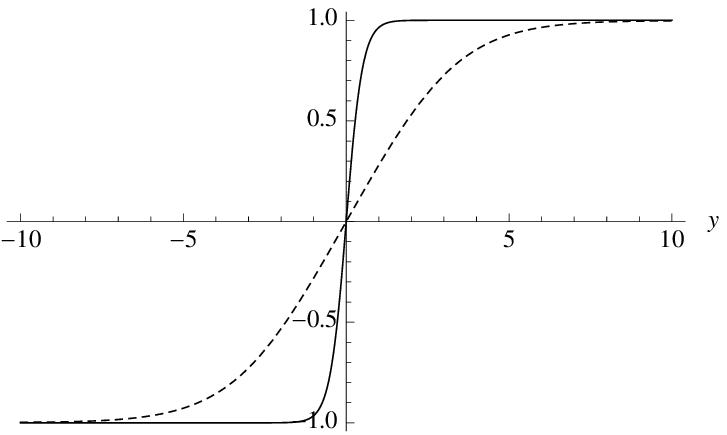}  
\end{tabular}
\caption[Left panel: profiles of Higgs fields and neutral scalar field with finite gauge coupling. Right panel: profile of neutral scalar field for finite and infinite gauge coupling.]{The left panel shows profiles of $H_{L}$ (solid line), 
$H_{R}$ (long-dashed line), 
and $\sigma$ (dashed-line) with finite 
gauge coupling ($g=0.5$). 
The right panel shows a plot of $\sigma$: 
dashed curve for finite ($g=0.5$) gauge coupling 
and solid curve for strong gauge coupling ($g=\infty$). 
The other parameters are $m=v=1$. }
\label{fig:dw_sample}
\end{center}
\end{figure}

Also notice that decomposition \refer{ch2:modulmatrix} of $H$ into the pair $(S,H_0)$ is not unique. For example, a pair $(VS,VH_0)$ with $V\in \mathbb{C}/\{0\}$ gives the same $H$. In consequence, a physically distinguishable pairs $(S,H_0)$ fall into equivalence classes induced by this transformation, called the $V$-transformation.
This equivalence restricts the number of physically distinguishable parameters of the moduli matrix $H_0$. In our case, we can use $V$-equivalence to set one of the moduli of $H_0$ to one, say $H_0 = (1,\Exp{2my_0}\Exp{\iunit \alpha})$, where we have rewritten the second moduli into the more convenient form with  $y_0,\alpha \in \mathbb{R}$.  It is easy to guess what these moduli represent. The $y_0$ is obviously a position of the domain wall in the $y$-direction and $\alpha$ is 
a phase difference of the left and right vacuum between which the domain wall interpolates.  
Using this notation, the master equation \refer{ch2:master} can be recast as
\begin{equation}\label{ch2:master2}
\frac{1}{2g^2}\partial_y\bigl(\Omega^{-1}\partial_y \Omega\bigr) = v^2\Bigl(1 -\Omega^{-1}\bigl(\Exp{2my}+\Exp{-2my+4my_0}\bigr)\Bigr)\,.
\end{equation}
Despite its apparent simplicity, no analytic solution of this equation is known. Therefore, it must be solved numerically. 
The generic solution is shown in Fig~\ref{fig:dw_sample} (left panel).

In Fig.~\ref{fig:dw_sample} (right panel) we show a comparison between cases with finite and infinitely large gauge coupling $g$. We see that there is almost no qualitative difference between these two cases. In fact, as we argued in Ch.~\ref{ch:2}, the low-energy effective action is determined solely from a symmetry breaking pattern and should be insensitive to value of $g$. Thus, at the level of approximation we are interested in, we can safely send $g\to \infty$. In this limit the master equation \refer{ch2:master2} becomes an algebraic equation for $\Omega$ with solution
\begin{equation}
\Omega = \Omega_0 = \Exp{2my}+\Exp{-2my+4my_0}\,.
\end{equation} 
Using  the axial gauge $A_y = 0$ we have $S=\Omega^{1/2}$ and $\sigma = \tfrac{1}{2}\Omega^{-1}\partial_y \Omega$. In terms of the original fields, we can summarize this domain wall solution in the infinite gauge coupling limit as
\begin{align}\label{ch2:sol1}
A_M & = 0\,, \\ \label{ch2:sol2}
\sigma & = m\tanh\bigl(2m(y-y_0)\bigr)\,, \\ \label{ch2:sol3}
H_L & = \frac{v}{\sqrt{1+\Exp{-4m(y-y_0)}}}\,, \\ \label{ch2:sol4}
H_R & = \frac{v\Exp{-2m(y-y_0)}\Exp{\iunit\alpha}}{\sqrt{1+\Exp{-4m(y-y_0)}}}\,.
\end{align}

Having this explicit solution, we can now calculate the low energy effective Lagrangian. For that we use the same procedure as in the chapter \ref{ch:2}, namely the moduli approximation. We promote all moduli into the four dimensional fields 
\begin{equation}
y_0 \to y_0(x^{\mu})\,, \hspace{5mm} \alpha \to \alpha(x^{\mu})\,.
\end{equation}
We plug the solution \refer{ch2:sol1}-\refer{ch2:sol4} with promoted moduli into the original Lagrangian \refer{ch2:abelian}
and take $g\to\infty$. After performing these steps we select the only surviving two derivative term $\oper{L}^{(2)}=
 D_\mu H D^\mu H^{\dagger}$ and integrate it over $y$ to obtain the low-energy effective Lagrangian $\oper{L}_{\mathrm{eff}}$  
\begin{equation}
\oper{L}_{\mathrm{eff}} = \lineint y\, \oper{L}^{(2)} = \lineint y\, D_\mu H D^\mu H^{\dagger}\,. 
\end{equation} 
Notice that here the gauge fields $A_{\mu}$ are non-dynamical and should be eliminated
\begin{equation}
\frac{\partial \oper{L}^{(2)}}{\partial A_{\mu}} = 0 \,, \hspace{5mm} \imply \hspace{5mm} A_{\mu} = -\frac{\iunit }{2v^2}\Bigl(
H\partial_{\mu}H^{\dagger}-\partial_{\mu}H H^{\dagger}\Bigr)\,.
\end{equation}
Inserting this back into $\oper{L}^{(2)}$ and introducing a matrix field $\oper{H}=H^{\dagger}H$, we arrive at the very concise form for the effective Lagrangian
\begin{equation}
\oper{L}_{\mathrm{eff}} = \frac{1}{2v^2}\lineint y\, \Tr\Bigl[\partial_{\mu}\oper{H}\partial^{\mu}\oper{H}\Bigr]\,, 
\end{equation} 
which will prove useful in the future. After straightforward calculations we obtain the result
\begin{equation}\label{ch2:efflagr}
\oper{L}_{\mathrm{eff}}=\frac{2mv^2}{2} \partial_\mu y_0\partial^\mu y_0 + \frac{v^2}{4m}\partial_{\mu}\alpha\partial^{\mu}\alpha\,.
\end{equation}
 
Let us make several observations about $\oper{L}_{\mathrm{eff}}$. First notice that $2mv^2$ is precisely the tension of the domain wall
\begin{equation}\label{ch4:tension}
T = \lineint y\, \oper{E} = \Bigl[v^2 \sigma -H\bigl(\sigma\mathbf{1}-M\bigr)H^{\dagger}\Bigr]_{-\infty}^{\infty} = 2 m v^2\,,
\end{equation}
where we used Eq.~\refer{ch2:bound} with the boundary values taken from Eqs.~\refer{ch2:sol2}-\refer{ch2:sol4}.
Secondly, interpretation of both terms in $\oper{L}_{\mathrm{eff}}$ is clear. The $y_0$ field represents a NG boson for spontaneously broken
translation symmetry in the $y$ direction, while $\alpha$ is NG boson corresponding to spontaneously broken $U(1)_A$ flavor symmetry. 
Most importantly, however, notice that there are no massless modes for gauge fields. On one hand, this is hardly surprising, since taking the $g\to\infty$ limit turned $A_{\mu}$ fields into auxiliary fields. On the other hand, as we already stressed in Ch.~\ref{ch:2}, the low-energy effective Lagrangian is determined from symmetry considerations alone and, as such, $\oper{L}_{\mathrm{eff}}$ should be insensitive to a particular value of $g$, including $g\to\infty$.   

Despite these observations, one may not be convinced that the lack of gauge fields zero modes in \refer{ch2:efflagr} is a genuine result and not an artifact of the strong gauge coupling limit. Let us, therefore, present a different argument, independent on this limit. We will consider the same model as in \refer{ch2:abelian} but we will gauge the global $U(1)_A$ symmetry. Thus, we introduce new gauge fields, say $W_M$, and we add the corresponding kinetic term to the Lagrangian \refer{ch2:abelian}. We also modify the covariant derivatives of the Higgs fields as
\begin{equation}
D_{M}H = \partial_M H + \iunit A_M H +\iunit W_M H Q\,, \hspace{5mm} 
Q = \begin{pmatrix}
1 & 0 \\
0 & -1
\end{pmatrix}\,,
\end{equation}       
where $Q$ is a matrix of charges of $H=(H_L,H_R)$ under $U(1)_A$. 
It is easy to see, that the background solution \refer{ch2:sol1}-\refer{ch2:sol4} remains unchanged, with addition of $W_M = 0$ to the list.
A derivation of the effective Lagrangian for this gauged Abelian-Higgs model follows the same pattern as in the ungauged case and we are not going to 
present it here (for detailed exposition see the Appendix A of \cite{Us1}).
The result is
\begin{equation}\label{ch2:efflagr2}
\oper{L}_{\mathrm{eff}}=-\frac{1}{4e^2}G_{\mu\nu}G^{\mu\nu}+\frac{2mv^2}{2} \partial_\mu y_0\partial^\mu y_0 + \frac{v^2}{4m}\bigl(\partial_{\mu}\alpha+W_\mu\bigr)
\bigl(\partial^{\mu}\alpha+W^\mu\bigr)\,,
\end{equation}
where $G_{\mu\nu} = \partial_{\mu}W_{\nu}-\partial_{\nu}W_{\mu}$.
We see, that gauging the global $U(1)_A$ symmetry presented itself  by the shift $\partial_{\mu}\alpha +W_{\mu}$. If we expand the third term, however, we discover that the gauge fields $W_{\mu}$ are massive, with the mass $\sqrt{\tfrac{v^2}{2m}}$. Therefore, they should not be considered as a part of the low-energy effective Lagrangian, which contains only zero modes. Thus, we returned to the effective Lagrangian \refer{ch2:efflagr}.

\begin{figure}
\begin{center}
{\small
\begin{tikzpicture}[scale=1.5]
\draw (0,-2) -- (0,2);
\draw (2,-2) -- (2,2);
\filldraw[color=blue,opacity=0.4] (0,-2) -- (0,2) -- (-3,2) -- (-3,-2) -- cycle;
\filldraw[color=blue,opacity=0.4] (2,-2) -- (2,2) -- (5,2) -- (5,-2) -- cycle;
\draw (1,2.0) node[above]{$brane$};
\draw (-2,0) node{$Higgs\, phase$};
\draw (4,0) node{$Higgs\, phase$};
\filldraw[black] (1,-1.5) circle (0.5mm) node[below=2mm]{$e^{-}$};
\filldraw[black] (1,1.5) circle (0.5mm) node[above=2mm]{$e^{+}$};
\draw[style=electron] (1,1.5) .. controls (1,1.2) and (0,1.2) .. (0,1.2);
\draw[style=electron] (1,1.5) .. controls (1,1.2) and (2,1.2) .. (2,1.2);
\draw[style=electron] (1,1.5) .. controls (1,1.5) and (0,1.5) .. (0,1.5);
\draw[style=electron] (1,1.5) .. controls (1,1.5) and (2,1.5) .. (2,1.5);
\draw[style=electron] (1,1.5) .. controls (1,1.8) and (0,1.8) .. (0,1.8);
\draw[style=electron] (1,1.5) .. controls (1,1.8) and (2,1.8) .. (2,1.8);
\draw[style=positron] (1,-1.5) .. controls (1,-1.2) and (0,-1.2) .. (0,-1.2);
\draw[style=positron] (1,-1.5) .. controls (1,-1.2) and (2,-1.2) .. (2,-1.2);
\draw[style=positron] (1,-1.5) .. controls (1,-1.5) and (0,-1.5) .. (0,-1.5);
\draw[style=positron] (1,-1.5) .. controls (1,-1.5) and (2,-1.5) .. (2,-1.5);
\draw[style=positron] (1,-1.5) .. controls (1,-1.8) and (0,-1.8) .. (0,-1.8);
\draw[style=positron] (1,-1.5) .. controls (1,-1.8) and (2,-1.8) .. (2,-1.8);
\draw (1,0.5) node{$F \sim \frac{\Exp{-mr}}{r^2}$};
\draw[<->] (0,-0.7) -- (2,-0.7) node[above,midway]{$1/m$};
\end{tikzpicture}
}
\end{center}
\caption[Schematic picture of brane surrounded by a bulk in the Higgs (superconducting) phase.]{\small Schematic picture of brane surrounded by a bulk in the Higgs (superconducting) phase. We see that although $U(1)$ gauge symmetry (here identified with electromagnetism for illustration) is not broken inside the brane, due to the Meissner effect, the electric field lines are pulled into the bulk, effectively making the electromagnetic force short-ranged. The mass of the photon generated by this effect is proportional to the inverse of the width of the brane.}
\label{fig:higgs}
\end{figure}
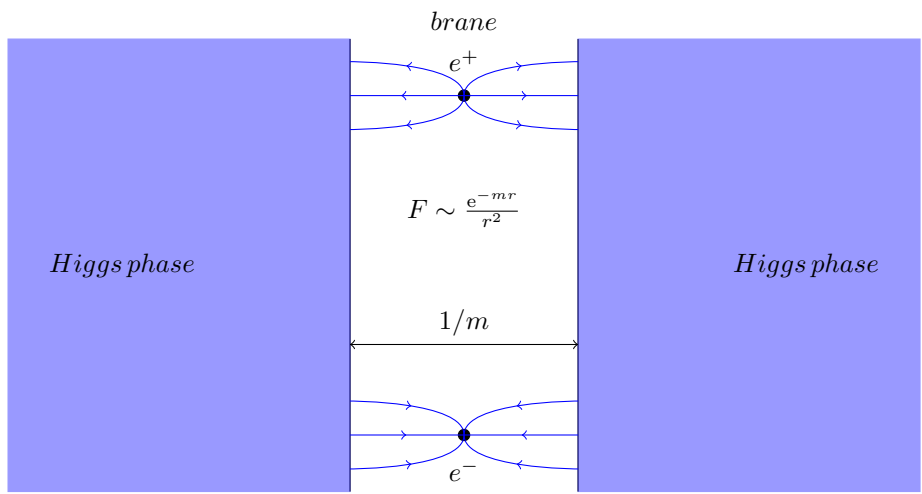

The qualitative reason, why the zero mode of a gauge field cannot be a part of the effective Lagrangian in either gauged or ungauged Abelian-Higgs model, is illustrated on Fig.~\ref{fig:higgs}. We depicted there a brane surrounded by a bulk in the
Higgs phase. To be concrete, let us identify the gauge symmetry of our model with electromagnetism. Then, we may say that the bulk is a superconductor.
Due to a well-known Meissner effect, electric field lines must terminate on the surface of a superconductor. Thus, if we put test charges inside the brane, the force between them will not follow Coulomb law $\sim \tfrac{1}{r^2}$, but rather Yukawa law $\sim \frac{\Exp{-mr}}{r^2}$. This is a signal
that mediating particles, photons, are massive and the force is short-ranged. The mass of the photon is proportional to the inverse of the width of the brane.

In conclusion, we verified that the naive approach, adopted in this section, does not lead to massless gauge fields localized on the domain wall. But we have also learned an important lesson, that we should not break the gauge symmetry in the bulk. Otherwise the Higgs phase outside the domain wall makes the
gauge fields inside massive by an analog of the Meissner effect from superconductivity.  In the next section, we are going to show that a possible solution of this problem is to consider a dual version of Fig.~\ref{fig:higgs}, where we replace a Higgs phase with a confining phase (see Fig.~\ref{fig:super}).

\section{Dvali-Shifman mechanism}


\begin{figure}
\begin{center}
{\small
\begin{tikzpicture}[scale=1.5]
\draw (0,-2) -- (0,2);
\draw (2,-2) -- (2,2);
\filldraw[color=blue,opacity=0.4] (0,-2) -- (0,2) -- (-3,2) -- (-3,-2) -- cycle;
\filldraw[color=blue,opacity=0.4] (2,-2) -- (2,2) -- (5,2) -- (5,-2) -- cycle;
\draw (1,2) node[above]{$brane$};
\draw (-2,0) node{$confining\, phase$};
\draw (4,0) node{$confining\, phase$};
\filldraw[black] (1,-1.5) circle (0.5mm) node[below=1mm]{$e^{-}$};
\filldraw[black] (1,1.5) circle (0.5mm) node[above=1mm]{$e^{+}$};
\draw[style=electron] (1,1.5) -- (1,-1.5);
\draw[style=electron] (1,1.5) .. controls (1.1,1.5) and (1.1,-1.5) .. (1,-1.5);
\draw[style=electron] (1,1.5) .. controls (0.9,1.5) and (0.9,-1.5) .. (1,-1.5);
\draw (0.5,0) node{$F \sim \frac{1}{r^2}$};
\end{tikzpicture}
}
\end{center}
\caption[Schematic picture of a brane surrounded by a confining phase]{\small Schematic picture of a brane surrounded by a confining phase. This time the dual analog of the Meissner effect pushes the electric field lines inside the brane, making the interaction of charges inside it long-ranged and exactly obeying the Coulomb law.}
\label{fig:super}
\end{figure}
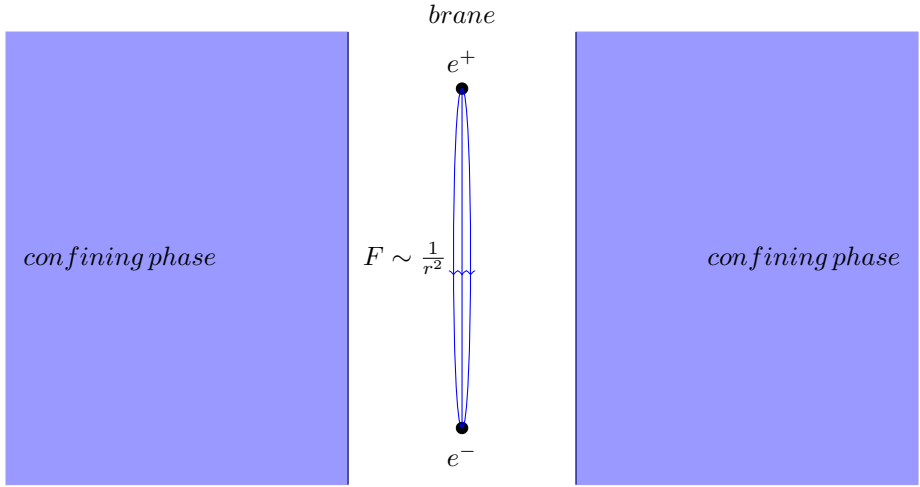

An idea, widely recognized as a plausible way how to localize gauge fields, was put forward in 1997 by  Dvali and Shifman \cite{Dvali2}. This idea
is based on a duality consideration. That is, instead of a model with the Higgs phase in the bulk as in Fig.~\refer{fig:higgs}, we should rather consider a dual scenario, depicted on Fig.~\refer{fig:super}, where the bulk is in the confining phase. The reason is that now the dual analog of the Meissner effect pushes field lines into the brane, making the interaction between
test charges exactly Coulomb-like.

In their paper \cite{Dvali2}, Dvali and Shifman considered an $SU(2)$ gauge theory in (3+1)-dimensions as a toy model. The matter content of this model was arranged in such a way, that the domain wall solution produced a spontaneous breakdown of $SU(2)$ symmetry down to $U(1)$ \emph{inside} the wall. Outside the wall, the bulk remained invariant under $SU(2)$ gauge symmetry, which is confining above some dynamical energy scale $\Lambda$. Dvali and Shifman showed, that if the relevant energy scales of the domain wall are much less than $\Lambda$, the massless $U(1)$ gauge field is trapped on the wall, since all states in the bulk are confined and, therefore, massive with the mass of order $\Lambda$.
From a point of view of an observer inside the brane, it would take an energy exceeding $\Lambda$, to figure out that his/hers Universe is actually surrounded by a confining bulk.

The major drawback of this idea is the fact that it is based on a non-perturbative phenomenon, confinement. The (4+1) dimensional Yang-Mills theory is non-renormalizable and whether it is confining in the same way as in four dimensions is really a conjecture.\footnote{Despite many decades of research, a generally accepted proposition that four-dimensional Yang-Mills theory is confining remains conjectural too.} Therefore, it is difficult to use the Dvali-Shifman mechanism for realistic model building of the brane world scenario without a certain amount of hand waving. Despite this technical complication, the Dvali-Shifman mechanism is still a sound candidate for such a model building (see for example \cite{George} and references therein). 

\section{Ohta-Sakai mechanism}\label{sec:otha}

A mechanism, which realizes confining phase of the bulk at the classical level, was introduced by Ohta and Sakai \cite{Otha} in 2010. The advantage of their approach over the Dvali-Shifman mechanism is a perturbative implementation of the confining phase, allowing for an explicit model building. 
The crux of the Ohta-Sakai mechanism is best explained using the analogy from classical electromagnetism. In usual dielectric medium, the external
electric field $\vec{E}$ is enhanced by induced polarization $\vec{P}$, so that the total electric flux density $\vec{D}$ is given as
\begin{equation}
\vec{D} = \varepsilon_0 \vec{E}+\vec{P} = \varepsilon \vec{E}\,,
\end{equation}
where $\varepsilon_0$ is the vacuum permeability. The dielectric permeability $\varepsilon$ is typically greater then $\varepsilon_0$, because typically an induced electric field points in the same direction as the external one. However, 
it has been recognized \cite{Susskind},\cite{Fukuda} that a classical representation of a vacuum in the confining phase could be described as a perfect dielectric medium $\varepsilon =0$.
This idea suggests, that we can reinterpret the situation of Fig.~\refer{fig:super} from the classical point of view as a continuous change of  dielectric properties of the vacuum. The dielectric permeability $\varepsilon\equiv \varepsilon(y)$ varies along the extra-dimensional coordinate $y$ in such a way that it rapidly decays away from the domain wall $\varepsilon \to 0$ as $y \to \pm\infty$, while staying nonzero near its center.
A relativistic description of dielectric medium with such a non-trivial dependence on $y$ is described by the Lagrangian \cite{Otha}
\begin{equation}
\oper{L} = -\frac{1}{4}\varepsilon(y)F_{MN}F^{MN}\,,
\end{equation}
where $F_{MN}=\partial_MA_N-\partial_NA_M$. But this is nothing else than position-dependent gauge coupling constant
\begin{equation}
\varepsilon(y) = \frac{1}{g(y)^2}\,.
\end{equation}
In their work \cite{Otha}, Ohta and Sakai demonstrated that a spatially varying gauge coupling constant
does localize massless modes of the gauge fields $A_{M}$, independently on details of the
 profile $g(y)$.

In order to localize massless gauge fields on the domain wall in this way, it is 
necessary to introduce some field-theoretical device, which create an 
appropriate spatial profile of the gauge coupling $g(y)$.
Surprisingly, Ohta and Sakai showed that such a device is already present in a certain kind of
theories and surprisingly still, these are exactly those we are interested in! More
specifically, they pointed out that generic properties of supersymmetric gauge theories 
in five dimensions
 naturally incorporates \emph{field}-dependent gauge couplings. Indeed, as we saw in Sec.~\refer{susy45d} and in particular in Eq.~\refer{ch2:slagr2}, the kinetic term for  gauge fields is multiplied by second derivatives of the prepotential. Since for five-dimensional SUSY theories the prepotential can be cubic polynomial \cite{Seiberg}, we see that an inverse of the square of the gauge coupling can depend linearly on the adjoint scalars fields $\Sigma$
\begin{equation}
\frac{1}{g_5^2} \sim \partial^2 f(\Sigma) = c^a\Sigma^a\,,
\end{equation}
where $a$ is running through all generators of the gauge group. If we consider a domain wall solution and study small fluctuations in the background, adjoint scalars usually develops some nontrivial dependence on the $y$-coordinate. In this way, the gauge coupling $g_5(\Sigma)$ becomes position-dependent through the dependence on the $\Sigma(y)$ fields. Moreover, as the term mostly contributing to the low-energy effective Lagrangian is given as
\begin{equation}
\lineint y\, \frac{1}{g_5^2(y)}F_{\mu\nu}F^{\mu\nu}\,,
\end{equation} 
the effective four-dimensional gauge coupling is given simply as
\begin{equation}\label{ch2:effc}
\frac{1}{g_4^2} = \lineint y\, \frac{1}{g_5^2(y)}\,.
\end{equation}
All the remaining work now is to arrange things in such a way that $1/g_5^2(y)$ has appropriate $y$-dependence, such that the integral \refer{ch2:effc} converge to a positive number. If this condition is satisfied, we obtain localized massless gauge fields on the domain wall.   

The actual workings of the Ohta-Sakai mechanism is perhaps best illustrated on concrete examples. In the remaining of this chapter, let us discuss two particular models realizing Ohta-Sakai mechanism. 

\subsection{Four-flavor model}  
 

The so-called four-flavor model is technically the simplest model where the Ohta-Sakai mechanism is realized. The name points to the fact that there are four flavors of Higgs fields. The Lagrangian is 
\begin{align}\label{ch4:lag4}
\oper{L}_{4F} & = -\frac{1}{4g_i^2}F_{i\, MN}F_i^{MN}-a(\sigma)G_{MN}G^{MN}+\frac{1}{2g_i^2}\partial_{\mu}\sigma\partial^{\mu}\sigma+D_M H_i D^M H_i^{\dagger} -V_i\,, \\
V_i & = \frac{g_i^2}{2}\Bigl(H_iH_i^{\dagger}-c_i\Bigr)^2+\abs{\sigma_i H_i-H_i M}^2\,,
\end{align}
where $i=1,2$.
Simply speaking, the four-flavor model is a double copy of the Abelian-Higgs model we discussed in section \refer{abelhiggs}, with additional $U(1)$ gauge symmetry with the field strength $G_{MN} = \partial_M W_N-\partial_N W_M$.
Correspondingly, the notation we use here is almost the same as we used for single Abelian-Higgs model. Notice that the only interaction between both copies comes through the field-dependent gauge coupling-like term $a(\sigma)$. As we discussed in Sec.~\refer{susy45d}, this term represents second derivatives of the prepotential and, as such, it is a linear function of scalar fields $\sigma_i$.
Since (apart from the $a(\sigma)$ factor) both sectors does not interact with each other, the domain wall solution of the Abelian-Higgs model \refer{ch2:sol1}-\refer{ch2:sol4} is a solution of the four-flavor model as well. More precisely, each sector allows this solution independently. If we consider the case, where there is a domain wall in both sectors, we have $\sigma_i = m\tanh(2m(y-y_i))$, where $y_i$ is the 
position of the $i$-th domain wall. Moreover, let us also choose the field-dependent gauge coupling to be 
\begin{equation}\label{ch4:prep}
a(\sigma)=\lambda (\sigma_1-\sigma_2)\,,
\end{equation} where $\lambda >0$.
Following the discussion at the start of this section, it is easy to write down the effective Lagrangian for the four-flavor model
\begin{equation}
\oper{L}_{\mathrm{eff}} = \oper{L}_{\mathrm{eff}\,, 1}+\oper{L}_{\mathrm{eff}\,, 2}- \frac{1}{4e_4^2}G_{\mu\nu}G^{\mu\nu}\,,
\end{equation}        
where
\begin{equation}
\oper{L}_{\mathrm{eff}\,, i}=\frac{2mc_i}{2} \partial_\mu y_i\partial^\mu y_i + \frac{c_i}{4m}\partial_{\mu}\alpha_i\partial^{\mu}\alpha_i
\end{equation}
is the effective Lagrangian of the $i$-th sector (compare with Eq.~\refer{ch2:efflagr}). The effective four-dimensional gauge coupling $e_4$ is given by integration of position-dependent gauge coupling $a(\sigma(y))$ over the extra-dimensional coordinate 
\begin{align}\label{ch2:effcoupl}
\frac{1}{4e_4^2} & = \lineint y\, a(\sigma(y)) = \lambda m\lineint y\, \biggl[\tanh\Bigl(2m(y-y_1)\Bigr) \nonumber \\
&-\tanh\Bigl(2m(y-y_2)\Bigr)\biggr] = \lambda (y_2-y_1)\,. 
\end{align}
Notice that to obtain positive (inverse of the square of the) gauge coupling we must assume $y_2 > y_1$. In effect, in this particular model, the Ohta-Sakai mechanism does not lead to the positive effective gauge coupling $e_4^2$ in the entire moduli space. This feature may not be desired. The problem is that both $y_1$ and $y_2$ are not parameters of the theory, but rather moduli subjected to dynamics. It may happen that during finite time of evolution $y_2-y_1$ flip sign, leading to the ill-defined kinetic term of localized gauge fields. This observation suggests that the four-flavor model might be unstable. Therefore, one would prefer to have the effective gauge coupling depend only on the parameters of the theory and not on moduli of the background solution. In the following, we will show that this can be achieved by considering slightly different model, the three-flavor model.

\subsection{Three-flavor model}

While the four-flavor model can be regarded as the most simple example of an Abelian gauge field localization, the three-flavor model, while achieving the same, is more economic. In the infinitely strong gauge coupling limit,  which we take to obtain background domain wall solution, each gauge group acts as a constraint on hypermultiplet scalars. The Ohta-Sakai mechanism is based on a gauge group containing a subgroup $U(1)\times U(1)$. Thus, we have at least two constraints, making the minimum amount of Higgs fields to form a non-trivial background to be at least three.

The Lagrangian of the three-flavor model reads
\begin{align}
\oper{L}_{3F} & = -\frac{1}{4g_i^2}F_{i\, MN}F_i^{MN}-a(\sigma)G_{MN}G^{MN}+\frac{1}{2g_i^2}\partial_M\sigma_i\partial^M\sigma_i \nonumber \\
& +\abs{D_M H_1}^2+\abs{(D_M-\iunit A_{M})H_2}^2+\abs{(\partial_M+\iunit A_M)H_3}^2-V\,,  \\
V & = (\sigma_1-m)^2\abs{H_1}^2+(\sigma_1-\sigma_2)^2\abs{H_2}^2+\sigma_2^2\abs{H_3}^2\,, 
\end{align}
where $i=1,2$. Here the covariant derivatives are expanded as $D_M = \partial_M+\iunit W_M+\iunit Z_M$ and the field strengths as $F_{1\, MN} = \partial_M W_N-\partial_N W_M$, $F_{2\, MN} = \partial_M A_N-\partial_N A_M$ and $G_{MN}=\partial_M Z_N-\partial_N Z_M$. In the three-flavor model we choose $a(\sigma) = \lambda \sigma_2$.

We would like to verify, that this model leads to a positive four-dimensional effective gauge coupling, which does not depend on the moduli of the background solution.
This amounts to find a domain wall solution and, in particular, calculate the profile of $\sigma_2$.
The BPS equations for the domain wall are (here we take $Z_M = 0$)
\begin{align}
D_y H_1 + (\sigma-m)H_1 & = 0\,, \\
D_y H_2 + (\sigma-\sigma_2-\iunit A_y)H_2 & = 0\,, \\
\partial_y H_3 + (\sigma_2+\iunit A_y)H_3 & = 0\,,
\end{align}
\begin{align}
\frac{1}{g_1^2}\partial_y\sigma_1 & = c_1 - \abs{H_1}^2-\abs{H_2}^2\,, \\
\frac{1}{g_2^2}\partial_y\sigma_2 & = c_2 +\abs{H_2}^2-\abs{H_3}^2\,.
\end{align}    
It can be checked that a domain wall solution, satisfying above equations in the strong gauge coupling limit $g_1^2, g_2^2 \to \infty$, is
\begin{align}
A_M & = W_M = Z_M = 0\,, \\
H_1 & = \sqrt{c_1}\frac{1}{\sqrt{1+\Exp{\eta}/e_0}}\,, \\
H_2 & = \sqrt{c_1}\,\frac{\Exp{\iunit \alpha}}{\sqrt{1+e_0/\Exp{\eta}}}\,, \\
H_3 & = \sqrt{c_2}\,\Exp{-\eta/2}\,, \\
\sigma_1 & = m+\frac{1}{2}\partial_y \ln\Bigl(1+\Exp{\eta}/e_0\Bigr)\,, \\ \label{ch2:soleta}
\sigma_2 & = \partial_y \eta\,,
\end{align} 
where we have denoted $e_0 = \Exp{2m(y-y_0)}$, and where $\eta$ is a solution to the equation\footnote{The correct root of the quadratic equation \refer{ch2:eta} is selected by demanding $\Exp{-\eta}\to 1$ as $c_1\to 0$.} 
\begin{equation}\label{ch2:eta}
\Exp{-\eta} = 1+\frac{c_1/c_2}{1+e_0\,\Exp{-\eta}}\,.
\end{equation}
It is clear from Eq.~\refer{ch2:soleta} that the effective four-dimensional gauge coupling depends only on boundary values of $\eta(y)$.
These can be read off from Eq.~\refer{ch2:eta} leading to the result
\begin{equation}\label{ch4:effcoupl2}
\frac{1}{4e_4^2} = \lineint y\, a(\sigma(y)) = \lambda\lineint y\, \sigma_2(y) = \lambda\lineint y\, \partial_y \eta = \lambda \ln\Bigl(1+\frac{c_1}{c_2}\Bigr)\,.  
\end{equation}
Thus, in contrast with \refer{ch2:effcoupl}, we see that in the three-flavor model the effective four-dimensional gauge coupling depends only  on parameters of the model. Interestingly, it can be shown that $\ln\bigl(1+c_1/c_2\bigr)$ is proportional to the width of the domain wall. 

\subsection{Conclusion}

We have discussed two models as an illustration of the Ohta-Sakai mechanism, where  a successful localization of Abelian gauge fields took place. 
In the four-flavor model the gauge fields are localized between two non-interacting domain walls and the strength of the effective gague coupling naturally depends on their mutual separation. However, as Eq.~\refer{ch2:effcoupl} shows, the ordering of the walls is important, pointing to  a potential instability of this model. In the three-flavor model, the result \refer{ch4:effcoupl2} shows the dependence of the effective gauge coupling on the width of the wall. In contrast to the four-flavor model, the four-dimensional gauge coupling is manifestly positive.
Let us stress, however, that these two models are particular examples of a whole class of models realizing the Otha-Sakai mechanism (see the discussion in \cite{Otha}).
In all these models, however, matter fields and gauge fields are localized without any mutual interaction. In the next chapter we will discuss how to improve on this feature. In particular, we will extend both the four-flavor and the three-flavor models to non-Abelian gauge models and we will show how to localize non-Abelian gauge fields together with minimally interacting matter fields on the walls.

\chapter{Non-Abelian matter fields on the domain wall}\label{ch:5} 




In this chapter we study a non-Abelian extension of the four-flavor model 
of Ch.~\ref{ch:4}, which we call the chiral model. 
In this model, two new features arise in  the effective Lagrangian, which have not so far appeared 
in any other model presented in this text. First, massless non-Abelian gauge fields are localized on the 
domain wall and second, the localized scalar fields are non-trivially interacting under localized
non-Abelian gauge symmetry.

Organization of this chapter is as follows. In the first section we discuss the model together with the background domain wall solution.
Having large gauge and flavor symmetry groups and degenerate masses, the chiral model posses multiple discrete vacua and multiple domain walls. As we want to preserve most of the flavor symmetry, 
we will take two most symmetric vacua and a very special domain wall solution, 
which leaves most of the flavor symmetry unbroken. Such special solution is called \emph{coincident} domain wall solution, since it
can be viewed as composed of multiple, indistinguishable domain walls, sitting at the same position.

In the second section we gauge the unbroken flavor symmetry and show that localized scalar fields interact minimally with localized gauge fields. 
Both in this chapter and in the following chapter, we will pay a lot of attention to the derivation of the effective Lagrangian, since 
the precise character of interactions between localized zero modes has phenomenological significance. We devote the third section to these matters. 
As a conclusion, in the fourth section we discuss the geometrical Higgs mechanism,  a  unique feature of the effective theory, resembling the similar mechanism of D3-branes from string theory.

Most of this chapter, content and notation alike, is adopted from the work \cite{Us1}.

\section{The domain walls in the chiral model}

As a natural extension of the four-flavor model of Ch.~\ref{ch:4}, we consider the 
Yang-Mills-Higgs model 
with $SU(N)_{c}\times U(1)$ gauge symmetry and 
$S[U(N)_{L}\times U(N)_{R}] = SU(N)_{L}\times SU(N)_{R} 
\times U(1)_{A}$ flavor symmetry. 
In order to localize the gauge fields, 
we again introduce two sectors $\oper{L}_1$ and $\oper{L}_2$, but 
only the former is extended to 
Yang-Mills-Higgs system and the latter is the same form 
as \refer{ch2:abelian}.
The second sector couples to the first sector through 
the field dependent coupling of Eq.~\refer{ch4:prep}, which after gauging the flavor symmetry, plays a role of position dependent gauge coupling, localizing massless modes of gauge fields on the domain wall.
The matter content is summarized in Tab.~\ref{table:SY}. 
Since the presence of two factors of $SU(N)$ global 
symmetry resembles the chiral symmetry of QCD, we call 
this Yang-Mills-Higgs system the chiral model. 
\begin{table}
\begin{center}
\begin{tabular}{c|cccccccc}
\hline
  & $SU(N)_{c}$ & $U(1)_1$ & $U(1)_2$ & $SU(N)_{L}$ 
& $SU(N)_{R}$ & $U(1)_{1A}$ & $U(1)_{2A} $ & mass\\ \hline
$H_{1L}$ & $\square$ & 1 & 0 & $\square$ & {\bf 1} 
& 1 & 0 & $m_1{\bf 1}_{N}$\\ 
$H_{1R}$ & $\square$ &  1 & 0 & {\bf 1} 
& $\square$ & $-1$ & 0 & $- m_1{\bf 1}_{N}$ \\ 
$\Sigma_1$ & ${\rm adj}\oplus{\bf 1}$ & 0 & 0 & {\bf 1} 
& {\bf 1} & 0 & 0 & 0\\
$H_{2L}$ & {\bf 1} & 0 & 1 & {\bf 1} 
& {\bf 1} & 0 & $1$  & $m_2$\\ 
$H_{2R}$ & {\bf 1} & 0 & 1 & {\bf 1} 
& {\bf 1}& 0 & $-1$ & $- m_2$ \\ 
$\Sigma_2$ & {\bf 1} & 0 & 0 & {\bf 1} 
& {\bf 1} & 0 & 0 & 0\\
\hline
\end{tabular}
\end{center}
\caption[Quantum numbers of the domain wall sectors in the 
chiral model.]{Quantum numbers of the domain wall sectors in the 
chiral model. Here $\square$ denotes that a particular field belongs to the fundamental representation of the given group, similarly ${\bf 1}$ denotes singlets
and adj denotes adjoint representation.}
\label{table:SY}
\end{table}

The Lagrangian is given by
\begin{align}
\oper{L} &= \oper{L}_1+\oper{L}_2\,, 
\label{eq:Lag_chi_sum}\\
\oper{L}_1 &= \Tr\left[-\frac{1}{2g_1^2}(F_{1MN})^2 
+ \frac{1}{g_1^2}(D_M\Sigma_1)^2 
+ \left|D_M H_1\right|^2 \right] - V_1\,,
\label{eq:Lag_chi}\\
V_1 &= \Tr\left[ \frac{g_1^2}{4} \left(H_1 H_1^\dagger 
- v_1^2 {\bf 1}_{N}\right)^2 + \left|\Sigma_1 H_1
-H_1M_1\right|^2\right]\,,
\label{eq:YM_Higgs_Lag}
\end{align}
with $H_1 = \left(H_{1L},\ H_{1R}\right)$ and
$\oper{L}_2$ being equal to the Lagrangian of Eq.~\refer{ch2:abelian}.
Gauge fields of $U(N)_c=(SU(N)_c\times U(1)_1)/\mathbb{Z}_{\bf N}$ 
are denoted as $W_{1M}$, and adjoint scalar as 
$\Sigma_{1}$. 
The covariant derivatives and the field strength are denoted as
\begin{align} 
D_M\Sigma_1 & = \partial_M \Sigma_1 + \iunit \left[W_{1M},\Sigma_1\right]\,, \\ 
D_M H_1 & = \partial_M H_1 + \iunit W_{1M} H_1\,,
\end{align} 
and $F_{1MN} = \partial_M W_{1N} - \partial_N W_{1M} 
+ \iunit \left[W_{1M},W_{1N}\right]$. 
The mass matrix is given by 
$M_1 = \mathrm{diag}\left(m_1 {\bf 1}_N, -m_1{\bf 1}_N\right)$. 
Let us note that the chiral model reduces to the (ungauged) four-flavor 
model in the limit of $N\to1$, by deleting all of the
$SU(N)$ groups. 
The second sector is needed to realize the 
field-dependent gauge coupling function, which we will discuss in the 
subsequent section. 
In the rest of this section, we focus only on the 
first sector ($i=1$) and suppress the index $i$.
The symmetry transformations act on the fields as
\begin{eqnarray}
H = \left(H_{L},H_{R}\right) &\to& U_{c}\left(H_{L},H_{R}\right)
\left(
\begin{array}{cc}
U_{L}\,\Exp{\iunit\alpha} &  \\
 & U_{R}\,\Exp{-\iunit\alpha}
\end{array}
\right)\,,
\label{eq:transf_H}
\\
\Sigma &\to& U_{c}\Sigma U_{c}^\dagger\,,
\label{eq:transf_sigma}
\end{eqnarray}
with $U_{c} \in U(N)_{c}$, $U_{L} \in SU(N)_{L}$, 
$U(N)_{R} \in SU(N)_{R}$ and $e^{\iunit \alpha} \in U(1)_{A}$.

There exist $N+1$ vacua in which the fields develop the 
following VEV
\begin{eqnarray}
H &=& (H_{L}, H_{R}) = v\left(
\begin{array}{cc|cc}
{\bf 1}_{N-r} & & {\bf 0}_{N-r} & \\
& {\bf 0}_{r} & & {\bf 1}_{r}
\end{array}
\right)\,,\\
\Sigma &=& m \left(
\begin{array}{cc}
{\bf 1}_{N-r} & \\
& -{\bf 1}_{r}
\end{array}
\right)\,,
\end{eqnarray}
with $r=0,1,2,\cdots, N$. 
In the $r$-th vacuum, both the local gauge symmetry $U(N)_c$ 
and the global symmetry are broken, but a diagonal global 
symmetries are unbroken (color-flavor-locking) 
\begin{eqnarray}
&&U(N)_c \times SU(N)_L \times SU(N)_R \times U(1)_{A} 
\to 
\nonumber \\
&&SU(N-r)_{L+c}\times SU(r)_L \times 
SU(r)_{R+c}\times SU(N-r)_R\times U(1)_{A+c}\,.
\label{eq:symmetry_r_vacuum}
\end{eqnarray}

We obtain the BPS equations  through the 
Bogomol'nyi completion of the energy density with the 
assumption that all the fields depend on only the
fifth coordinate $y$ and $W_{\mu} = 0$:
\begin{eqnarray}
{\cal E} &=& \Tr\left[
\frac{1}{g^2}\left(D_y\Sigma 
- \frac{g^2}{2}\left(v^2 {\bf 1}_N 
- HH^\dagger\right)\right)^2
+\left|D_yH+\Sigma H-HM\right|^2
\right]\nonumber \\
&+& \partial_y \left\{ \Tr \left[v^2 \Sigma 
- \left(\Sigma H-HM\right)H^\dagger\right]\right\} \nonumber \\
&\geq &  \partial_y \left\{ \Tr \left[v^2 \Sigma 
- \left(\Sigma H-H M \right)H^\dagger\right]\right\}\,.
\end{eqnarray}
This bound is saturated when the following BPS equations 
are satisfied 
\begin{eqnarray}
D_y\Sigma - \frac{g^2}{2}\left(v^2 {\bf 1}_{N} 
- H H^\dagger\right) = 0\,,\label{eq:BPS_SY1}\\
D_y H+\Sigma H-H M = 0\,.\label{eq:BPS_SY2}
\end{eqnarray}
The tension of the domain wall is given by 
\begin{eqnarray}
T &=& \lineint y\, \partial_y 
\left\{\Tr \left[v^2\Sigma 
- \left(\Sigma H-H M\right)H^\dagger\right]\right\} \nonumber\\
&=& v^2 \, {\rm Tr} \left[\Sigma(+\infty) 
- \Sigma(-\infty)\right].
\label{eq:tension_SY}
\end{eqnarray}

We will concentrate on the domain wall which connects the 
0-th vacuum at $y\to \infty$ and the $N$-th vacuum at 
$y \to -\infty$. Its tension is 
\begin{equation}
T_{coin} = 2N v^2m\,,
\end{equation}
which follows from
Eq.~(\ref{eq:tension_SY}). Notice that $T_{coin}$ is $N$ times the tension 
of single domain wall of the Abelian-Higgs model \refer{ch4:tension}.
Thus, we interpret such a domain wall as being composed of $N$ different (but
indistinguishable) 
domain walls.
In the simplest case, all domain walls have the same position,
which corresponds to making an 
ansatz that $H_{L}$, $H_{R}$, $\Sigma$ and $W_{y}$ 
are all proportional to the unit matrix. 
Then the BPS equations (\ref{eq:BPS_SY1}) and 
(\ref{eq:BPS_SY2}) can be identified with 
the BPS equations in  the 
Abelian-Higgs model.
The coincident domain wall solution is
\begin{align}
H_{L} &= v \Exp{-\frac{\psi}{2}}\Exp{m y}~{\bf 1}_{N}\,, \label{eq:sol1_SY}\\
H_{R} &= v \Exp{-\frac{\psi}{2}}\Exp{-m y}~{\bf 1}_{N}\,, \label{eq:sol2_SY}\\
\Sigma + \iunit W_{y} &= \frac{1}{2}\partial_y\psi{\bf 1}_{N}\, \label{eq:sol3_SY}
\end{align}
where $\psi$ is the solution of the master equation
\begin{equation}
\frac{1}{2g^2}\partial_y^2\psi = v^2 \left(1 - \Exp{-\psi} 
 \Exp{2m(y-y_0)} \right)\,.
\end{equation} 
Eq.~(\ref{eq:symmetry_r_vacuum}) shows that the unbroken 
global symmetry for $N$-th vacuum ($H_{L} = 0$, 
$H_{R} = v{\bf 1}_{N}$ and $\Sigma = -m{\bf 1}_{N}$) 
at the left infinity $y \to -\infty$ is 
$SU(N)_L \times SU(N)_{R+c} \times U(1)_{A+c}$, 
whereas that for the $0$-th vacuum ($H_{L} = v{\bf 1}_{N}$, 
$H_{R} = 0$ and $\Sigma = m{\bf 1}_{N}$) 
at the right infinity $y\to\infty$ is 
$SU(N)_{L+c} \times SU(N)_R \times U(1)_{A+c}$.

The domain wall solution further breaks these unbroken 
symmetries because it interpolates the two vacua.
The breaking pattern by the domain wall is
\begin{equation}
U(N)_c \times SU(N)_L \times SU(N)_R \times U(1)_{A} 
\to SU(N)_{L + R+c}.
\end{equation}
This spontaneous breaking of the global symmetry gives 
NG modes on the domain wall as massless degrees of freedom 
valued on the coset similarly to the chiral symmetry breaking 
in QCD\footnote{
Local gauge symmetry $SU(N)_c$ contains global 
symmetry as a constant gauge transformation, which is 
displayed in the above symmetry breaking pattern. 
However, Nambu-Goldstone modes only come from the 
genuine global symmetry which is not locally gauged. 
Hence we do not count the $SU(N)_c$ transformations.}
\begin{equation}
\frac{SU(N)_{L} \times SU(N)_R}{SU(N)_{L + R}} 
\times U(1)_{A}.
\label{eq:NGmodes_wall}
\end{equation}
Since our model can be embedded into a supersymmetric 
field theory, these NG modes ($U(N)$ chiral fields) appear as complex scalar fields
accompanied with additional $N^2$ pseudo-NG modes.\footnote{
One of them is actually a genuine NG mode corresponding to 
the broken translation.}

\subsection{Localization of the matter fields}
\label{sec:chiral}

In the following, we will present 
the low-energy effective Lagrangian, where the massless moduli fields (the matter fields) are localized.
This serves the double purpose of estabilishing our notation and determining the spectrum of 
massless modes of matter fields on the coincident domain wall.

We  employ the moduli matrix formalism 
\begin{align}
H_{L} & = v \Exp{my}S^{-1}\,,\label{eq:solU1_SY}\\
H_{R} & = v \Exp{-my} S^{-1}\Exp{\phi}\,,\label{eq:solU2_SY}\\
\Sigma + \iunit W_{y} & = S^{-1}\partial_y S\,,
\label{eq:solU3_SY}
\end{align}
where $S\in GL(N,{\bf C})$ and $\Omega=S S^\dagger$ is the 
solution of the following 
master equation 
\begin{eqnarray}
\frac{1}{g^2}\partial_y\left( \Omega^{-1}\partial_y \Omega\right) = v^2 
  \left(\mathbf{1}_N - \Omega^{-1}\Omega_0\right)\,, 
\label{eq:wll:master-eq-wall}
\end{eqnarray}
where
\begin{equation*}
\Omega_0 
= \Exp{2my}\mathbf{1}_N+\Exp{-2my}\Exp{\phi}\Exp{\phi^{\dagger}}\,.
\label{eq:omega0}
\end{equation*}
We have used the $V$-transformation to identify 
the moduli $\Exp{\phi}$, which is a complex $N$ by $N$ matrix. 
It can be parametrized by an $N \times N$ hermitian matrix 
$\hat x$ 
and a unitary matrix $U$ as 
\begin{eqnarray}
\Exp{\phi} = \Exp{\hat x}U^{\dagger}, 
\label{eq:xu_decomp}
\end{eqnarray}
where $U$ is nothing but the $U(N)$ chiral fields associated with
the spontaneous symmetry breaking Eq.~(\ref{eq:NGmodes_wall}) and
components of $\hat x$ are the pseudo-NG modes whose existence we promised above.

In the strong gauge coupling limit $g \to \infty$, 
solution of master equation is simply $\Omega = \Omega_0$. 
After fixing the $U(N)_c$ gauge, 
we obtain
\begin{eqnarray}
S=\Exp{\hat x/2}\sqrt{2\cosh (2my-\hat x)}\,. 
\end{eqnarray} 
Let us denote, for brevity 
\begin{equation}
\hat y= 2my-\hat x\,, 
\end{equation}
the Higgs fields are then given as 
\begin{align}
H_{L} & = v \frac{\Exp{\hat y/2}}{\sqrt{2\cosh \hat y}}\,,
\label{eq:solU1_SY3}\\
H_{R} & = v \frac{\Exp{-\hat y/2}}{\sqrt{2\cosh \hat y}} 
U^\dagger\,.
\label{eq:solU2_SY3}
\end{align}
From this solution, one can easily recognize that eigenvalues of $\hat x$
correspond to the positions of the $N$ domain walls in the $y$ direction.
Now we  promote moduli parameters $\hat x$ and 
$U$ to fields on the domain wall world volume, namely 
functions of world volume coordinates $x^\mu$. 
We plug the domain wall solutions 
$H_{L,R}(y; \hat x(x^\mu),U(x^\mu))$ 
into the original Lagrangian $\oper{L}$ in 
Eq.(\ref{eq:Lag_chi}) at $g \to \infty$ 
and pick up the terms quadratic in the derivatives. 
Thus the low energy effective Lagrangian is given by
\begin{equation}
\oper{L}_{\mathrm{eff}} = \lineint y\,\Tr\left[
\partial_\mu H_{L}\partial^\mu H_{L}^\dagger 
+ \partial_\mu H_{R}\partial^\mu H_{R}^\dagger
-v^2 W_\mu W^\mu
\right]\,,
\end{equation}
where
\begin{equation}
 W_\mu=\frac{\iunit}{2v^2}\left[\partial_\mu H_LH_L^\dagger 
-H_L\partial_\mu H_L^\dagger+(L\leftrightarrow R)\right].
\end{equation}
Here we have eliminated the massive gauge field $W_\mu$ 
by using the equation of motion.
Using the solutions for $H_L$ and $H_R$ 
it is possible to find a closed formula for the effective Lagrangian 
up to the second order of derivatives with full nonlinear 
interactions involving moduli fields $\hat x$ and $U$.
The result is the same as in Eq.~\refer{eq:result}, only with covariant deivatives
replaced with partial derivatives. 
Detailed derivation is given in the third section of this chapter.

Here we show the result only in the leading orders of 
$U-\mathbf{1}_N$ and $\hat x$:
\begin{equation}
\oper{L}_{\rm eff} = \frac{v^2}{2m}\Tr
\Bigl(\partial_{\mu}U^{\dagger}\partial^{\mu}U
+\partial_{\mu}\hat x\partial^{\mu}\hat x\Bigr)+\ldots
\end{equation}  
When $N = 1$ and with the redefinitions $U = \Exp{\iunit\alpha}$, 
and $\hat x = 2m y$, this coincides with the effective 
Lagrangian of Eq.~\refer{ch2:efflagr} of the Abelian-Higgs model.
As already discussed, $U$ is $N\times N$ unitary field, which we identify as NG bosons of spontaneously
broken flavor symmetry and $\hat x$ is $N\times N$ Hermitian matrix of pseudo-NG bosons.
We can identfy the ``pion'' field of our model usign the ansatz 
\begin{equation}
\frac{1}{f_\pi}\partial_\mu \pi = \iunit\, U\partial_\mu U^{\dagger}\,,
\end{equation}
where the pion decay constant $f_\pi = \sqrt{\tfrac{v^2}{2m}}$. The effective Lagrangian now reads
\begin{equation}\label{ch5:chirallagr}
\Tr\Bigl(\partial_\mu \pi^{\dagger}\partial^{\mu}\pi\Bigr)
+\frac{v^2}{2m}\Tr\Bigl(\partial_{\mu}\hat x\partial^{\mu}\hat x\Bigr)+\ldots
\end{equation}

\section{Localization of gauge fields}

Let us next introduce gauge fields which are to be 
localized on the domain walls. 
As we learned in the previous chapter, the associated gauge 
symmetry should not be broken by the domain walls. 
Therefore, the symmetry which we can gauge is the unbroken 
symmetry $SU(N)_{L+R}$ itself. 

Let us gauge $SU(N)_{L+R}\equiv SU(N)_V$ and let $A_{\mu}^a$ be the $SU(N)_{V}$ gauge fields.
The  Higgs fields are in the bi-fundamental representation of $U(N)_c$ and $SU(N)_{V}$.
The covariant derivatives of the Higgs fields are modified as
\begin{align}
\tilde D_M H_{1L} & = \partial_M H_{1L} + \iunit W_{1M} H_{1L} 
- \iunit H_{1L}A_{M}\,,\label{eq:covd1}\\
\tilde D_M H_{1R} & = \partial_M H_{1R} + \iunit W_{1M} H_{1R} 
- \iunit H_{1R}A_{M}\label{eq:covd2}\,.
\end{align}
The quantum numbers are summarized in Tab.~\ref{table:gaugedSY}.
\begin{table}
\begin{center}
\begin{tabular}{c|ccccccc}
\hline
  & $SU(N)_{c}$& $U(1)_1$ & $U(1)_2$ &$SU(N)_{V}$  & $U(1)_{1A}$ & $U(1)_{2A}$ & mass\\ \hline
$H_{1L}$ & $\square$ & 1 & 0 & $\square$ & 1  & 0 & $m_1{\bf 1}_{N}$\\ 
$H_{1R}$ & $\square$ & 1 & 0 & $\square$ & $-1$ & 0 & $- m_1{\bf 1}_{N}$ \\ 
$\Sigma_1 $ & ${\rm adj}\oplus{\bf 1}$  & 0 & 0 & {\bf 1} & 0 & 0 & 0 \\
$H_{2L}$ & {\bf 1} & 0 & 1 & {\bf 1} & 0 & 1  & $m_2$\\ 
$H_{2R}$ & {\bf 1} & 0 & 1 & {\bf 1} & 0 & $-1$ & $- m_2$ \\ 
$\Sigma_2 $ & {\bf 1} & 0 & 0 & {\bf 1} & 0 & 0 & 0  \\

\hline
\end{tabular}
\end{center}
\caption[Quantum numbers of the domain wall sectors in gauged chiral model.]{Quantum numbers of the domain wall sectors in gauged chiral model.
The notation is explained in Tab.~\ref{table:SY}.}
\label{table:gaugedSY}
\end{table}

We now introduce a field-dependent gauge coupling function, which is a direct analog
to that of the four-flavor model. 
\begin{equation}
\frac{1}{2e^2(\Sigma)} = \frac{\lambda}{2}
\left(\frac{\Tr(\Sigma_1)}{Nm_1} 
- \frac{\Sigma_2}{m_2}\right)\,.
\label{Lag_gChi2}
\end{equation}
The Lagrangian is given by 
\begin{equation}
\oper{L} = \tilde{\oper{L}}_1+\oper{L}_2
 - \frac{1}{2e^2(\Sigma)}
\Tr\left[G_{MN}G^{MN}\right]\,. 
\label{Lag_gChi}
\end{equation}
The $\tilde {\oper{L}}_1$ in Eq.~(\ref{Lag_gChi}) 
is given by Eq.~(\ref{eq:Lag_chi}) with the covariant 
derivatives replaced with those in Eqs.~(\ref{eq:covd1}) 
and (\ref{eq:covd2}).

First we wish to find the domain wall solutions in this 
extended model.
As before, we make ansatz that all the fields depend on 
only $y$ and $W_{\mu} = A_{\mu} =0$.
Let us first look at the equation of motion of the new 
gauge field $A_M$, which is of the form
\begin{equation}
D_M G^{MN} = J^N,
\label{eq:eom_A}
\end{equation}
where $J_{M}$ stands for the current of $A_{M}$.
Note that the current $J_{M}$ is zero, by definition, if 
we plug the domain wall solution
in the chiral model before gauging the $SU(N)_{L+R}$. 
This is because the domain wall 
configurations do not break $SU(N)_{L+R}$.
Therefore, $A_{M}=0$ is a solution of Eq.~(\ref{eq:eom_A}).

Then, we are left with equations of motion with $A_{M}=0$, 
which are identical to those in the ungauged chiral model 
in the previous sections. 
Therefore, the gauged chiral model admits 
the same domain wall solutions 
as those (Eqs.~(\ref{eq:solU1_SY3}) and (\ref{eq:solU2_SY3})) in the ungauged chiral model.

The next step is to derive the low energy effective theory on 
the domain wall world-volume 
in the moduli approximation as in the previous subsection.
Again, we promote the moduli parameters as the fields on the 
domain wall world-volume 
and pick up the terms up to the quadratic order of the 
derivative $\partial_\mu$.
Similarly to subsection~\ref{sec:chiral}, we utilize 
the strong gauge coupling limit $g_i\to \infty$. 
Let us emphasize that we keep the field-dependent 
gauge coupling function $e(\Sigma)$ finite. 
The spectrum of massless NG modes is unchanged 
by switching on the $SU(N)_{L+R}$ gauge interactions.\footnote{
Tree level mass spectra are unchanged even though the chiral 
symmetry $SU(N)_L \times SU(N)_R$ is 
broken by the $SU(N)_{L+R}$ gauge interactions. 
}

We just repeat the similar computation to those in subsection~\ref{sec:chiral}. 
Again we shall focus on the first sector ${\cal L}_1$ and suppress the index $i=1$ of fields.
Since color gauge fields $W_\mu$ becomes auxiliary fields 
and eliminated through their equations of motion, it is 
convenient to define the covariant derivative only for the 
flavor $SU(N)_{L+R}$ gauge interactions as 
\begin{equation}\label{ch5:covs}
\hat D_\mu H = \partial_\mu H - \iunit H A_{\mu}\,.
\end{equation}

Then we obtain the effective Lagrangian of the first sector as
\begin{eqnarray}
\oper{L}_{1,\rm eff} &=& \lineint y \,
\mathrm{Tr}\Bigl[\hat D_{\mu}H_{L}(\hat D^{\mu}H_{L})^{\dagger}
+\hat D_{\mu}H_{R}(\hat D^{\mu}H_{R})^{\dagger}
-v^2W_{\mu}W^{\mu} \nonumber \\ \label{ch5:lag}
&&-\frac{1}{2e^2(\Sigma)} G_{MN}G^{MN}\Bigr]\,,
\end{eqnarray}
with
\begin{equation}
 W_{\mu}=\frac{\iunit}{2v^2}\left[\hat{D}_\mu H_{L}H_{L}^\dagger
-H_{L}(\hat{D}_\mu H_{L})^\dagger+(L\leftrightarrow R)\right]\,.
\end{equation}
Eliminating $W_\mu$, we obtain 
the following expression for the integrand of the 
effective Lagrangian after some simplifications 
\begin{equation}\label{eq:lagr1}
\oper{L}_{\rm eff} = \frac{1}{2v^2}\int_{-\infty}^{\infty} 
\diff y\, \Tr\Bigl[\oper{D}_{\mu}H_{ab}\oper{D}^{\mu}H_{ba}
\Bigr]\,,
\end{equation}
where we defined fields $H_{ab}$ with the label $ab$ of the adjoint 
representation of the flavor gauge group $SU(N)_{L+R+c}$. The covariant derivatives are given as
\begin{equation}
 \oper{D}_\mu H_{ab}=\partial_{\mu}H_{ab}+\iunit [A_\mu, H_{ab}]\,, \hspace{5mm} H_{ab}\equiv H_a^\dagger H_b\,, \hspace{5mm} a,b=L,R.
\end{equation}
We will describe the procedure to derive the effective 
Lagrangian fully in the next section. 
Here we merely state the result:
\begin{align}
 \oper{L}_{1,{\rm eff}} & = \frac{v^2}{2m}
\Tr\biggl[\oper{D}_{\mu}\hat x\,
 \frac{\cosh(\oper{L}_{\hat x})-1}{\oper{L}_{\hat x}^2
\sinh(\oper{L}_{\hat x})}
\ln\biggl(\frac{1+\tanh(\oper{L}_{\hat x})}
{1-\tanh(\oper{L}_{\hat x})}\biggr)(\oper{D}^{\mu}\hat x) 
\nonumber \\
& +U^{\dagger}\oper{D}_{\mu}U\,
\frac{\cosh(\oper{L}_{\hat x})-1}{\oper{L}_{\hat x}
\sinh(\oper{L}_{\hat x})}
\ln\biggl(\frac{1+\tanh(\oper{L}_{\hat x})}
{1-\tanh(\oper{L}_{\hat x})}\biggr)(\oper{D}^{\mu}\hat x) 
\nonumber \\
&+\frac{1}{2}\oper{D}_{\mu}U^{\dagger}U\frac{1}
{\tanh(\oper{L}_{\hat x})}\ln\biggl(
\frac{1+\tanh(\oper{L}_{\hat x})}{1-\tanh(\oper{L}_{\hat x})}
\biggr)
(U^{\dagger}\oper{D}^{\mu}U)\biggr]\,,
\label{eq:result}  
\end{align}
where 
\begin{equation}
{\cal L}_A(B)=[A,B]
\label{eq:lie_derivative}
\end{equation} 
is a Lie derivative with respect to $A$.
The covariant derivative $\oper{D}_\mu $  is defined by 
\begin{equation}
\oper{D}_\mu U = \partial_\mu U + \iunit \left[A_{\mu}, U\right].
\end{equation}

The above result suggests that the chiral fields $U(x^\mu)$ 
and hermitian fields $\hat x(x^{\mu})$ are 
in the adjoint representation of $SU(N)_{L+R}$. We will prove this assertion in the last section 
of this chapter.

\section{Effective Lagrangian on the domain wall}\label{app2}

In this section we derive our main result (\ref{eq:result}) 
of the effective Lagrangian for the gauged chiral model. 

\subsection{Compact form of gauged nonlinear model}

Our starting point is the Lagrangian \refer{ch5:lag}. We adopt the Einstein summation 
convention for $a=\{L, R\}$ and, for brevity, we ignore the kinetic term for the localized gauge fields $A_{\mu}$.
Thus, we have 
\begin{equation}\label{eq:lagr}
\oper{L}_{\rm eff} =  \int_{-\infty}^{\infty} \diff y\,
\Tr\Bigl[\hat D_{\mu}H_a\hat D^{\mu}H_a^{\dagger}
-v^2W_{\mu}W^{\mu}\Bigr]\,,
\end{equation}
with the covariant derivatives given in Eq.~\refer{ch5:covs} and with the Higgs fields obeying the constraint 
\begin{equation}\label{eq:constraint}
H_aH_a^{\dagger} = v^2\mathbf{1}_N\,,
\end{equation}
coming from the strong gauge coupling limit. We see that this model is actually a nonlinear sigma model.
Let us first eliminate the gauge fields $W_\mu$ to make this fact manifest. 
Using equation of motion we arrive at
\begin{equation}
W_{\mu} = \frac{\iunit}{2v^2}\left[\hat D_{\mu}H_a  
H_a^{\dagger}-H_a \hat D_{\mu} H_a^{\dagger}\right] \,,
\end{equation}
and
\begin{equation}
\hat D_{\mu} H = \partial_{\mu}H - \iunit H A_{\mu}\,.
\end{equation} 
The effective Lagrangian \refer{eq:lagr} can be simplified by using 
the following identities
\begin{gather}
H_a \hat D_{\mu} H_b^{\dagger} = \partial_{\mu}
(H_aH_b^{\dagger})-\hat D_{\mu}H_a H_b^{\dagger}\,, \\
H_a^{\dagger}\hat D_{\mu} H_b = -\hat D_{\mu} H_a^{\dagger}H_b 
+ \oper{D}_{\mu}H_{ab}\,, 
\end{gather}
where
\begin{equation}
\oper{D}_{\mu} H_{ab} = \partial_{\mu}H_{ab} 
+\iunit\comm{A_{\mu}}{H_{ab}}\,, \hspace{1cm} 
H_{ab} \equiv H_a^{\dagger}H_b\,.
\end{equation}
After some algebra we find:
\begin{equation*}
W_{\mu}W^{\mu} = \tfrac{1}{v^2}\hat D_{\mu} H_a 
\hat D^{\mu}H_a^{\dagger} - 
\tfrac{1}{2v^4}\oper{D}_{\mu}H_{ab}\oper{D}^{\mu}
H_{ba}\,.
\end{equation*} 
Plugging above expression back into the \refer{eq:lagr} 
we conclude
\begin{equation}\label{eq:lagr2}
\oper{L}_{\rm eff} 
= \frac{1}{2v^2}\int_{-\infty}^{\infty} \diff y\, 
\Tr\Bigl[\oper{D}_{\mu}H_{ab}
\oper{D}^{\mu}H_{ba}\Bigr]\,.
\end{equation}

\subsection{Effective Lagrangian}\label{ch5:subsec:efflagr}

Now we are ready to perform the integration over the extra-dimensional coordinate. 
The background solution is given as:
\begin{equation}
H =  (H_L, H_R) = \biggl(\frac{v}{\sqrt{2}}
\frac{\Exp{\hat y/2}}{\sqrt{\cosh(\hat y)}},
\frac{v}{\sqrt{2}}\frac{\Exp{-\hat y/2}
U^{\dagger}}{\sqrt{\cosh(\hat y)}}\biggr)\,,
\end{equation}
where $\hat y = my\mathbf{1}_N-\hat x$.
The composite fields $H_{ab} = H_a^{\dagger}H_b$ are easily shown to be
\begin{align}
H_{LL} & = \frac{v^2}{2}\frac{\Exp{\hat y}}{\cosh(\hat y)}\,, \\
H_{LR} & = \frac{v^2}{2}\frac{1}{\cosh(\hat y)}U^{\dagger}
=H_{RL}^\dagger\,, 
\\
H_{RR} & = \frac{v^2}{2}U\frac{\Exp{-\hat y}}{\cosh(\hat y)}
U^{\dagger}\,.
\end{align}
Using these expansions we arrive at the form
\begin{gather}
\oper{L}_{\rm eff} = \frac{v^2}{4}\int_{-\infty}^{\infty} 
\diff y\,\Tr\biggl\{\oper{D}_{\mu}
\frac{\Exp{\hat y}}{\cosh(\hat y)}\oper{D}^{\mu}
\frac{\Exp{\hat y}}{\cosh(\hat y)}
+\oper{D}_{\mu}\frac{1}{\cosh(\hat y)}\oper{D}^{\mu}
\frac{1}{\cosh(\hat y)} \nonumber \\+
U^{\dagger}\oper{D}_{\mu}U\comm{\frac{\Exp{-\hat y}}
{\cosh(\hat y)}}
{U^{\dagger}\oper{D}^{\mu}\Bigl(U\frac{\Exp{-\hat y}}
{\cosh(\hat y)}\Bigr)}
+U^{\dagger}\oper{D}_{\mu}U\comm{\frac{1}
{\cosh(\hat y)}}{\oper{D}^{\mu}\frac{1}
{\cosh(\hat y)}} \nonumber \\+
 \oper{D}_{\mu}U^{\dagger}\oper{D}^{\mu}U
\frac{1}{\cosh^2(\hat y)}\biggr\}\,. \label{master}
\end{gather}
In the following, we would like to carry out the integration
over the extra-dimensional coordinate $y$. 
This can be done in two steps. 
First, we must factorize all quantities depending 
on $y$ (or on $\hat y$) to one term inside the trace, 
effectively reducing our problem to match the general form
\begin{equation}
\int_{-\infty}^{\infty} \diff y\, 
\Tr\Bigl[f(my\mathbf{1}_N-\hat x)M\Bigr]\,,
\end{equation} 
where $M$ is some matrix, independent of $y$ and $f$ 
is some function. In the second step, we diagonalize $\hat x$
\begin{equation*}
\hat x = P^{-1}\mathrm{diag}(\lambda_1,\ldots ,\lambda_N)P\,
\end{equation*}
and use the fact that $f(P^{-1}\hat y P)=P^{-1}f(\hat y)P$. 
This transformation leads to
\begin{equation*}
\int_{-\infty}^{\infty} \diff y\,
\Tr\Bigl[f\bigl(my\mathbf{1}_N
-\mathrm{diag}(\lambda_i)\bigr)PMP^{-1}\Bigr]
= \int_{-\infty}^{\infty} \diff y\,
\sum_{i=1}^{\lambda}f(my-\lambda_i)(PMP^{-1})_{ii}\,.
\end{equation*}
For every term in the sum we can perform a substitution 
$\tilde y = my-\lambda_i$. The key observation is 
that in each term the integration 
is independent on a particular value 
of $\lambda_i$. This allow us to conclude
\begin{equation}\label{eq:ident}
\int_{-\infty}^{\infty} \diff y\, \Tr\Bigl[f(\hat y)M\Bigr] 
= \frac{1}{m}\Tr (M)\int_{-\infty}^{\infty} \diff \tilde y\, 
f(\tilde y)\,.
\end{equation}
It appears as if we have just made a substitution 
$\hat y = \tilde y\mathbf{1}_N$. 
This is possible, of course, only thanks to the 
diagonalization trick and properties of the trace. 
In the subsequent subsections, however, we will refer 
to this procedure as 
if it is just a `substitution', for brevity.

Let us decompose the effective Lagrangian \refer{master} 
into three pieces 
\begin{equation}
\oper{L}_{{\rm eff}} = \oper{T}_{\hat x} + \oper{T}_{U}
+\oper{T}_{mixed}
\end{equation}
and see the outlined procedure for each term.

\subsection{Kinetic term for $U$}

First, let us concentrate only on terms containing double 
derivatives of $U$, which we denote $\oper{T}_{U}$:
\begin{equation*}
\oper{T}_{U} = \frac{v^2}{4}\int_{-\infty}^{\infty} \diff y\, 
\Tr\biggl\{\oper{D}_{\mu}U^{\dagger}\oper{D}^{\mu}U
\frac{1}{\cosh^2(\hat y)}
+\oper{D}_{\mu}U^{\dagger}U\comm{\frac{\Exp{\hat y}}
{\cosh(\hat y)}}{U^{\dagger}\oper{D}^{\mu}U}
\frac{\Exp{-\hat y}}{\cosh(\hat y)}\biggr\}\,,
\end{equation*}
where we have used a fact that inside the commutator 
it is possible to freely interchange
\begin{equation*}
\frac{\Exp{-\hat y}}{\cosh(\hat y)}\to 
-\frac{\Exp{\hat y}}{\cosh(\hat y)}\,,
\end{equation*}
since the difference is just a constant matrix. 
In this way we made $\oper{T}_{U}$ manifestly invariant
under exchange $\hat y \to -\hat y$.

Since in the first factor of $\oper{T}_{U}$ all $\hat y$-dependent quantities are on the right side, we can, according to our 
previous discussion, make use of the identity \refer{eq:ident} and carry out the integration:
\begin{gather*}
\frac{v^2}{4}\int_{-\infty}^{\infty} \diff y\, \Tr\Bigl[\oper{D}_{\mu}U^{\dagger}\oper{D}^{\mu}U\frac{1}{\cosh^2(\hat y)}\Bigr]   
= \frac{v^2}{2m}\Tr\Bigl[\oper{D}_{\mu}U^{\dagger}\oper{D}^{\mu}U\Bigr]\,.
\end{gather*}
For the second term, however, we first use the identity:
\begin{equation}\label{id1}
\comm{f(A)}{B} = \sum_{k=1}^{\infty}\frac{1}{k!}\oper{L}_{A}^k(B)f^{(k)}(A)\,,
\end{equation}
where $\oper{L}_A(B) = \comm{A}{B}$ is a Lie derivative with respect to $A$. Thus
\begin{equation*}
\comm{\frac{\Exp{\hat y}}{\cosh(\hat y)}}{U^{\dagger}\oper{D}^{\mu}U}=
\sum_{k=1}^{\infty}\frac{(-1)^k}{k!}\oper{L}_{\hat x}^k(U^{\dagger}\oper{D}^{\mu}U)\biggl(\frac{\Exp{\hat y}}{\cosh(\hat y)}\biggr)^{(k)}\,.
\end{equation*}
Now all $\hat y$-dependent factors are standing on the right and we can formally exchange $\hat y \to \tilde y$. 
The summation can be carried out easily
\begin{equation}
\sum_{k=1}^{\infty}\frac{(-1)^k}{k!}\oper{L}_{\hat x}^k(U^{\dagger}\oper{D}^{\mu}U)\biggl(\frac{\Exp{\tilde y}}{\cosh(\tilde y)}\biggr)^{(k)} = 
\frac{\Exp{\tilde y-\oper{L}_{\hat x}}}{\cosh(\tilde y-\oper{L}_{\hat x})}(U^{\dagger}\oper{D}^{\mu}U)-
\frac{\Exp{\tilde y}}{\cosh(\tilde y)} U^{\dagger}\oper{D}^{\mu}U\,.
\end{equation}
The formula for $\oper{T}_{U}$ now reads:
\begin{equation}
\oper{T}_{U} = \frac{c}{4m}\int_{-\infty}^{\infty} \diff \tilde y\, \Tr\biggl[\frac{\Exp{-\oper{L}_{\hat x}}}{\cosh(\tilde y-\oper{L}_{\hat x})\cosh(\tilde y)}
(U^{\dagger}\oper{D}^{\mu}U)\oper{D}_{\mu}U^{\dagger}U\biggr]\,.
\end{equation}
Since we started with $\oper{T}_{U}$ invariant under the transformation
$\hat y \to -\hat y$, we should take only even part of the above formula (with respect to the exchange $\oper{L}_{\hat x} \to -\oper{L}_{\hat x}$) 
as the final result
\begin{equation}
\oper{T}_{U} = \frac{c}{4m}\int_{-\infty}^{\infty} \diff \tilde y\, \Tr\biggl[\frac{\cosh(\oper{L}_{\hat x})}{\cosh(\tilde y-\oper{L}_{\hat x})\cosh(\tilde y)}
(U^{\dagger}\oper{D}^{\mu}U)\oper{D}_{\mu}U^{\dagger}U\biggr]\,.
\end{equation}
Using the primitive function 
\begin{gather*}
\int\frac{\diff y }{\cosh(y-\alpha)\cosh(y)} = \frac{1}{\sinh(\alpha)}
\ln\frac{1}{1-\tanh(\alpha)\tanh(y)}\,
\end{gather*}
 we obtain the result to all orders in $\hat x$ 
\begin{equation}
\oper{T}_{U} = \frac{v^2}{4m}\Tr\biggl[\oper{D}_{\mu}U^{\dagger}U\frac{1}{\tanh(\oper{L}_{\hat x})}\ln\biggl(\frac{1+\tanh(\oper{L}_{\hat x})}{1-\tanh(\oper{L}_{\hat x})}\biggr)
(U^{\dagger}\oper{D}^{\mu}U)\biggr]\,.
\end{equation}
Performing the Taylor-expansion of the function 
\begin{equation}
 \frac{1}{\tanh(x)}\ln\biggl(\frac{1+\tanh(x)}{1-\tanh(x)}\biggr) = 2+\frac{2 x^2}{3}-\frac{2 x^4}{45}+\frac{4 x^6}{945}-\frac{2 x^8}{4725}+\frac{4 x^{10}}{93555}+O(x^{12}) \; , 
\end{equation}
we can easily read off coefficients of terms beyond the leading one. For example, the first three terms reads:
\begin{multline}
\oper{T}_{U}  = \frac{v^2}{2m}\Tr\Bigl(\oper{D}_{\mu}U^{\dagger}\oper{D}^{\mu}U\Bigr)-\frac{v^2}{6m}
\Tr\Bigl(\comm{\hat x}{U^{\dagger}\oper{D}_{\mu}U}\comm{\hat x}{\oper{D}^{\mu}U^{\dagger}U}\Bigr)\\
-\frac{v^2}{90m}\Tr\Bigl(\comm{\hat x}{\comm{\hat x}{U^{\dagger}\oper{D}_{\mu}U}}\comm{\hat x}{\comm{\hat x}{\oper{D}^{\mu}U^{\dagger}U}}\Bigr)
+\ldots 
\end{multline}

\subsection{Mixed term}

Mixed term between $\hat x$ and $U$ is given by
\begin{equation*}
\oper{T}_{mixed} = \frac{v^2}{4}\int_{-\infty}^{\infty} \diff y\, \Tr\biggl\{U^{\dagger}\oper{D}_{\mu}U\biggl( \comm{\frac{1}{\cosh(\hat y)}}{\oper{D}^{\mu}\frac{1}{\cosh(\hat y)}}
-\comm{\frac{\Exp{\hat y}}{\cosh(\hat y)}}{\oper{D}^{\mu}\frac{\Exp{-\hat y}}{\cosh(\hat y)}}\biggr)\biggr\}\,.
\end{equation*}
With use of the identity \refer{id1} and
\begin{equation}
\label{id2} \oper{D}_{\mu}f(\hat x) = \sum_{k=0}^{\infty}\oper{L}_{\hat x}^k(\oper{D}_{\mu}\hat x)\frac{f^{(k+1)}(\hat x)}{(k+1)!}\,,
\end{equation}
one can prove the following:
\begin{equation*}
\label{id3} \comm{f(\hat x)}{\oper{D}_{\mu}g(\hat x)} = \sum_{n=2}^{\infty}\frac{1}{n!}\oper{L}_{\hat x}^{n-1}(\oper{D}_{\mu}\hat x)
\Bigl[\Bigl(f(\hat x)g(\hat x)\Bigr)^{(n)}-f^{(n)}(\hat x)g(\hat x)-f(\hat x)g^{(n)}(\hat x)\Bigr]\,.
\end{equation*}
We can use this result to factorize all $\hat y$-dependent quantities to the right and make the substitution $\hat y = \tilde y\mathbf{1}_N$
\begin{multline*}
\oper{T}_{mixed} = \frac{v^2}{4m}\int_{-\infty}^{\infty} \diff \tilde y\, \sum_{n=2}^{\infty}\frac{(-1)^n}{n!}\Tr\Bigl[U^{\dagger}\oper{D}_{\mu}U\oper{L}_{\hat x}^{n-1}
(\oper{D}^{\mu}\hat x)\Bigr]\\ \times\biggl[\biggl(\frac{\Exp{\tilde y}}{\cosh(\tilde y)}\biggr)^{(n)}\frac{\Exp{-\tilde y}}{\cosh(\tilde y)}
+\biggl(\frac{\Exp{-\tilde y}}{\cosh(\tilde y)}\biggr)^{(n)}\frac{\Exp{\tilde y}}{\cosh(\tilde y)}
-2\biggl(\frac{1}{\cosh(\tilde y)}\biggr)^{(n)}\frac{1}{\cosh(\tilde y)}\biggr]\,.
\end{multline*}
Now we are free to perform summation and integration to obtain
\begin{equation}
\oper{T}_{mixed} = \frac{v^2}{2m}\Tr\biggl[U^{\dagger}\oper{D}_{\mu}U\,\frac{\cosh(\oper{L}_{\hat x})-1}{\oper{L}_{\hat x}\sinh(\oper{L}_{\hat x})}
\ln\biggl(\frac{1+\tanh(\oper{L}_{\hat x})}{1-\tanh(\oper{L}_{\hat x})}\biggr)(\oper{D}^{\mu}\hat x)\biggr]\,.
\end{equation}
Performing the Taylor-expansion of the function 
\begin{equation}
\frac{\cosh(x)-1}{x\sinh(x)}\ln\biggl(\frac{1+\tanh(x)}{1-\tanh(x)}\biggr) =
x-\frac{x^3}{12}+\frac{x^5}{120}-\frac{17 x^7}{20160}+\frac{31 x^{9}}{362880}+O(x^{11}) ,
\end{equation}
we can easily read off coefficients of terms beyond the 
leading order in the expansion 
\begin{equation}
\oper{T}_{mixed} = \frac{v^2}{2m}\Tr\Bigl[U^{\dagger}\oper{D}_{\mu}U\comm{\hat x}{\oper{D}^{\mu}\hat x}\Bigr]
-\frac{v^2}{24m}\Tr\Bigl[U^{\dagger}\oper{D}_{\mu}U\comm{\hat x}{\comm{\hat x}{\comm{\hat x}{\oper{D}^{\mu}\hat x}}}\Bigr]+\ldots 
\end{equation}

\subsection{Kinetic term for $\hat x$}

Kinetic term for $\hat x$ is given by
\begin{equation}
\oper{T}_{\hat x} = \frac{v^2}{4}\int_{-\infty}^{\infty} \diff y\, \Tr\biggl\{
\oper{D}_{\mu}\frac{1}{\cosh(\hat y)}\oper{D}^{\mu}\frac{1}{\cosh(\hat y)}
-\oper{D}_{\mu}\frac{\Exp{\hat y}}{\cosh(\hat y)}\oper{D}^{\mu}\frac{\Exp{-\hat y}}{\cosh(\hat y)}
\biggr\}\,.
\end{equation}
We are going to need the identity
\begin{multline}
\label{id4} \Tr\Bigl[\oper{D}_{\mu}f(\hat x)\oper{D}^{\mu}g(\hat x)\Bigr] = \sum_{n=2}^{\infty}\frac{(-1)^n}{n!}\Tr\biggl\{
\oper{L}_{\hat x}^{n-2}(\oper{D}_{\mu}\hat x)\oper{D}^{\mu}\hat x\\\times
\Bigl[\Bigl(f(\hat x)g(\hat x)\Bigr)^{(n)}-f^{(n)}(\hat x)g(\hat x)-f(\hat x)g^{(n)}(\hat x)\Bigr]\biggr\}\,.
\end{multline}
With the aid of this we arrive at
\begin{multline*}
\oper{T}_{\hat x} = \frac{v^2}{4m}\int_{-\infty}^{\infty} \diff \tilde y\,\sum_{n=2}^{\infty}\frac{1}{n!}
\Tr\Bigl[\oper{L}_{\hat x}^{n-2}(\oper{D}_{\mu}\hat x)\oper{D}^{\mu}\hat x\Bigr] \\
\times\biggl[\biggl(\frac{\Exp{\tilde y}}{\cosh(\tilde y)}\biggr)^{(n)}\frac{\Exp{-\tilde y}}{\cosh(\tilde y)}
+\biggl(\frac{\Exp{-\tilde y}}{\cosh(\tilde y)}\biggr)^{(n)}\frac{\Exp{\tilde y}}{\cosh(\tilde y)}
-2\biggl(\frac{1}{\cosh(\tilde y)}\biggr)^{(n)}\frac{1}{\cosh(\tilde y)}\biggr]\,,
\end{multline*}
where we have again employed the diagonalization trick and the identity \refer{eq:ident}. After performing the summation and the integration we obtain
\begin{equation}
\oper{T}_{\hat x} = \frac{v^2}{2m}\Tr\biggl[\oper{D}_{\mu}\hat x\,\frac{\cosh(\oper{L}_{\hat x})-1}{\oper{L}_{\hat x}^2\sinh(\oper{L}_{\hat x})}
\ln\biggl(\frac{1+\tanh(\oper{L}_{\hat x})}{1-\tanh(\oper{L}_{\hat x})}\biggr)(\oper{D}^{\mu}\hat x)\biggr]\,,
\end{equation}
leading to the power series
\begin{equation}
\oper{T}_{\hat x} = \frac{v^2}{2m}\Tr\Bigl[\oper{D}_{\mu}\hat x\oper{D}^{\mu}\hat x\Bigl]+
\frac{v^2}{24m}\Tr\Bigl[\comm{\hat x}{\oper{D}_{\mu}\hat x}\comm{\hat x}{\oper{D}^{\mu}\hat x}\Bigl]+
\ldots 
\end{equation}

Putting all pieces together as 
$\oper{L}_{{\rm eff}} = \oper{T}_{\hat x} + \oper{T}_{U}
+\oper{T}_{mixed}$, we confirmed our final result 
(\ref{eq:result}). 

\section{Discussion}

Let us illustrate nonlinear 
interactions in $\oper{L}_{1,\rm eff}$ up to fourth orders in the 
fluctuations $\hat x$ and $U- \mathbf{1}_{N}$ 
\begin{eqnarray}
\oper{L}_{1,\rm eff}&=&\frac{v^2}{2m}
\Tr\Bigr(\oper{D}_{\mu}U^{\dagger}\oper{D}^{\mu}U
+\oper{D}_{\mu}\hat x\oper{D}^{\mu}\hat x 
+U^{\dagger}\oper{D}_{\mu}U\comm{\hat x}{\oper{D}^{\mu}\hat x}
\nonumber \\
&&
-{1 \over 12}\comm{\oper{D}_{\mu}\hat x}{\hat x}
\comm{\hat x}{\oper{D}^{\mu}\hat x}
+\frac{1}{3}[\oper{D}_{\mu}U^\dagger U, \hat x]
[\hat x, U^\dagger \oper{D}^{\mu}U]
+\cdots \Bigr)\,,
\label{eq:4thorder_nonlinear}
\end{eqnarray}
where $\oper{D}_{\mu}U = \partial_\mu U +\iunit \comm{A_{\mu}}{U}$ and $\oper{D}_{\mu}\hat x = \partial_\mu \hat x +\iunit \comm{A_{\mu}}{\hat x}$. Our result suggests that $U$ and $\hat x$ are in the adjoint representation of the  $SU(N)_{L+R}$ flavor gauge group.
Let us now examine the transformation properties of 
$U$ and $\hat x$ in order to 
demonstrate that they are, indeed, in the adjoint representation. 
The domain wall solution only preserves the diagonal subgroup 
$SU(N)_{L+R+c}$. Eqs.~(\ref{eq:transf_H}) and 
(\ref{eq:transf_sigma}) shows 
the fields transform under the $SU(N)_{L+R+c}$ 
transformations ${\cal U}$ as 
\begin{equation}
H'_L={\cal U} H_L {\cal U}^\dagger\,, \hspace{5mm} 
H'_R={\cal U} H_R {\cal U}^\dagger\,, \hspace{5mm} 
\Sigma'={\cal U} \Sigma {\cal U}^\dagger\,. 
\end{equation} 
Eqs.~(\ref{eq:solU1_SY}) and (\ref{eq:solU2_SY}) show that 
\begin{equation}
S'={\cal U} S {\cal U}^\dagger\,, 
\hspace{5mm} \Exp{\phi'}={\cal U}\Exp{\phi} {\cal U}^\dagger\,, \hspace{5mm}
\Omega'={\cal U}\Omega {\cal U}^\dagger\,.
\end{equation}
The complex moduli $\Exp{\phi}$ is decomposed into hermitian part 
$\Exp{\hat x}$ and unitary part $U$ in Eq.~(\ref{eq:xu_decomp}). 
Since we can express 
$\Exp{2\hat x}=\Exp{\phi} \Exp{{\phi}^\dagger}$, 
and $U=\Exp{-\phi}\Exp{\hat x}$, 
we find that they transform as the adjoint representation 
\begin{equation}
\Exp{2\hat x'}
={\cal U}\Exp{2\hat x} {\cal U}^\dagger\,, 
\hspace{5mm} U'={\cal U} U {\cal U}^\dagger\,. 
\end{equation}

\begin{figure}[ht]
\begin{center}
\includegraphics[width=8cm]{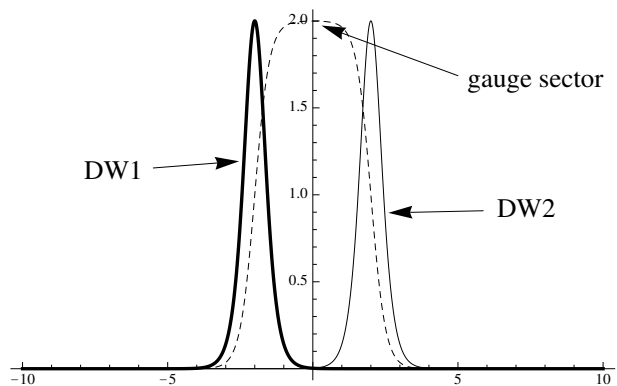}
\caption[The wave functions of the zero modes.]{The wave functions of the zero modes. 
DW1 and DW2 stand for the wave functions 
of the massless matter fields of the $i=1$ 
domain wall and $i=2$ domain wall, respectively 
for strong gauge coupling limit $g_i=\infty$ and 
$m_i=1$. 
The gauge fields are localized between the domain walls.}
\label{fig:schematic}
\end{center}
\end{figure}

Similarly to Eq.~\refer{ch2:effcoupl}, we 
can define the (3+1)-dimensional non-Abelian gauge 
coupling $e_{4}$ 
by integrating (\ref{Lag_gChi2}) and find 
\begin{equation}
\frac{1}{2e_{4}^2} 
=\int \diff y\, \frac{1}{2e^2(\Sigma)} 
= \lambda(y_2-y_1)\,,
\label{eq:4d_NA_gauge_coupling}
\end{equation}
where $y_i$ is the wall position for the $i$-th domain wall 
sector. 
Summarizing, we have obtained the following effective Lagrangian
\begin{equation}
\oper{L}_{\rm eff} = 
\oper{L}_{1, {\rm eff}} + 
\oper{L}_{2, {\rm eff}}- \frac{1}{2e_{4}^2}
\Tr \Bigl[G_{\mu\nu}G^{\mu\nu}\Bigr]\,,
\label{eq:full_eff_Lag}
\end{equation}
where $\oper{L}_{2, {\rm eff}}$ is given in Eq.~\ref{ch2:efflagr}.
This is the main result of this chapter.
We have succeeded in constructing the low-energy 
effective theory in which the matter fields 
(the chiral fields) and the non-Abelian gauge 
fields are localized with the non-trivial interaction.
We show the profile of "wave functions" of localized 
massless gauge field and massless matter fields 
as functions of the coordinate $y$ of the extra 
dimension in Fig.~\ref{fig:schematic}.

As we see from Eq.~(\ref{eq:4thorder_nonlinear}), 
the flavor gauge symmetry $SU(N)_{L+R+c}$ is further 
(partly) broken and the corresponding gauge field 
$A_\mu$ becomes massive, when the fluctuation 
$\phi = \Exp{\hat x} U$ develops non-zero vacuum expectation values.
Especially, $\hat x$ is interesting because 
its non-vanishing (diagonal) values of the fluctuation 
has the physical meaning as the separation 
between walls away from the coincident case. 
For instance, if all the walls are separated, 
$SU(N)_{L+R+c}$ is spontaneously broken to
the maximal $U(1)$ subgroup $U(1)^{N-1}$. 
However, if $r$ walls are still coincident 
and all other walls are separated, we have an unbroken 
gauge symmetry $SU(r)\times U(1)^{N-r+1}$.
Then, a part of the pseudo-NG modes $\hat x$ turn to NG modes associated with
the further symmetry breaking $SU(N)_{L+R+c} \to SU(r) \times U(1)^{N-r+1}$, so that
the total number of zero modes is preserved.
These new NG modes, called the non-Abelian cloud, spread between the separated domain walls \cite{Eto2}.
The flavor gauge fields eat the non-Abelian cloud and get masses which are proportional
to the separation of the domain walls. This is the Higgs mechanism in our model.
This geometrical understanding of the Higgs mechanism is quite similar to D-brane systems
in superstring theory. So our domain wall system provides a genuine prototype
of field theoretical D3-branes.



One more issue remains to be addressed. 
That is the question of sign of the gauge kinetic term. 
In our present model, the positivity of the gauge 
coupling function is assured only when positions of 
walls are properly ordered (see 
Eq.~(\ref{eq:4d_NA_gauge_coupling})), namely only 
in half of the moduli space.
More economical models such as extensions of the three-flavor model of Ch.~\ref{ch:4} 
may not have such moduli and, therefore, the effective 
gauge coupling may be always positive. The purpose of the next chapter is
to demonstrate, that this is indeed the case.



\chapter{Generalized three-flavor model}\label{ch:6} 




In the previous chapter we have considered a non-Abelian extension of the four-flavor model of 
Ch.~\ref{ch:4}. We have found, among other things, that 
the effective four-dimensional gauge coupling $e_4$ of Eq.~\refer{eq:4d_NA_gauge_coupling} 
depends on moduli of the background domain wall solution and that in some part of the moduli 
space $e_4^2$ is negative. This points to the potential instability of the kinetic term of localized gauge 
fields. 
As this is presumably not a desired feature, in this chapter we consider a new model with a
 strictly positive 
effective gauge coupling, which is a function of parameters of the model alone.
From the last subsection of Ch.~\ref{ch:4} we know that the three-flavor model
 has exactly this property in the case of Abelian gauge symmetry. Therefore, 
 in ths chapter we are going to present its non-Abelian extension, 
 which we correspondingly call the generalized three-flavor model.

Apart from positivity of the effective gauge coupling, the generalized three-flavor model shares 
all other properties with the chiral model of the previous chapter with additional interesting 
modifications. In consequence, the structure of this chapter  will be very similar to the previous
 one. 
In the first section we will introduce the model, the domain wall solution and establish the main 
claim of this chapter. 
The second section is devoted 
to derivation of the effective Lagrangian. Compared to the chiral model, 
 the full non-linear interaction between moduli fields in generalized three-flavor model turns out to be much more complicated.
 As we were unable to find a close formula, we will discuss the effective Lagrangian using the approximation, where we take the thickness of the domain wall to be very small.  Interestingly, the effective Lagrangian of the chiral model is contained within the effective Lagrangian of the generalized three-flavor model as its zero order approximation.
We will conclude this chapter with the third section, where we discuss other models, leading to positive effective gauge couplings.

Most of this chapter, content and notation alike, is adopted from the work \cite{Us2}.

\section{Localization of matter and gauge fields}
\label{sc:threeflavor}

In this section we present a model allowing the domain 
wall solution with unbroken non-Abelian global symmetry. 
As the second step, we introduce non-Abelian gauge fields 
for the unbroken symmetry and consider field-dependent gauge coupling 
similar to Eq.~\refer{ch4:effcoupl2}
to localize non-Abelian gauge fields. We will then demonstrate that four-dimensional effective
gauge coupling is determined solely from parameters of the model and that it is strictly positive.

\subsection{Lagrangian with global symmetry 
and domain wall solutions}

Let us consider a five-dimensional 
$SU(N)_c\times U(1)_1\times U(1)_2$ gauge theory 
and $N$ scalar fields $H_1$ ($H_2$) in the fundamental 
representation with the degenerate mass $m$ ($-m$), 
together with a singlet scalar field $H_3$. 
The global symmetry is
$SU(N)_L\times SU(N)_R\times U(1)_A$.
In addition, we introduce adjoint and singlet scalars 
$\Sigma$ and $\sigma$ associated with the gauge group 
$SU(N)_c\times U(1)_1$ and $U(1)_2$, respectively.
We summarize charge 
assignments of matter fields in Tab.~\ref{ch6:table01}.  

\begin{table}
\begin{center}
\begin{tabular}{c|ccc|ccc|c}
\hline
  & $SU(N)_{c}$ & $U(1)_1$ & $U(1)_2$ & $SU(N)_{L}$ 
& $SU(N)_{R}$ & $U(1)_{A}$ & mass\\ \hline
$H_{1}$ & $\square$ & 1 & 0 & $\square$ & {\bf 1} 
& 1 & $m{\bf 1}_{N}$\\ 
$H_{2}$ & $\square$ &  1 & $-1$ & {\bf 1} 
& $\square$ & $-1$ & 0 \\ 
$H_{3}$ & {\bf 1} & 0 & 1 & {\bf 1} 
& {\bf 1} & 0  & 0\\ 
$\Sigma$ & ${\rm adj}\oplus{\bf 1}$ & 0 & 0 & {\bf 1} 
& {\bf 1} & 0 & 0\\
$\sigma$ & ${\bf 1}$ & 0 & 0 & {\bf 1} & {\bf 1} & 0 & 0 \\
\hline
\end{tabular}
\end{center}
\caption[Quantum numbers of fields of the ungauged generalized three-flavor model.]{Quantum numbers of fields of the ungauged generalized three-flavor model. The notation is explained in Tab.~\ref{table:SY}.
}
\label{ch6:table01}
\end{table}

We assume the following Lagrangian 
\begin{align}
\oper{L}   
& =-\frac{1}{2g^2}\Tr\Bigl(G_{MN}G^{MN}\Bigr) 
-\frac{1}{4e^2}F_{MN}F^{MN} 
+\frac{1}{g^2}\Tr\Bigl(D_M\Sigma\Bigr)^2
 +\frac{1}{2e^2}(\partial_M\sigma)^2\nonumber \\ 
\label{eq:lag1}
 & +\Tr\abs{D_MH_1}^2+\Tr\abs{(D_M-\iunit A_M)H_2}^2
+\abs{(\partial_M+i A_M)H_3}^2 -V\,, \\
V & = \Tr\abs{(\Sigma-m\mathbf{1}_N)H_1}^2  
+ \Tr\abs{(\Sigma-\sigma\mathbf{1}_N)H_2}^2
+\abs{\sigma H_3}^2 
\nonumber \\ 
\label{eq:pot1}
& +\frac{1}{4}g^2\Tr\Bigl(c_1\mathbf{1}_N-H_1H_1^{\dagger}
-H_2H_2^{\dagger}\Bigr)^2 + 
\frac{1}{2}e^2\Bigl(c_2+\Tr(H_2H_2^{\dagger})
-\abs{H_3}^2\Bigr)^2\,.
\end{align}
The $U(N)_c = SU(N)_c\times U(1)_1$ gauge coupling and 
gauge fields are denoted by $g$ and an $N\times N$ matrix 
$W_M$ with $M=0,1,2,3,4$.  
The $U(1)_2$ gauge coupling and gauge fields are denoted 
by $e$ and $A_M$. 
Covariant derivatives and field strengths are defined by 
\begin{equation}
\label{eq:cov1} 
D_M H_{1,2}  = \partial_M H_{1,2} +\iunit W_{M} H_{1,2}, 
\quad 
D_M \Sigma  = \partial_M \Sigma +\iunit \comm{W_M}{\Sigma}\,, 
\end{equation}
and $G_{MN} = \partial_MW_N-\partial_NW_M+\iunit \comm{W_M}{W_N}$, 
$F_{MN} = \partial_M A_N-\partial_NA_M$. 

The global symmetry and 
the local gauge symmetry 
act on the fields as 
\begin{gather}
\label{eq:symm1} H_1 \to e^{\iunit \alpha} U_c H_1 U_L \,, \hspace{0.5cm} H_2 \to e^{-\iunit (\alpha+\beta)}U_c H_2 U_R \,, \hspace{0.5cm}
H_3 \to e^{\iunit \beta} H_3\,, \\
\label{eq:symm2} \Sigma \to U_c\Sigma U_c^{\dagger}\,, \hspace{1cm} \sigma \to \sigma\,,
\end{gather}
where $U_L \in SU(N)_L, U_R \in SU(N)_R$, 
$e^{\iunit \alpha} \in U(1)_A$ and $U_c \in U(N)_c$,
$e^{\iunit \beta} \in U(1)_2$.

Let us note that the Lagrangian \refer{eq:pot1} can be 
embedded into a five-dimensional $N=1$ 
supersymmetric gauge theory with $8$ supercharges. 
This fact allows us to obtain BPS domain wall solution 
which we consider in the next subsection. 

Without loss of generality, we can assume the mass 
parameter $m$ to be positive. 
We also assume $c_1 > 0$ and $c_2 > 0$. 
Then, there exist $N+1$ discrete vacua, where scalar fields 
 develop the following vacuum expectation values (VEVs) 
\begin{gather}\label{eq:vev}
H_1 = \sqrt{c_1}\begin{pmatrix}
\mathbf{1}_{N-r} & \\
& \mathbf{0}_{r}\end{pmatrix}\,, \hspace{0.5cm}
H_2 = \sqrt{c_1}\begin{pmatrix}
\mathbf{0}_{N-r} & \\
& \mathbf{1}_{r}\end{pmatrix}\,, \hspace{0.5cm} 
H_3 = \sqrt{c_2+rc_1}\,, \\
\Sigma = m \begin{pmatrix}
\mathbf{1}_{N-r} & \\
 & \mathbf{0}_r\end{pmatrix}\,, \hspace{1cm} \sigma = 0\,,
\end{gather}
where $r =0,1,\,\ldots\,, N$.
The local gauge symmetry is completely broken and only a subgroup 
of the global symmetry remains 
in these vacua. 
The breaking patterns in the $r=0$ and $r=N$ vacua are 
\begin{align*}
&U(N)_c\times SU(N)_L\times SU(N)_R\times U(1)_2\times 
U(1)_A \qquad & \\
& \qquad \qquad \xrightarrow[0-\mathrm{th\ vacuum}]{}  
SU(N)_{L+c}\times SU(N)_R\times U(1)_{A+c}\,, \\
& \qquad \qquad \xrightarrow[N-\mathrm{th\ vacuum}]{}  
SU(N)_{R+c}\times SU(N)_L\times U(1)_{A-c}\,.
\end{align*}

Let us consider domain wall solutions connecting $N$-th 
 ($0$-th) vacuum at left (right) infinity 
$y=-\infty~(y=\infty)$. 
The coincident domain wall solution preserve 
the diagonal subgroup as the largest global\footnote{
Local gauge symmetry $SU(N)_c$ contains global 
symmetry as a constant gauge transformation, which is 
displayed in the above symmetry breaking pattern of 
$r$-th and $N$-th vacuua. 
However, Nambu-Goldstone modes only come from the 
genuine global symmetry which is not locally gauged. 
Hence we do not count the $SU(N)_c$ transformations 
to preserve the color-flavor-locked vacua \refer{eq:vev}. 
} symmetry 
$SU(N)_{L+R}$, so the Nambu-Goldstone modes 
associated with this breaking take values in the coset 
\begin{equation}
\frac{SU(N)_L\times SU(N)_R\times U(1)_A}  
{SU(N)_{L+R}}\,.
\label{eq:nambu_goldstone_mode}
\end{equation}
To find a corresponding domain wall solution we assume that fields depend only on extra-dimensional 
coordinate $y$ and that all gauge fields, except $W_y$ 
and $A_y$, vanish. 
The energy density can be ``completed to a square'' as follows 
\begin{align}
\oper{E} & = \frac{1}{g^2}\Tr\biggl[D_y\Sigma
-\frac{g^2}{2}\Bigl(c_1\mathbf{1}_N-H_1H_1^{\dagger}
-H_2H_2^{\dagger}\Bigr)\biggr]^2 +
\Tr\abs{D_y H_1+(\Sigma-m\mathbf{1}_N)H_1}^2\nonumber \\
& +\frac{1}{2e^2}\biggl(\partial_y\sigma-e^2\Bigl(c_2
+\Tr(H_2H_2^{\dagger}-\abs{H_3}^2)\Bigr)\biggr)^2+
\Tr\abs{D_yH_2+(\Sigma-(\sigma+i A_y)\mathbf{1}_N)H_2}^2 
\nonumber \\
& +\abs{\partial_y H_3 + (\sigma+i A_y)H_3}^2 
+c_2\partial_y \sigma \nonumber \\
& +\partial_y\Tr\biggl[c_1\Sigma -H_1H_1^{\dagger}(\Sigma
-m\mathbf{1}_N)
-H_2H_2^{\dagger}(\Sigma-\sigma\mathbf{1}_N)\biggr]\,.
\end{align}
Thus, we obtain the Bogomol'nyi bound for 
the total energy (per unit volume) 
\begin{equation}
E = \lineint y \,\oper{E} \ge T 
= \lineint y\, \left[c_1\partial_y \Tr(\Sigma) 
+c_2\partial_y \sigma \right]= Nmc_1\,,
\end{equation}
where $T$ is the tension of the domain wall. 
This bound is saturated when the following BPS equations 
are satisfied 
\begin{align}
\label{eq:h1} \partial_yH_1+(\Sigma +\iunit W_y - m\mathbf{1}_N)H_1 & = 0\,, \\
\label{eq:h2} \partial_yH_2+\Bigl(\Sigma +\iunit W_y - (\sigma+\iunit A_y)\mathbf{1}_N\Bigr)H_2 & = 0\,, \\
\label{eq:h3} \partial_yH_3+(\sigma+\iunit A_y)H_3 & = 0\,,
\end{align}
\begin{align}
\label{eq:s1} D_y\Sigma & = \frac{1}{2}g^2\Bigl(c_1\mathbf{1}_N-H_1H_1^{\dagger}-H_2H_2^{\dagger}\Bigr)\,, \\
\label{eq:s2} \partial_y\sigma & = e^2\Bigl(c_2+\Tr(H_2H_2^{\dagger})-\abs{H_3}^2\Bigr)\,.
\end{align}

As usual, we employ the moduli-matrix 
formalism by introducing the moduli fields 
$S(y)\in GL(N,\mathbb{C})$ and 
$\psi(y) \in \mathbb{C}$ 
\begin{equation}
\label{eq:s01}\Sigma+\iunit W_y 
 = S^{-1}\partial_y S, 
\quad 
\sigma+\iunit A_y 
 = \frac{1}{2}\partial_y \psi\,.
\end{equation}
The hypermultiplet part (\ref{eq:h1})-(\ref{eq:h3}) 
can be solved by  
\begin{align}
\label{eq:h01}H_1 & = \Exp{my}S^{-1}H_1^0\,, \\
\label{eq:h02}H_2 & = \Exp{\frac{1}{2}\psi}S^{-1}H_2^0\,, \\
\label{eq:h03}H_3 & = \Exp{-\frac{1}{2}\psi}H_3^0\,,
\end{align}
with complex constant $N\times N$ matrices 
$H_1^0,H_2^0$ and a complex constant $H_3^0$, which 
describe moduli of the  
solution.
The vector multiplet part (\ref{eq:s1})-(\ref{eq:s2}) turns into the master equations for 
gauge-invariant Hermitian fields $\Omega \equiv SS^{\dagger}$ and $\eta \equiv \mathrm{Re}(\psi)$. 
\begin{align}
\label{eq:master01}\partial_y(\partial_y\Omega\Omega^{-1}) 
& = \frac{1}{2}g^2\Bigl(c_1\mathbf{1}_N
-(\Exp{2my}H_1^0H_1^{0\, \dagger}
+\Exp{\eta}H_2^0H_2^{0\, \dagger})\Omega^{-1}\Bigr)\,,\\
\label{eq:master02}\frac{1}{2}\partial_y^2 \eta & 
= e^2\Bigl(c_2+\Exp{\eta}\Tr(H_2^0H_2^{0\, \dagger}\Omega^{-1})
-\Exp{-\eta}\abs{H_3^0}^2\Bigr)\,.
\end{align}

Moduli matrices related by the following 
 $V$-transformations give identical physical fields 
\begin{equation}
(S,\psi,H_1^0,H_2^0,H_3^0) \to (VS,\psi+v, VH_1^0, 
VH_2^0\Exp{-\frac{1}{2}v},H_3^0\Exp{\frac{1}{2}v})\,,
\end{equation}
where $V\in GL(N,\mathbb{C})$ and $v \in \mathbb{C}$.
The equivalence class quotiented by this $V$-transformation 
defines the moduli space of domain walls. 
We can use this freedom to choose the form of the moduli 
matrices
\begin{equation}
H_1^0 = \sqrt{c_1}\mathbf{1}_N\,, \hspace{1cm} 
H_3^0 = \sqrt{c_2}\,.\label{eq:ms}
\end{equation}
Let us also decompose $H_2^0$ as 
\begin{equation}\label{eq:decomp}
H_2^0 = \sqrt{c_1}\Exp{\phi}U^{\dagger}\,,
\end{equation}
where $\phi$ is a Hermitian $N\times N$ matrix and $U$ is 
a unitary $N\times N$ matrix. 
With this choice, the master equations (\ref{eq:master01}) 
and (\ref{eq:master02}) become 
\begin{align}
\label{eq:master1}\partial_y(\partial_y\Omega\Omega^{-1}) 
& = \frac{c_1}{2}g^2\Bigl(\mathbf{1}_N-
\Omega_0\Omega^{-1}\Bigr)\,,\\
\label{eq:master2}\frac{1}{2}\partial_y^2 \eta & 
= e^2\Bigl(c_2+c_1\Exp{\eta}\Tr(e^{2\phi}\Omega^{-1})
-\Exp{-\eta}c_2\Bigr)\,,
\end{align}
where $\Omega_0 = \Exp{2my}\mathbf{1}_N+\Exp{2\phi}\Exp{\eta}$.

No analytic solution of this system of the differential 
equations is known in general. 
However, one can study essential features of solutions, 
if one takes the strong gauge coupling limit 
$g^2, e^2 \to \infty$. 
Eqs. (\ref{eq:master1}) and (\ref{eq:master2}) reduce to 
a system of algebraic equations in this limit:
\begin{align}
\label{eq:strong1} \Omega & = \Exp{2my}\mathbf{1}_N
+\Exp{2\phi}\Exp{\eta}\,, \\
\label{eq:strong2} c_2 & = \Exp{-\eta}c_2
-c_1\Exp{\eta}\Tr(\Exp{2\phi}\Omega^{-1})\,.
\end{align}
It turns out that the effective theory describing massless 
excitations localized on the background solution of the 
equations of this system precisely coincides (at least in 
the lowest order of approximation) with the one obtained from 
(\ref{eq:master1}) and (\ref{eq:master2}) (see the detailed 
discussion in Ref. \cite{Us1}). 
It is therefore sufficient just to study solutions of 
(\ref{eq:strong1}) and (\ref{eq:strong2}). 

If we use the fact that the moduli $\phi$ can be diagonalized by a unitary 
matrix $P$ 
\begin{equation}
\phi = m P^{-1}\mathrm{diag}(y_1,\ldots,y_N)P\,,
\end{equation} 
we can recast Eq.~\refer{eq:strong2} into a polynomial equation of 
the order $N+1$ for $x:= \Exp{-\eta}$ 
\begin{equation}\label{eq:x}
x = 1 + \frac{c_1}{c_2}\sum_{i=1}^N \frac{1}{1+e_i x},\quad
e_i \equiv \Exp{2m(y-y_i)}.
\end{equation} 
If this equation is solved, one can supply its solution 
into Eq.\refer{eq:strong1} to obtain $\Omega$.

In the simplest case, where all walls are coincident $\phi = my_0\mathbf{1}_N$, we can solve equation  \refer{eq:x} 
explicitly ($e_0 \equiv \Exp{2m(y-y_0)}$) to find 
\begin{align}
\label{coin1} \Exp{-\eta} & = \frac{1}{2e_0}\Bigl(e_0-1+\sqrt{(1-e_0)^2+4(1+Nc_1/c_2)e_0}\Bigr)\,, \\
\label{coin2} \Omega & = (\Exp{2my} + \Exp{2my_0}\Exp{\eta})\mathbf{1}_N\,. 
\end{align}
Physical fields can be expressed in terms of $\Omega$ and 
$\sigma$ as 
\begin{align}
\label{eq:coin1} H_1 & = \sqrt{c_1}\frac{\mathbf{1}_N}{\sqrt{1+\Exp{-2m(y-y_0)+\eta}}}\,, \\
\label{eq:coin2} H_2 & = \sqrt{c_1}\frac{U^{\dagger}}{\sqrt{1+\Exp{2m(y-y_0)-\eta}}}\,, \\
\label{eq:coin3} H_3 & = \sqrt{c_2}\Exp{-\eta/2}\,, \\
\label{eq:coin4} \Sigma & = \frac{1}{2}\partial_y \ln\Omega\,, \\
\label{eq:coin5} \sigma & = \partial_y \eta\,, \\
\label{eq:coin6} W_y & = A_y = 0\,,
\end{align}
where we fixed the gauge such that 
$S = \Omega^{1/2}$ and $\mathrm{Im} (\psi) = 0$.

\begin{figure}
\begin{center}
\begin{minipage}[b]{0.49\linewidth}
\centering
\includegraphics[width=\textwidth]{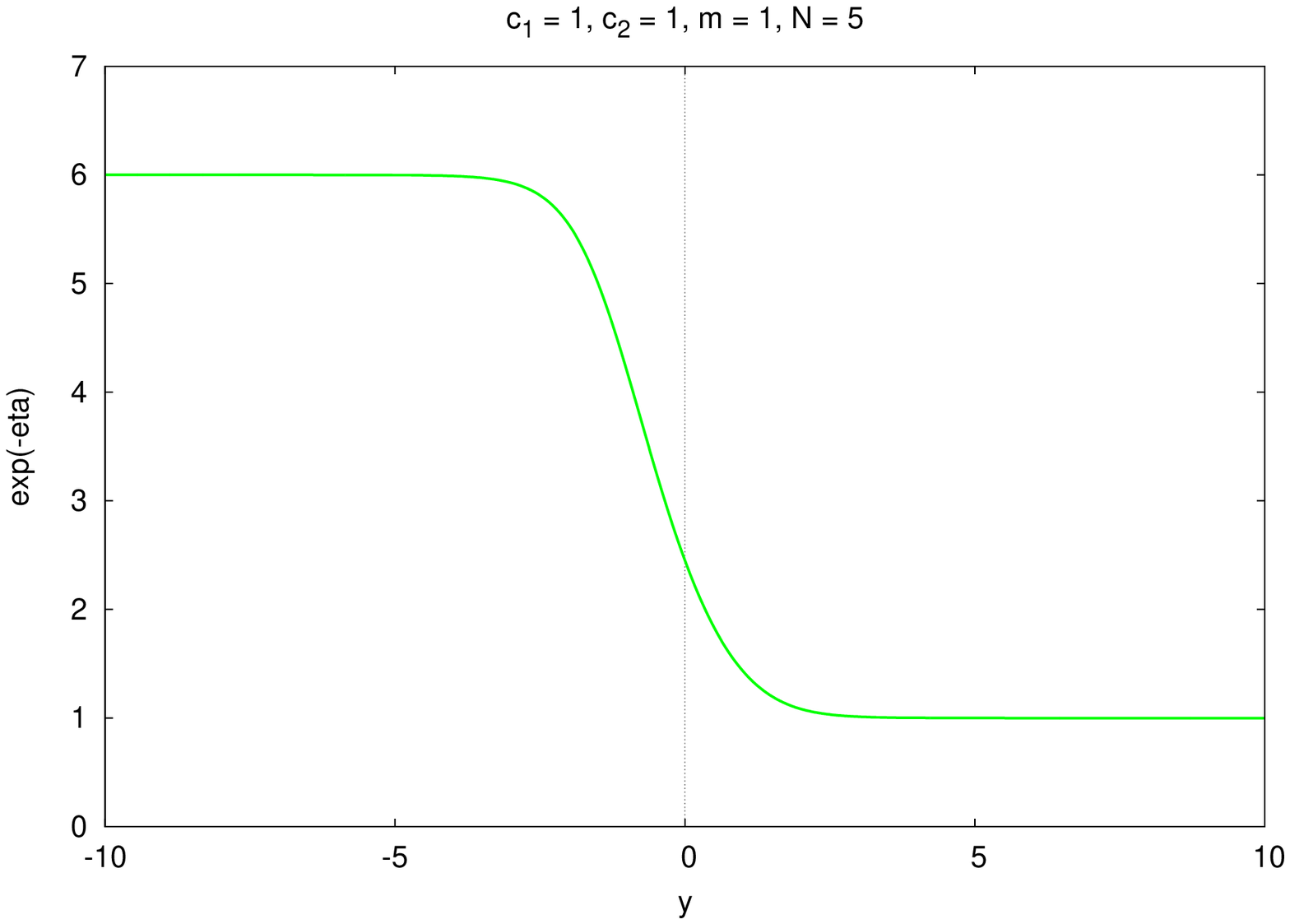}
\end{minipage}
\begin{minipage}[b]{0.49\linewidth}
\centering
\includegraphics[width=\textwidth]{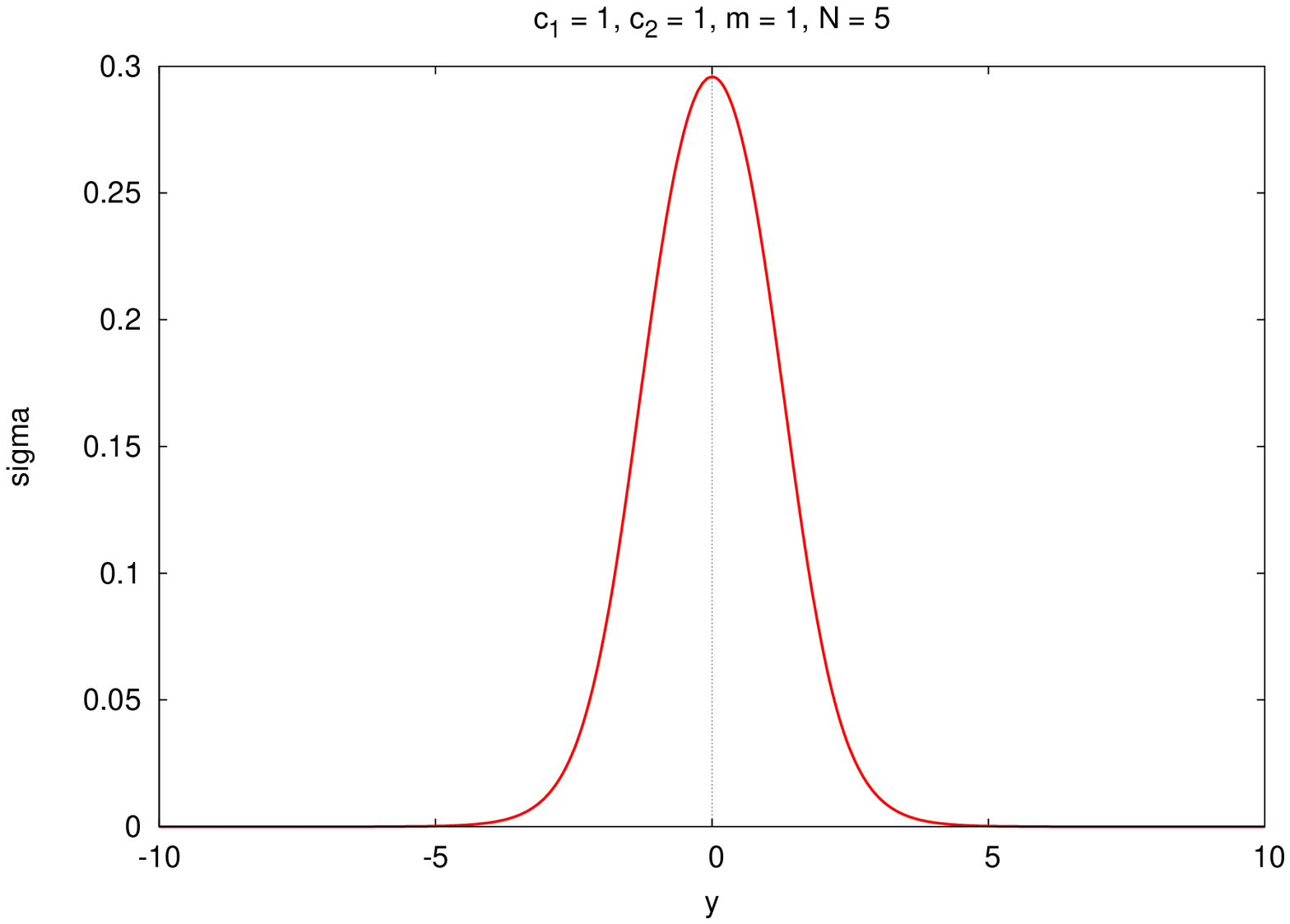}
\end{minipage}
\caption[Profiles of $\Exp{-\eta}$ and $\sigma$ in the 
coincident case.]{Profiles of $\Exp{-\eta}$ and $\sigma$ in the 
coincident case. 
The parameters of the plot are given above the picture. 
Positions of all domain walls are centered at the origin.}
\label{fig:02}
\end{center}
\end{figure}

This solution is not invariant under the 
symmetry transformations \refer{eq:symm1} in general. 
In the case of $U = \mathbf{1}_N$, however, 
the fieds (\ref{eq:coin1})-(\ref{eq:coin6}) 
 do not change under the action of the diagonal global 
symmetry $SU(N)_{L+R+c}$. 
We show the $y$-dependence of $\Exp{-\eta}$ and of 
$\sigma = \partial_y \eta/2$ for the 
coincident case in Fig.~\ref{fig:02}.

\begin{figure}
\begin{center}
\begin{minipage}[b]{0.49\linewidth}
\centering
\includegraphics[width=\textwidth]{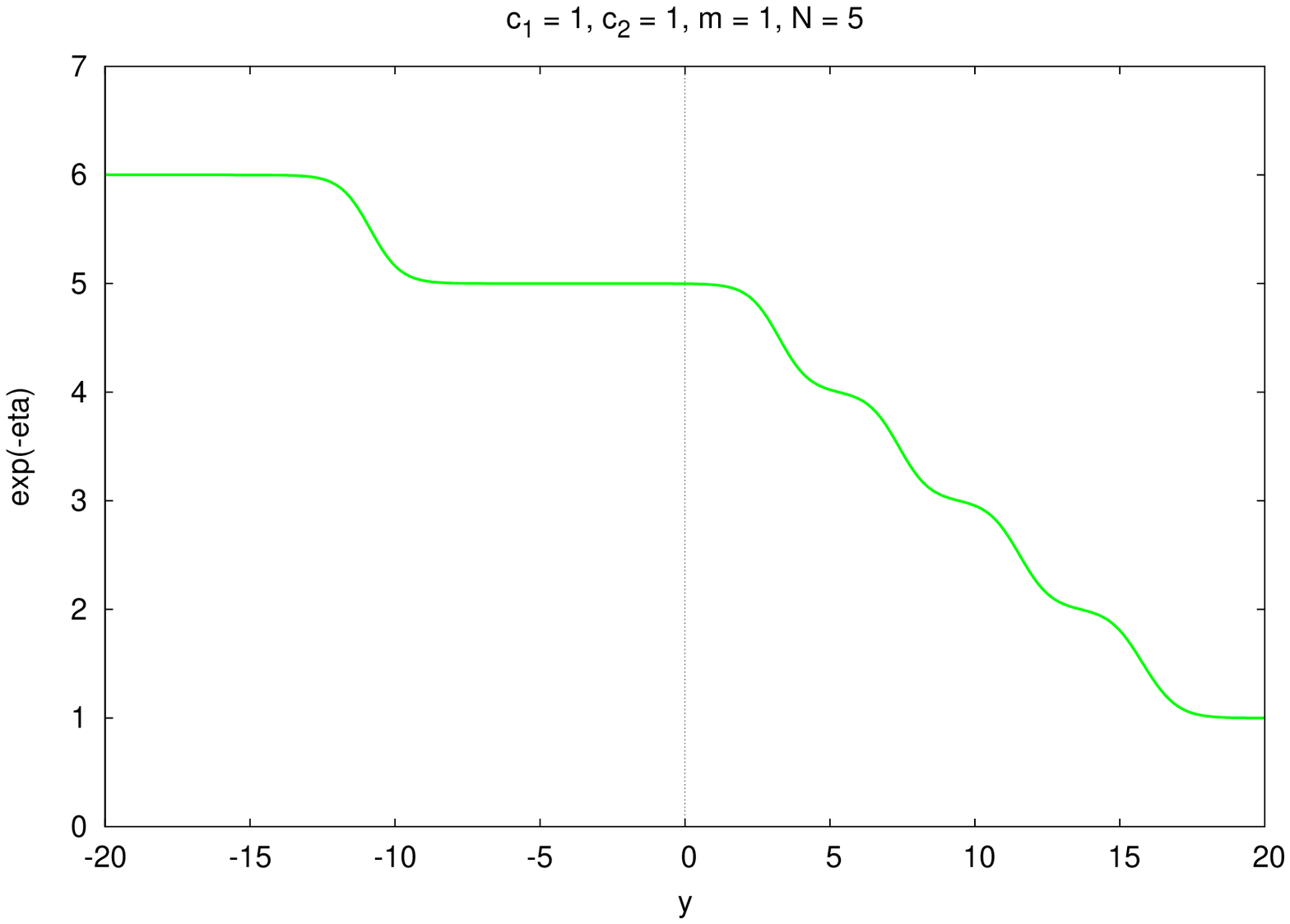}
\end{minipage}
\begin{minipage}[b]{0.49\linewidth}
\centering
\includegraphics[width=\textwidth]{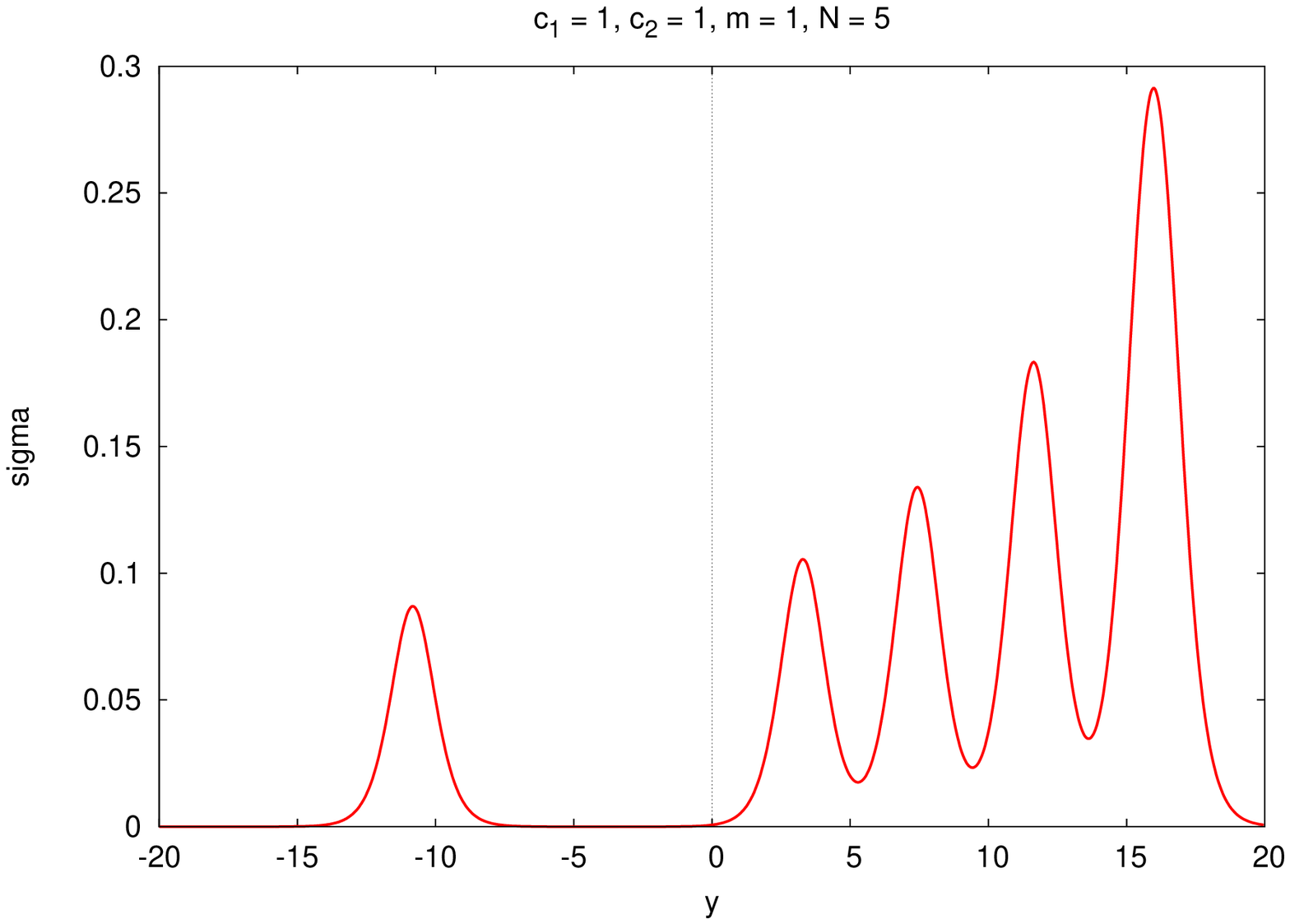}
\end{minipage}
\caption[Profiles of $\Exp{-\eta}$ and $\sigma$ in the 
non-coincident case.]{Profiles of $\Exp{-\eta}$ and $\sigma$ in the 
non-coincident case. 
The parameters of the plot are given above the picture. 
Positions of domain walls are $y_1 = -10, y_2 = 4, 
y_3 = 8, y_4 = 12$ and $y_5 = 16$.}
\label{fig:03}
\end{center}
\end{figure}

For a more general case, such as 
non-coincident walls, the dependence of $\Exp{-\eta}$ and 
$\sigma$ on $y$ is more complicated. 
Furthermore, the equation \refer{eq:x} cannot be solved 
in closed form in general, except for first few values 
of $N$. 
Thus, one has to use numerical techniques. 
In Fig. \ref{fig:03} we present an example of five 
non-coincident walls.

\subsection{Localization of non-Abelian gauge fields}

In order to obtain massless gauge fields localized on the 
domain wall, we need to introduce new gauge symmetry, 
which is not broken in the bulk.
As we have seen in the previous subsections, the 
coincident domain wall solutions 
(\ref{eq:coin1})-(\ref{eq:coin6}) do not 
break a large part of the global symmetry.
Let us gauge $SU(N)_{L+R}\equiv SU(N)_V$ and denote 
new gauge fields as $V_{M}$. 
Then the fields $H_1$ and $H_2$ are in the bi-fundamental 
representation of $SU(N)_c\times SU(N)_V$ and the 
covariant derivatives \refer{eq:cov1} are modified to 
\begin{equation}\label{eq:cov2}
\tilde D_M H_{1,2}  = \partial_M H_{1,2} +\iunit W_{M} H_{1,2} -\iunit H_{1,2}V_{M}\,.
\end{equation}
Quantum numbers of the gauged model are summarized in Tab. \ref{table:02}

\begin{table}
\begin{center}
\begin{tabular}{c|cccc|c|c}
\hline
 & $SU(N)_{c}$ & $U(1)_1$ & $U(1)_2$ & $SU(N)_{V}$ 
 & $U(1)_{A}$ & mass\\ \hline
$H_{1}$ & $\square$ & 1 & 0 & $\square$  
& 1 & $m{\bf 1}_{N}$\\ 
$H_{2}$ & $\square$ &  1 & $-1$ & $\square$ 
 & $-1$ & 0 \\ 
$H_{3}$ & {\bf 1} & 0 & 1 & {\bf 1} 
 & 0  & 0\\ 
$\Sigma$ & ${\rm adj}\oplus{\bf 1}$ & 0 & 0 & {\bf 1} 
 & 0 & 0\\
$\sigma$ & ${\bf 1}$ & 0 & 0 & {\bf 1}  & 0 & 0 \\
\hline
\end{tabular}
\end{center}
\caption[Quantum numbers of the gauged  
generalized three-flavor model.]{Quantum numbers of the gauged  
generalized three-flavor model. The notation is explained in Tab.~\ref{table:SY}.}
\label{table:02}
\end{table}

We introduce the field-dependent gauge coupling for $V_M$ as
\begin{equation}\label{eq:pdg}
\frac{1}{2\tilde g^2(\sigma)} = \lambda \sigma\,,
\end{equation}
where we assume that $\lambda$ is a positive constant $\lambda >0$.
If we denote the field strength for the new gauge fields 
as $\tilde G_{MN}$, the Lagrangian for the gauged model is 
given by 
\begin{equation}\label{eq:gaugedlagr}
\oper{L} = \tilde{\oper{L}}-\frac{1}{2\tilde g^2(\sigma)}
\Tr\Bigl[\tilde G_{MN}\tilde G^{MN}\Bigr]\,,
\end{equation} 
where $\tilde{\oper{L}}$ is the same as in \refer{eq:lag1}, 
except that the covariant derivatives \refer{eq:cov1} are 
replaced by \refer{eq:cov2}.
It is not hard to see that the solution 
(\ref{eq:coin1})-(\ref{eq:coin6}) in the ungauged model 
is equally valid in the gauged model.
If we write down the equation of motion for the new gauge 
fields $V_M$ we have 
\begin{equation}
\partial_M \tilde G^{MN} = J^{N}\,,
\label{eq:EOM_gauge_field}
\end{equation}
where $J_M$ stands for the current of $V_M$. 
Since the solution preserves $SU(N)_V$, the current vanishes 
for the domain wall solution 
(\ref{eq:coin1})-(\ref{eq:coin6}), and $V_M = 0$ is a valid 
solution to the equation of motion (\ref{eq:EOM_gauge_field}). 
Then the other equations of motion of the gauged model 
reduce
to those of the ungauged model because of $V_M = 0$. 
Therefore, we see that (\ref{eq:coin1})-(\ref{eq:coin6}), 
in addition to the condition $V_M = 0$, solves the whole set 
of equations of motion of the gauged model.

\subsection{Positivity of the position-dependent gauge coupling}

We wish to show that the field-dependent gauge coupling 
\refer{eq:pdg} assures the positive definiteness of the 
position-dependent gauge coupling for any configurations 
of the domain wall. 
Since we do not need 
the effective theory in full,  
we will derive the rest of the effective Lagrangian  
in the next section. 
For the moment, it is sufficient to know, that  
the field-dependent gauge coupling $1/\tilde g^2(\sigma)$ 
is given by its value in the background solution 
\begin{equation}
\frac{1}{\tilde g^2(\sigma)}\bigg|_{\mathrm{domain~wall}} 
\equiv  \frac{1}{\tilde g^2(y)} 
= \lambda\partial_y \eta = - \lambda \partial_y \ln x\,,
\end{equation}
where $x = \Exp{-\eta} \ge 0$ is the solution to \refer{eq:x}. 
Differentiating \refer{eq:x} we find 
\begin{equation}
\frac{1}{x}\partial_y x = - \frac{c_1}{c_2}\sum_{i=1}^N
\frac{e_i }{(1+e_i x)^2}\biggl(2m+\frac{1}{x}
\partial_y x\biggr)\,.
\end{equation}
This leads to the formula
\begin{equation}
\frac{1}{2\tilde g^2(y)} = \frac{\lambda c_1}{c_2}
\sum_{i=1}^N \frac{e_i}{(1+e_ix)^2}\bigg/\biggl(1
+\frac{c_1}{c_2}\sum_{i=1}^N \frac{e_i}{(1+e_ix)^2}\biggr)\,,
\end{equation} 
which is indeed positive in the whole range of 
$y$-coordinate.

We obtain the effective gauge coupling in 
$3+1$-dimensional world volume by
tntegrating $1/\tilde g^2(y)$ over the extra-dimensional 
coordinate $y$ 
\begin{equation}
\frac{1}{g_4^2} = \lambda \lineint y\, 
\partial_y \eta 
= \lambda [\eta(\infty)-\eta(-\infty)]\,.
\label{eq:eff_coupling}
\end{equation}
The asymptotic values of $\eta$ are found from 
(\ref{coin1}) as 
\begin{equation}
\eta(\infty) = 0\,, \hspace{0.5cm} \eta(-\infty) 
= -\ln\left(1+N{c_1 \over c_2} \right)\,.
\end{equation}
The
easiest way to see these asymptotic values is to note 
that 
in (\ref{coin2}) we can take the limits of \refer{eq:x} 
to obtain
\begin{equation}
x = 1 + \frac{c_1}{c_2}\sum_{i=1}^N \frac{1}{1+e_i x} 
\longrightarrow 
\Bigg\{\begin{array}{lcr}
1 & \mathrm{if} & y \to \infty\,, \\
1+\tfrac{Nc_1}{c_2} & \mathrm{if} & y \to -\infty\,,
\end{array}
\end{equation}
since $\eta$ is finite  at both infinities.\footnote{
We can just look at \refer{eq:h03} with 
$H_3^0 = \sqrt{c_2}$ and recall \refer{eq:vev} 
to obtain the same result.
}

Thus, the effective gauge coupling is given as 
\begin{equation}\label{eq:effcoupl2}
\frac{1}{g_4^2} = \lambda \ln\biggl(1
+N\frac{c_1}{c_2}\biggr)\,.
\end{equation}
It is interesting to observe that the 
effective gauge coupling is proportional to the width 
of the domain wall $\ln(1+N{c_1}/{c_2})$, 
which is not a modulus, but is 
fixed by parameters of the theory. 
This feature is in sharp contrast to that of Eq.~\refer{eq:4d_NA_gauge_coupling}, 
 where the effective gauge 
coupling constant is proportional to the separation of walls in each sector, 
which is a modulus undetermined by the theory. 
We have now confirmed the stability of the gauge kinetic 
term (by choosing the parameters of the theory as 
$\lambda, m, c_1,c_2 > 0$).

\begin{figure}
\begin{center}
\begin{minipage}[b]{0.45\linewidth}
\centering
\includegraphics[width=\textwidth]{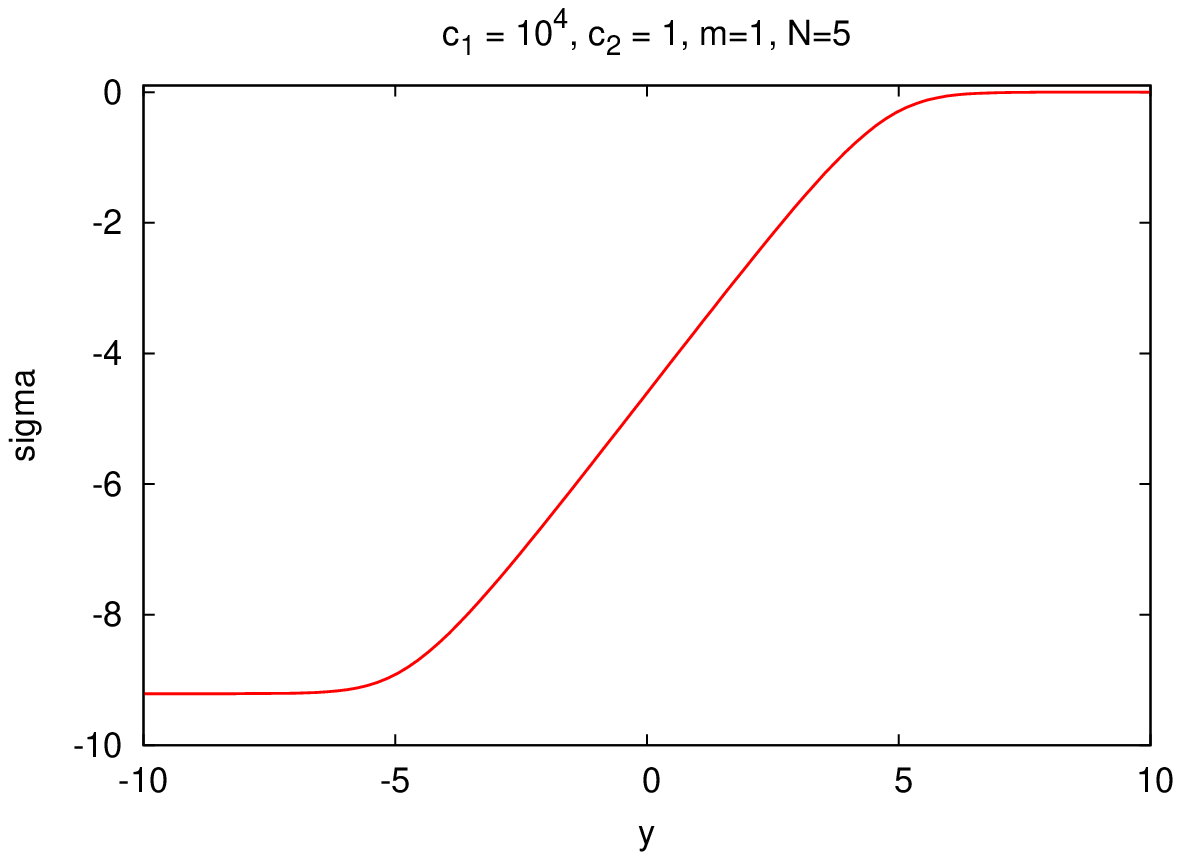}
\end{minipage}
\begin{minipage}[b]{0.45\linewidth}
\centering
\includegraphics[width=\textwidth]{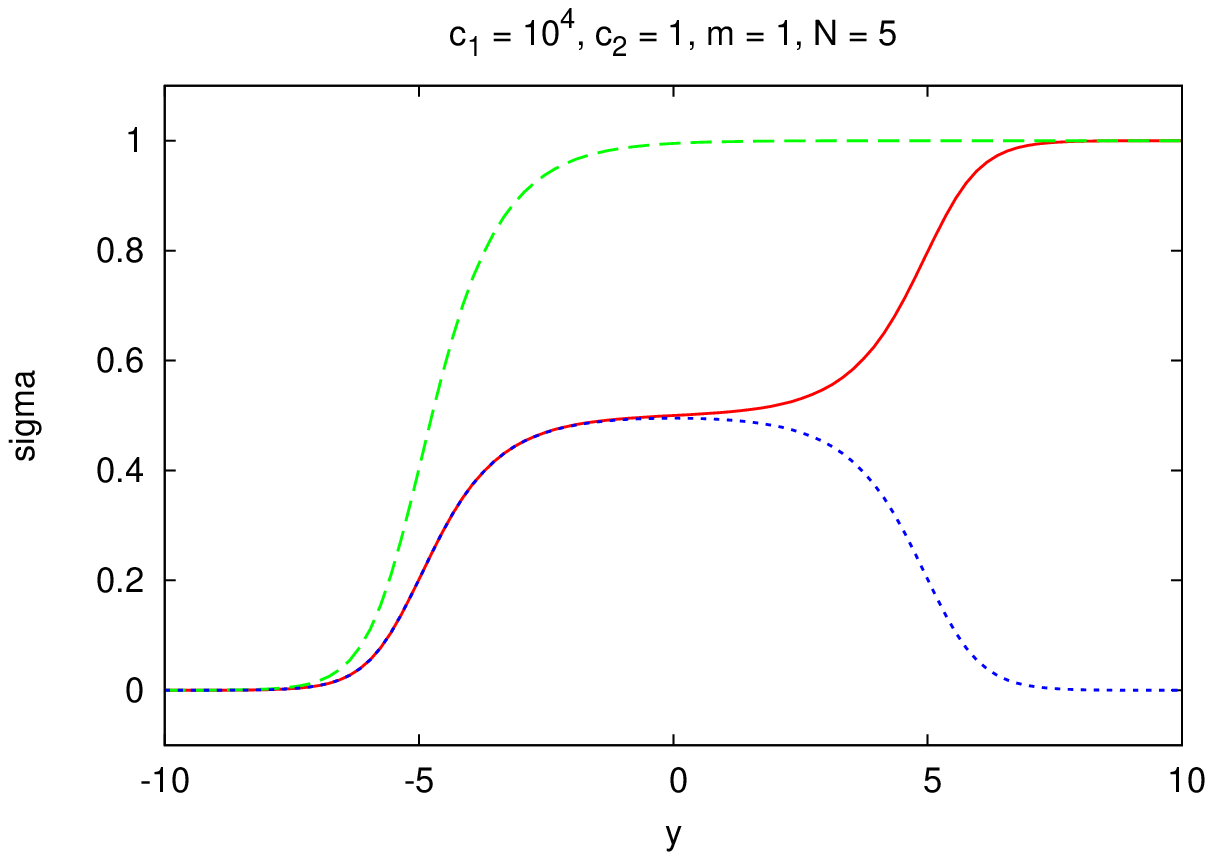}
\end{minipage}
\caption[Profile of $\eta$-kink in the left panel, $\Tr(\Sigma)$, $\Tr(\Sigma)-\sigma$ and 
$\sigma$ are shown in the right panel.]
{Profile of $\eta$-kink is shown in the left 
panel for  the coincident case.
In the right panel, 
plots of $\Tr(\Sigma)$ (green dashed curve),
$\Tr(\Sigma)-\sigma$ (red solid curve), and
$\sigma$ (blue dotted curve) are shown.
}
\label{fig:eta}
\end{center}
\end{figure}

Eq.~(\ref{eq:eff_coupling}) shows that the effective 
four-dimensional gauge coupling constant $g_4$ is 
determined only by the boundary conditions at infinity. 
This can be interpreted as a kind of topological charge 
of the $\eta$-kink, whose profile is shown 
in Fig.~\ref{fig:eta}.

Note that we have considered only the BPS solutions so 
far, and confirmed the stability of the model.
The anti-BPS solutions are also stable, since 
they can are obtained by the parity transformation 
$y \to - y$.

\section{Effective Lagrangian}

In this section we calculate the effective Lagrangian on the coincident background domain wall solution.  We first consider 
the simple case, where only $U$ moduli is promoted to a field on the domain wall's world-volume, to discuss possible connections with phenomenology. Then we attempt 
to tackle the general case, which we, however, will be able to do only in certain approximation.

\subsection{Preliminaries}

In the spirit of the
low-energy approximation, let us pick up only two derivative terms as a lowest order approximation for the effective Lagrangian:\footnote{We do not explicitly write down the kinetic term for localized gauge fields until the end of this section, for simplicity.}
\begin{equation}\label{ch6:eff}
\oper{L}^{(2)} = \Tr\abs{D_{\mu}H_1}^2+\Tr\abs{(D_\mu-\iunit A_\mu)H_2}^2+\abs{(\partial_\mu+\iunit A_\mu)H_3}^2\,,
\end{equation} 
with covariant derivatives given as
\begin{equation}
D_{\mu}H_{1,2} = \hat D_{\mu}H_{1,2}+\iunit W_{\mu}H_{1,2} = \partial_{\mu}H_{1,2}-\iunit H_{1,2}V_{\mu}+\iunit W_{\mu}H_{1,2}\,.
\end{equation}
Here we have singled out covariant derivatives $\hat D_{\mu}$ containing $V_{\mu}$ fields 
associated with the gauged flavor symmetry $SU(N)_{L+R}$.
In the strong gauge coupling limit there are two constrains among the Higgs fields:
\begin{align}
\label{ch6:d1} H_1H_1^{\dagger}+H_2H_2^{\dagger} & = c_1\mathbf{1}_N\,, \\
\label{ch6:d2} \abs{H_3}^2-\Tr(H_2H_2^{\dagger}) & = c_2\,.
\end{align} 

First we eliminate $W_{\mu}$ through corresponding EoM
\begin{equation}
W_{\mu} = -\frac{\iunit}{2c_1}\Bigl[H_{a}\hat D_{\mu}H_a^{\dagger}-\hat D_{\mu}H_aH_a^{\dagger}+2\iunit A_{\mu}H_2H_2^{\dagger}\bigr]
=: \hat W_{\mu}+\frac{1}{c_1}A_{\mu}H_2H_2^{\dagger}\,,
\end{equation}
where we employ Einstein summation convention for the index $a=1,2$. Plugging this back into 
\refer{ch6:eff}, we obtain
\begin{align}
\oper{L}^{(2)} & = \Tr(\hat D_{\mu}H_a\hat D^{\mu}H_a^{\dagger}) -c_1\Tr(\hat W_{\mu}\hat W^{\mu})
 +\iunit A_{\mu}\Bigl(H_3\partial^{\mu}H_3^{\dagger}-\partial^{\mu}H_3H_3^{\dagger}-
\Tr(H_2\hat D^{\mu}H_2^{\dagger}\nonumber \\
& -\hat D^{\mu}H_2H_2^{\dagger})\Bigr) -2A_{\mu}\Tr(\hat W^{\mu}H_2H_2^{\dagger})+A_{\mu}A^{\mu}\Bigl(\abs{H_3}^2+\Tr(H_2H_2^{\dagger})-\frac{1}{c_1}\Tr\bigl[(H_2H_2^{\dagger})^2\bigr]\Bigr)\nonumber \\
& +\partial_\mu H_3^{\dagger}\partial^{\mu}H_3\,.
\end{align}
Before eliminating remaining auxiliary fields $A_{\mu}$ we first rewrite some 
terms using the constraint \refer{ch6:d1}
\begin{align}
\Tr\bigl[(H_2H_2^{\dagger})^2\bigr] & = c_1\Tr(H_2H_2^{\dagger})-\Tr(H_2H_2^{\dagger}H_1H_1^{\dagger})\,, \\
-2\Tr(\hat W^{\mu}H_2H_2^{\dagger}) & = \iunit \Tr(H_2\hat D_{\mu}H_2^{\dagger}-\hat D_{\mu}H_2H_2^{\dagger})
 +\frac{\iunit}{c_1}\varepsilon_{ab}\Tr\bigl[(H_a\hat D_{\mu}H_a^{\dagger}\nonumber \\
&-\hat D_{\mu}H_aH_a^{\dagger})H_bH_b^{\dagger}\bigr]\,.
\end{align}
Using these identities $A_{\mu}$ can be expressed as
\begin{equation}
A_{\mu} = -\frac{\iunit}{2}\frac{H_3\partial_{\mu}H_3^{\dagger}-\partial_{\mu}H_3H_3^{\dagger}
+\tfrac{1}{c_1}\varepsilon_{ab}\Tr\bigl[(H_a\hat D_{\mu}H_a^{\dagger}-\hat D_{\mu}H_aH_a^{\dagger})H_bH_b^{\dagger}\bigr]}
{\abs{H_3}^2+\tfrac{1}{c_1}\Tr(H_2H_2^{\dagger}H_1H_1^{\dagger})}\,.
\end{equation} 
Next we use the following identity 
\begin{equation}
\Tr\Bigl[\hat D_{\mu}H_a\hat D^{\mu}H_a^{\dagger} -c_1\hat W_{\mu}\hat W^{\mu}\Bigr]
= \frac{1}{2c_1} 
\Tr\Bigl[\oper{D}_{\mu}\oper{H}_{ab}\oper{D}^{\mu}\oper{H}_{ab}^{\dagger}\Bigr]\,,
\end{equation}
where $\oper{H}_{ab} := H_a^{\dagger}H_b$.
Let us stress that covariant derivatives $\hat D_{\mu}$ act on $H_a^{\dagger}H_b$ as if they were in the adjoint representation:
\begin{equation}
\hat D_{\mu} (H_a^{\dagger}H_b) = \partial_{\mu}(H_a^{\dagger}H_b) +\iunit V_{\mu}H_a^{\dagger} H_b -\iunit H_a^{\dagger}H_bV_{\mu} =
\partial_{\mu}(H_a^{\dagger}H_b)+\iunit\comm{V_{\mu}}{H_a^{\dagger}H_b}\,.
\end{equation}
Nevertheless, to make this fact explicit, we use for covariant derivatives of composite fields different notation: $\oper{D}_{\mu}$.
In terms of composite fields $\oper{H}_{ab}$ we can rewritte $A_{\mu}$ as
\begin{equation}\label{ch6:amu}
A_{\mu} = -\frac{\iunit}{2}\frac{H_3\partial_{\mu}H_3^{\dagger}-\partial_{\mu}H_3H_3^{\dagger}
+\tfrac{1}{c_1}\Tr\bigl[\oper{H}_{12}^{\dagger}\oper{D}_{\mu}\oper{H}_{12}-\oper{D}_{\mu}(\oper{H}_{12}^{\dagger})\oper{H}_{12}\bigr]}
{\abs{H_3}^2+\tfrac{1}{c_1}\Tr\bigl[\oper{H}_{12}^{\dagger}\oper{H}_{12}\bigr]}\,,
\end{equation} 
where we used the fact that
\begin{equation}
\varepsilon_{ab}\Tr\bigl[(H_a\hat D_{\mu}H_a^{\dagger}-\hat D_{\mu}H_aH_a^{\dagger})H_bH_b^{\dagger}\bigr]
= \Tr\bigl[\oper{H}_{12}^{\dagger}\oper{D}_{\mu}\oper{H}_{12}-\oper{D}_{\mu}(\oper{H}_{12}^{\dagger})\oper{H}_{12}\bigr]\,.
\end{equation}
Putting all pieces together we arrive at the effective Lagrangian 
\begin{align}
 \oper{L}_{\rm eff} &= \lineint y\, \biggl[\frac{1}{2c_1}\Tr\Bigl(\oper{D}_{\mu}\oper{H}_{ab}
 \oper{D}^{\mu}\oper{H}_{ab}^{\dagger}\Bigr)
 +\partial_{\mu}H_3\partial^{\mu}H_3^{\dagger} -A_{\mu}A^{\mu}\Bigl(\abs{H_3}^2 \nonumber \\ \label{ch6:eff2}
&+\tfrac{1}{c_1}\Tr(\oper{H}_{12}^{\dagger}\oper{H}_{12})\Bigr)\biggr]\,.
\end{align}

\subsection{A simple case}

Let us calculate the effective Lagrangian on the background of the coincident domain walls
\begin{align}
H_1 & = \sqrt{c_1}\Exp{my}\Omega^{-1/2}\,, \\
H_2 & = \sqrt{c_1}\Exp{\eta/2}\Exp{my_0}\Omega^{-1/2}U^{\dagger}\,, \\
H_3 & = \sqrt{c_2}\Exp{-\eta/2}\,,
\end{align} 
with $\Omega$ and $\eta$ given as in Eqs.~\refer{coin1}-\refer{coin2}. 
In this subsection we consider, for simplicity, the case where positions of all walls are 
fixed at $y=y_0$ and only relative phases and center of mass position are allowed to fluctuate. 
Thus the moduli are promoted to world-volume fields as
\begin{equation}
U^{\dagger} = U^{\dagger}(x^{\mu})\,, \hspace{5mm}
\phi = y_0(x^{\mu})\mathbf{1}_N\,.
\end{equation}
Composites $\oper{H}_{ab}$ and their covariant derivatives are given as
\begin{align}
\oper{H}_{11} & = c_1\frac{e_0\Exp{-\eta}}{1+e_0\Exp{-\eta}}\mathbf{1}_N\,, \\
\oper{H}_{12} & = c_1\frac{e_0^{1/2}\Exp{-\eta/2}}{1+e_0\Exp{-\eta}}U^{\dagger}\,, \\
\oper{H}_{22} & = c_1\frac{1}{1+e_0\Exp{-\eta}}\mathbf{1}_N\,.
\end{align}
\begin{align}
\oper{D}_{\mu}\oper{H}_{11} & = -\frac{2c_2}{N}\Exp{-\eta}\sigma \partial_{\mu}y_0 \mathbf{1}_N\,, \\
\oper{D}_{\mu}\oper{H}_{12} & = \frac{2c_2\sigma}{N}\Exp{-\eta} \sinh\bigl(m(y-y_0)-\eta/2\bigr)U^{\dagger}\partial_{\mu}y_0 +\frac{c_1}{2}
\frac{\oper{D}_{\mu}U^{\dagger}}{\cosh\bigr(m(y-y_0)-\eta/2\bigl)}\,, \\
\oper{D}_{\mu}\oper{H}_{22} & = \frac{2c_2}{N}\Exp{-\eta}\sigma \partial_{\mu}y_0 \mathbf{1}_N\,,
\end{align}
where 
\begin{equation}
\frac{\sigma}{m} = \Bigl(1+\frac{c_2}{Nc_1}(1+e_0\Exp{-\eta})^2/e_0\Bigr)^{-1}\,.
\end{equation}
Putting this into Eq.~\refer{ch6:amu} we obtain
\begin{equation}
A_{\mu} = -\frac{\iunit\sigma}{2Nm}\Tr[U\oper{D}_{\mu}U^{\dagger}-U^{\dagger}\oper{D}_{\mu}U]
\end{equation}
and applying it into \refer{ch6:eff2} we have 
\begin{align}
\oper{L}_{\mathrm{eff}} & = \frac{c_2}{Nm}\lineint y\, \frac{\sigma}{1-\sigma/m}\Exp{-\eta}\Tr\bigl[\oper{D}_{\mu}U^{\dagger}\oper{D}^{\mu}U\bigr]
+mc_2\lineint y\,\Exp{-\eta}\sigma\partial_{\mu}y_0\partial^{\mu}y_0
\nonumber \\ & +\frac{c_2}{N^2m^2}\lineint y\, 
\frac{\sigma^2}{1-\sigma/m}\Exp{-\eta}
\Tr\bigl[U\oper{D}_{\mu}U^{\dagger}\bigr]\Tr\bigl[U\oper{D}^{\mu}U^{\dagger}\bigr]\,.
\end{align}
Let us now compute integration factors in the above expression. This is most easily done by substituting $x = \Exp{-\eta}$ and using the identity:
\begin{equation}
\frac{\sigma}{m} = \frac{(x-1)\Bigl(1-\tfrac{c_2}{Nc_1}(x-1)\Bigr)}{x+(x-1)
\Bigl(1-\tfrac{c_2}{Nc_1}(x-1)\Bigr)}\,,
\end{equation} 
leading to the result
\begin{align}
\oper{L}_{\rm eff} &= \frac{c_1}{2m}\biggl[ (\alpha+1)\Tr\bigl[\oper{D}_{\mu}U^{\dagger}\oper{D}^{\mu}U\bigr] + 
\frac{\alpha}{N}\, \Tr\bigl[U\oper{D}_{\mu}U^{\dagger}\bigr]\Tr\bigl[U\oper{D}^{\mu}U^{\dagger}\bigr]
\biggr] \nonumber \\ \label{ch6:result} &+\frac{Nmc_1}{2}\partial_{\mu}y_0\partial^{\mu}y_0\,,
\end{align}
where
\begin{equation}
\alpha  = \frac{1}{2}+\frac{c_2}{Nc_1}-\frac{c_2}{Nc_1}\left(1+\frac{c_2}{Nc_1}\right)\ln \biggl(1+\frac{Nc_1}{c_2}\biggr)\,.
\end{equation}

Let us reinterpret our result in terms of pion field, which we define as
\begin{equation}
\frac{1}{f_\pi}\oper{D}_{\mu}\hat \pi = \iunit \Bigl[U\oper{D}_{\mu}U^{\dagger}-\frac{\mathbf{1}_N}{N}\Tr\Bigl(U\oper{D}_{\mu}U^{\dagger}\Bigr)\Bigr]\,.
\end{equation}
Also let us define
\begin{equation}\label{ch6:eta1}
\frac{1}{f_\eta}\partial_{\mu}\eta := \iunit\, \Tr\Bigl(U\oper{D}_{\mu}U^{\dagger}\Bigr)\,.
\end{equation}
Setting decay constants $f_\pi$ and $f_\eta$ to have canonical kinetic terms, we can recast Eq.~\refer{ch6:result} as (including the kinetic term for gauge fields)
\begin{equation}\label{ch6:eta2}
\oper{L}_{\rm eff} = \Tr\Bigl(\oper{D}_{\mu}\hat\pi\oper{D}^{\mu}\hat\pi\Bigr)+\frac{1}{2}\partial_{\mu}\eta\partial^{\mu}\eta
+\frac{Nmc_1}{2}\partial_{\mu}y_0\partial^{\mu}y_0-\frac{1}{2g_4^2}\Tr\Bigl[\tilde G_{\mu\nu}\tilde G^{\mu\nu}\Bigr]\,,
\end{equation}
with
\begin{equation}
f_{\pi} = \sqrt{\frac{c_1(\alpha+1)}{2m}}\,, \hspace{5mm} f_\eta = \sqrt{\frac{c_1}{2m}}\,.
\end{equation}
A new feature of  
\refer{ch6:eta2}
compared with Eq.~\refer{ch5:chirallagr} is that now $f_\pi$ of the adjoint field is larger
 than 
$f_\eta$ of the singlet field by 
the factor $\sqrt{\alpha+1}$.

\subsection{General case I: formulating the problem}

Let us consider the case where both $U$ and $\phi$ are promoted to full four-dimensional fields.
To obtain the effective Lagrangian, we need to repeat 
the same procedure as in the previous subsection, but here the covariant derivatives acting on functions of 
matrices require more care (see e.g. 
subsection \ref{ch5:subsec:efflagr}) and 
causes difficulty when deriving the closed form for
the effective Lagrangian. 
However, we have a convenient parameter to expand the 
effective Lagrangian, the ratio $c_1/c_2$ whose 
logarithm has a physical meaning as the width of 
the domain wall (see Eq.~(\ref{eq:effcoupl2})). 

The background solution with promoted moduli fields is in the form
\begin{align}
H_1 & = \sqrt{c_1}\Exp{my}\Omega^{-1/2}\,, \\
H_2 & = \sqrt{c_1}\Exp{\eta/2}\Omega^{-1/2}\Exp{\phi}U^{\dagger}\,, \\
H_3 & = \sqrt{c_2}\Exp{-\eta/2}\,,
\end{align} 
where
\begin{align}
\Omega^{-1} & = \frac{\Exp{-\eta}\Exp{-2\phi}}{\mathbf{1}_N+\Exp{2my}\Exp{-2\phi}\Exp{-\eta}}\,, \\
\Exp{-\eta} & = 1+\frac{c_1}{c_2}\Tr\left(\frac{\mathbf{1}_N}{\mathbf{1}_N+\Exp{2my}\Exp{-2\phi}\Exp{-\eta}}\right)\,.
\end{align}
Composites $\oper{H}_{ab}$ are given as
\begin{align}
\oper{H}_{11} & = c_1\frac{\mathbf{1}_N}{\mathbf{1}_N+\Exp{-2my}\Exp{2\phi}\Exp{\eta}}\,, \\
\oper{H}_{12} & = c_1\frac{\Exp{my}\Exp{-\phi}\Exp{-\eta/2}}{\mathbf{1}_N+\Exp{2my}\Exp{-2\phi}\Exp{-\eta}}U^{\dagger}\,, \\
\oper{H}_{22} & = c_1U\frac{\mathbf{1}_N}{\mathbf{1}_N+\Exp{2my}\Exp{-2\phi}\Exp{-\eta}}U^{\dagger}\,.
\end{align}

Let us now list some partial results. The derivatives of $\eta \equiv \eta(y,x^{\mu})$ are
\begin{align}
\frac{1}{2}\partial_y \eta &  = \frac{mc_1}{c_2}\Tr\biggl[\frac{\Exp{2my}\Exp{-2\phi}}{\bigl(\mathbf{1}_N+\Exp{2my}\Exp{2\phi}\Exp{-\eta}\bigr)^2} \biggr]
\Bigg/\Biggl(1+\frac{c_1}{c_2}\Tr\biggl[\frac{\Exp{2my}\Exp{-2\phi}}{\bigl(\mathbf{1}_N+\Exp{2my}\Exp{2\phi}\Exp{-\eta}\bigr)^2} \biggr]\Biggr)\,, \\
\frac{1}{2}\partial_{\mu}\eta &  = -\frac{c_1}{4c_2}\Exp{\eta}
\left(1-\frac{\sigma}{m}\right)\Tr\biggl[\frac{1}{\cosh^2(\hat y)}\partial_{\mu}\phi\biggr]\,,
\end{align}
where we abbreviated $\hat y := (my-\eta/2)\mathbf{1}_N-\phi$.
Covariant derivatives of composites are given as
\begin{align}
\oper{D}_{\mu}\oper{H}_{11} &= -c_1\frac{\Exp{-2\hat y}}{(\mathbf{1}_N+\Exp{-2\hat y})^2}\partial_{\mu}\eta-c_1\sum_{k=0}^{\infty}\frac{1}{(k+1)!}
\left(\frac{\mathbf{1}_N}{\mathbf{1}_N+\Exp{-2\hat y}}\right)^{(k+1)}\oper{L}_{\phi}^k(\oper{D}_{\mu}\phi)\,, \\
\oper{D}_{\mu}\oper{H}_{12} & = \frac{c_1}{4}\frac{\tanh(\hat y)}{\cosh(\hat y)}U^{\dagger}\partial_{\mu}\eta+\frac{c_1}{2}\frac{1}{\cosh(\hat y)}\oper{D}_{\mu}U^{\dagger} \nonumber \\
&-\frac{c_1}{2}\sum_{k=0}^{\infty}\frac{1}{(k+1)!}
\left(\frac{\mathbf{1}_N}{\cosh(\hat y)}\right)^{(k+1)}\oper{L}_{\phi}^k(\oper{D}_{\mu}\phi)U^{\dagger}\,, \\
\oper{D}_{\mu}\oper{H}_{22} & = U\oper{D}_{\mu}\oper{H}_{11}({-\hat y})U^{\dagger}
+\oper{D}_{\mu}U\frac{\mathbf{1}_N}{\mathbf{1}_N+\Exp{2\hat y}}U^{\dagger}+
U\frac{\mathbf{1}_N}{\mathbf{1}_N+\Exp{2\hat y}}\oper{D}_{\mu}U^{\dagger}\,, 
\end{align} 
where we have used the identity
\begin{equation}
\oper{D}_{\mu}f(\hat y) = -\frac{1}{2}f^{\prime}(\hat y)\partial_{\mu}\eta-\sum_{k=0}^{\infty}\frac{1}{(k+1)!}f^{(k+1)}(\hat y)\oper{L}_{\phi}^k(\oper{D}_{\mu}\phi)\,.
\end{equation}

Using above results and recalling \refer{ch6:amu} we may now compute last two terms appearing in the effective Lagrangian \refer{ch6:eff2}
\begin{equation}\label{ch6:part1}
\partial_{\mu}H_3\partial^{\mu}H_{3}^{\dagger}  = \frac{c_1^2}{16c_2}\Exp{\eta}\left(1-\frac{\sigma}{m}\right)^2
\Tr\Bigl[\frac{1}{\cosh^2(\hat y)}\partial_{\mu}\phi\Bigr]\Tr\Bigl[\frac{1}{\cosh^2(\hat y)}\partial^{\mu}\phi\Bigr]\,,
\end{equation}
\begin{multline} \label{ch6:part2}
-A_{\mu}A^{\mu}\Bigl(\abs{H_3}^2+\tfrac{1}{c_1}\Tr(\oper{H}_{12}^{\dagger}\oper{H}_{12})\Bigr)  \\= 
\frac{c_1^2}{16c_2}\Exp{\eta}\left(1-\frac{\sigma}{m}\right)
\Tr\Bigl[\frac{1}{\cosh^2(\hat y)}\oper{D}_{\mu}U^{\dagger}U\Bigr]\Tr\Bigl[\frac{1}{\cosh^2(\hat y)}\oper{D}^{\mu}U^{\dagger}U\Bigr]\,,
\end{multline}
Remaining terms can be organized as follows:
\begin{equation}
\frac{1}{2c_1}\Tr\Bigl[\oper{D}_{\mu}\oper{H}_{ab}\oper{D}^{\mu}\oper{H}_{ab}^{\dagger}\Bigr]  = \oper{T}_{\phi}(\hat y)+\oper{T}_{mix}(\hat y)+\oper{T}_{U}(\hat y)\,,
\end{equation}
where 
\begin{align}
\oper{T}_{\phi}(\hat y) & = \frac{c_1}{4}\Tr\biggl\{
\oper{D}_{\mu}\frac{1}{\cosh(\hat y)}\oper{D}^{\mu}\frac{1}{\cosh(\hat y)}
+\oper{D}_{\mu}\tanh(\hat y)\oper{D}^{\mu}\tanh(\hat y)
\biggr\}\,, \\
\oper{T}_{mix}(\hat y) & = \frac{c_1}{4} \Tr\biggl\{U^{\dagger}\oper{D}_{\mu}U\biggl( \comm{\frac{1}{\cosh(\hat y)}}{\oper{D}^{\mu}\frac{1}{\cosh(\hat y)}}
+\Bigl[\tanh(\hat y),\oper{D}^{\mu}\tanh(\hat y)\Bigr]\biggr)\biggr\}\,, \\
\oper{T}_{U}(\hat y) & = \frac{c_1}{4} \Tr\biggl\{\oper{D}_{\mu}U^{\dagger}\oper{D}^{\mu}U\frac{1}{\cosh^2(\hat y)}
+\oper{D}_{\mu}U^{\dagger}U\comm{\frac{\Exp{\hat y}}{\cosh(\hat y)}}{U^{\dagger}\oper{D}^{\mu}U}\frac{\Exp{-\hat y}}{\cosh(\hat y)}\biggr\}\,.
\end{align}
Using the identity
\begin{multline}
\Tr\Bigl[\oper{D}_{\mu}f(\hat y)\oper{D}^{\mu}f(\hat y)\Bigr]  = \frac{1}{4}\partial_{\mu}\eta\partial^{\mu}\eta\, 
\Tr\Bigl[\Bigl(f^{\prime}(\hat y)\Bigr)^2\Bigr]+\Tr\Bigl[\oper{D}_{\mu}\phi\Bigr(f^{\prime}(\hat y)\Bigl)^2\Bigr]\partial^{\mu}\eta\\ +\sum_{n=2}^{\infty}
\frac{1}{n!}\Tr\biggl\{\oper{L}_{\phi}^{n-2}\Bigl(\oper{D}_{\mu}\phi\Bigr)\oper{D}^{\mu}\phi\Bigl[\Bigl(f^2(\hat y)\Bigr)^{(n)}-2f^{(n)}(\hat y)f(\hat y)
\Bigr]\biggr\}
\end{multline}
we can write down $\oper{T}_{\phi}(\hat y)$ as
\begin{multline}\label{ch6:tphi}
\oper{T}_{\phi}(\hat y)  = \frac{c_2}{4}\frac{\sigma/m-2}{1-\sigma/m}\Exp{-\eta}\partial_{\mu}\eta\partial^{\mu}\eta+\frac{c_1}{4}\Tr\Bigl[
\frac{1}{\cosh^2(\hat y)}\oper{D}_{\mu}\phi\oper{D}^{\mu}\phi\Bigr]\\
 - \frac{c_1}{2}\sum_{n=3}^{\infty}\frac{1}{n!}\Tr\biggl\{\oper{L}_{\phi}^{n-2}
\Bigl(\oper{D}_{\mu}\phi\Bigr)\oper{D}^{\mu}\phi\Bigl[\Bigl(\frac{1}{\cosh(\hat y)}\Bigr)^{(n)}\frac{1}{\cosh(\hat y)}+\Bigl(\tanh(\hat y)\Bigr)^{(n)}\tanh(\hat y)\Bigr]\biggr\}
\end{multline}
Additional identity
\begin{align}
\comm{f(\hat y)}{\oper{D}_{\mu}f(\hat y)} = \sum_{n=2}^{\infty}\frac{(-1)^n}{n!}\oper{L}_{\phi}^{n-1}\Bigl(\oper{D}_{\mu}\phi\Bigr)
\Bigl[\Bigl(f^2(\hat y)\Bigr)^{(n)}-2f^{(n)}(\hat y)f(\hat y)\Bigr]\,,
\end{align}
allow us to expand $\oper{T}_{mix}(\hat y)$ as
\begin{multline}\label{ch6:tmixed}
\oper{T}_{mix}(\hat y) = - \frac{c_1}{2}\sum_{n=2}^{\infty}\frac{(-1)^n}{n!}\Tr\biggl\{U^{\dagger}\oper{D}_{\mu}U\oper{L}_{\phi}^{n-1}
\Bigl(\oper{D}^{\mu}\phi\Bigr)\Bigl[\Bigl(\frac{1}{\cosh(\hat y)}\Bigr)^{(n)}\frac{1}{\cosh(\hat y)}\\+\Bigl(\tanh(\hat y)\Bigr)^{(n)}\tanh(\hat y)
\Bigr]\biggr\}\,.
\end{multline}
And finally using 
\begin{equation}
\comm{f(\hat y)}{M} = \sum_{n=1}^{\infty}\frac{(-1)^n}{n!}\oper{L}_{\phi}^n(M)f^{(n)}(\hat y)\,,
\end{equation}
we can express the last piece as
\begin{multline}\label{ch6:tu}
\oper{T}_{U}(\hat y) = \frac{c_1}{4} \Tr\biggl\{\oper{D}_{\mu}U^{\dagger}\oper{D}^{\mu}U\frac{1}{\cosh^2(\hat y)}
\\+\oper{D}_{\mu}U^{\dagger}U\sum_{n=1}^{\infty}\frac{(-1)^n}{n!}\oper{L}_{\phi}^n(U^{\dagger}\oper{D}^{\mu}U)
\biggl(\frac{\Exp{\hat y}}{\cosh(\hat y)}\biggr)^{(n)}
\frac{\Exp{-\hat y}}{\cosh(\hat y)}\biggr\}\,.
\end{multline}
Let us now put all our partial results \refer{ch6:part1}-\refer{ch6:part2} and \refer{ch6:tphi}-\refer{ch6:tu} together to clearly see the 
 work ahead of us:
\begin{gather}
\oper{L}_{\mathrm{eff}}  = \lineint y\, \Biggl\{\Tr\Bigl[\frac{c_1}{4\cosh^2(\hat y)}\oper{D}_{\mu}\phi\oper{D}^{\mu}\phi\Bigr]
-\frac{c_1^2}{16c_2}\Exp{\eta}\Bigl(1-\frac{\sigma}{m}\Bigr)\Tr\Bigl[\frac{\oper{D}_{\mu}\phi}{\cosh^2(\hat y)}\Bigr]
\Tr\Bigl[\frac{\oper{D}^{\mu}\phi}{\cosh^2(\hat y)}\Bigr]\Biggr\} \nonumber \\
 + \lineint y\, \Biggl\{\Tr\Bigl[\frac{c_1}{4\cosh^2(\hat y)}\oper{D}_{\mu}U^{\dagger}\oper{D}^{\mu}U\Bigr]
+\frac{c_1^2}{16c_2}\Exp{\eta}\Bigl(1-\frac{\sigma}{m}\Bigr)\Tr\Bigl[\frac{\oper{D}_{\mu}U^{\dagger} U}{\cosh^2(\hat y)}\Bigr]
\Tr\Bigl[\frac{\oper{D}^{\mu}U^{\dagger}U}{\cosh^2(\hat y)}\Bigr]\Biggr\} \nonumber \\
 -  \frac{c_1}{2}\sum_{n=3}^{\infty}\frac{1}{n!}\lineint y\,\Tr\biggl\{\oper{L}_{\phi}^{n-2}
\Bigl(\oper{D}_{\mu}\phi\Bigr)\oper{D}^{\mu}\phi\Bigl[\Bigl(\frac{1}{\cosh(\hat y)}\Bigr)^{(n)}\frac{1}{\cosh(\hat y)}+\Bigl(\tanh(\hat y)\Bigr)^{(n)}\tanh(\hat y)\Bigr]\biggr\} \nonumber \\
 +\frac{c_1}{4}\sum_{n=1}^{\infty}\frac{(-1)^n}{n!}\lineint y\,\Tr\biggl\{\oper{D}_{\mu}U^{\dagger}U\oper{L}_{\phi}^n(U^{\dagger}\oper{D}^{\mu}U)
\biggl(\frac{\Exp{\hat y}}{\cosh(\hat y)}\biggr)^{(n)}
\frac{\Exp{-\hat y}}{\cosh(\hat y)}\biggr\} \nonumber \\ 
- \frac{c_1}{2}\sum_{n=2}^{\infty}\frac{(-1)^n}{n!}\lineint y\,\Tr\biggl\{U^{\dagger}\oper{D}_{\mu}U\oper{L}_{\phi}^{n-1}
\Bigl(\oper{D}^{\mu}\phi\Bigr)\Bigl[\Bigl(\frac{1}{\cosh(\hat y)}\Bigr)^{(n)}\frac{1}{\cosh(\hat y)} \nonumber \\ \label{ch6:labor}
  +\Bigl(\tanh(\hat y)\Bigr)^{(n)}
\tanh(\hat y)\Bigr]\biggr\}\,.
\end{gather}

\subsection{General case II: method}

In this subsection we would like to solve the following general expression
\begin{equation}
\int\limits_{-\infty}^{\infty}\diff y\, \Tr\Bigl[f(\hat y)M\Bigr]\,,
\end{equation}
where $M$ is some matrix, independent of $y$ and $f(\hat y)$ is some function. We proceed as follows: first we make a substitution $x = my-\eta/2$.
Since $\diff x = m \Bigl(1-\sigma/m\Bigr)\diff y$ we obtain:
\begin{equation*}
\int\limits_{-\infty}^{\infty}\diff y\, \Tr\Bigl[f(\hat y)M\Bigr] = 
\int\limits_{-\infty}^{\infty}\frac{\diff x}{m}\, \Tr\Bigl[f(x\mathbf{1}_N-\phi)M\Bigr] +
\int\limits_{-\infty}^{\infty}\frac{\diff x}{m}\, \frac{\sigma/m}{1-\sigma/m}\Tr\Bigl[f(x\mathbf{1}_N-\phi)M\Bigr]\,.  
\end{equation*}
While the first  part can be solved easily by the diagonalization trick
\begin{equation}
\int\limits_{-\infty}^{\infty}\frac{\diff x}{m}\, \Tr\Bigl[f(x\mathbf{1}_N-\phi)M\Bigr] = 
\int\limits_{-\infty}^{\infty}\frac{\diff x}{m}\, f(x)\Tr(M)\,,
\end{equation}
we found that the second integral in the above equation is very hard to solve in general. To make progress, let us adopt the approximation $c_1/c_2 \ll 1$. Since 
$\ln\bigl(1+N c_1/c_2\bigr)$ is proportional to the width of the domain wall, we call this approximation the thin wall approximation.\footnote{To be precise $\ln\bigl(1+N c_1/c_2\bigr)$ represents the inner width of the wall. Therefore the word ``thin'' is not really justified as the ``outer skin'' of the domain wall, proportional to $1/m$, can be arbitrary large.} 
Thus, we now only aim to solve 
\begin{equation}
\frac{c_1}{4c_2}\int\limits_{-\infty}^{\infty}\frac{\diff x}{m}\,
\Tr\biggl[\frac{\mathbf{1}_N}{\cosh^2(x\mathbf{1}_N-\phi)}\biggr]
\Tr\Bigl[f(x\mathbf{1}_N-\phi)M\Bigr]\,.
\end{equation}
To evaluate an integral over expression involving two traces we use the following identity
\begin{equation}
\int\limits_{-\infty}^{\infty}\diff x\,\Tr\Bigl[f(x\mathbf{1}_N-\phi)M\Bigr]\Tr\Bigl[g(x\mathbf{1}_N-\phi)N\Bigr]
 = F(\partial_x)\Tr\Bigl[\Exp{x\phi}M\Bigr]\Tr\Bigl[\Exp{-x\phi}N\Bigr]\Bigg|_{x = 0}\,, \label{ch6:twotrace}
\end{equation}
where $F(x)$ is given by the convolution
\begin{equation}
F(\alpha-\beta) := \int\limits_{-\infty}^{\infty}\diff x\, f(x-\alpha)g(x-\beta)\,.
\end{equation}
Applying this procedure, we conclude this section with the identity 
\begin{equation}
\int\limits_{-\infty}^{\infty}\diff y\, \Tr\Bigl[f(\hat y)M\Bigr] = \frac{F^{0}}{m}\Tr[M]+\frac{c_1}{4c_2m}F^{1}(\partial_x)
\Tr\Bigl[\Exp{x\phi}M\Bigr]\Tr\Bigl[\Exp{-x\phi}\Bigr]\bigg|_{x=0}+O\Bigl((c_1/c_2)^2\Bigr)\,,
\end{equation}
where
\begin{equation}
F^{0} = \lineint y\, f(y)\,, \hspace{1cm} F^{1}(x) := \lineint y\, \frac{f(y)}{\cosh^2(y-x)}\,.
\end{equation}

\subsection{General case III: calculations}

Let us denote
\begin{equation}
\oper{T}_{U} : = \lineint y\, \oper{T}_{U}(\hat y) = \oper{T}_{U}^{0}+\oper{T}_{U}^{1}+c_1O\Bigl((c_1/c_2)^2\Bigr)\,,
\end{equation}
where\footnote{We included the pure kinetic term for $U$ into the summation as $n=0$ term in contrast to \refer{ch6:tu}.} ($\hat x := x\mathbf{1}_N-\phi$)
\begin{align}
\oper{T}_{U}^{0} := & \frac{c_1}{4m} \sum_{n=0}^{\infty}\frac{(-1)^n}{n!}\lineint x\,
\biggl(\frac{\Exp{x}}{\cosh x}\biggr)^{(n)}\frac{\Exp{-x}}{\cosh x}
\Tr\Bigl[\oper{D}_{\mu}U^{\dagger}U\oper{L}_{\phi}^n(U^{\dagger}\oper{D}^{\mu}U)\Bigr]\,, \\
\oper{T}_{U}^{1} := & \frac{c_1^2}{16c_2 m} \sum_{n=0}^{\infty}\frac{(-1)^n}{n!}\lineint x\, \Tr\Bigl[\frac{\mathbf{1}_N}{\cosh^2(\hat x)}\Bigr] \nonumber \\
& \times\Tr\biggl\{\oper{D}_{\mu}U^{\dagger}U\oper{L}_{\phi}^n(U^{\dagger}\oper{D}^{\mu}U)
\biggl(\frac{\Exp{\hat x}}{\cosh(\hat x)}\biggr)^{(n)}
\frac{\Exp{-\hat x}}{\cosh(\hat x)}\biggr\}\,.
\end{align}
First expression can be easily integrated (and the result is the same as the corresponding expression of  the four-flavor model):
\begin{equation}
\oper{T}_{U}^0 = \frac{c_1}{4m}\Tr\Bigl[\oper{D}_{\mu}U^{\dagger}UF_U^0(\oper{L}_{\phi})(U^{\dagger}\oper{D}^{\mu}U)\Bigr]\,,
\end{equation}
where
\begin{equation}
F_U^0(\oper{L}_{\phi}) = \lineint y\, \frac{\cosh(\oper{L}_{\phi})}{\cosh(y)\cosh(y-\oper{L}_{\phi})} = 
\frac{1}{\tanh(\oper{L}_{\phi})}\ln\biggl(\frac{1+\tanh(\oper{L}_{\phi})}{1-\tanh(\oper{L}_{\phi})}\biggr)\,.
\end{equation}
Using the technique developed in the previous subsection, the next order in $c_1/c_2$ is given as
\begin{equation}
\oper{T}_{U}^1 = \frac{c_1^2}{16c_2m}\Tr\Bigl[\oper{D}_{\mu}U^{\dagger}UF_U^1(\partial_x,\oper{L}_{\phi})(U^{\dagger}\oper{D}^{\mu}U)
\Exp{x\phi}\Bigr]\Tr\Bigl[\Exp{-x\phi}\Bigr]\bigg|_{x=0}\,,
\end{equation}  
with
\begin{equation}\label{ch6:closed}
F_U^1(x,\oper{L}_{\phi}) = \lineint y\, \frac{\cosh(\oper{L}_{\phi})}{\cosh^2(y-x)\cosh(y)\cosh(y-\oper{L}_{\phi})}\,.
\end{equation}
Closed form for this integral exists, but it is complicated. The value of \refer{ch6:closed} can be better seen in its Taylor expansion.
\begin{align}
F_U^1(0,\oper{L}_{\phi}) & = \frac{\sinh(2\oper{L}_{\phi})}{\sinh^3(\oper{L}_{\phi})}-2\frac{\oper{L}_{\phi}}{\sinh^3(\oper{L}_{\phi})}\,,\\
\partial_x F_U^1(0,\oper{L}_{\phi}) & = \frac{\sinh(3\oper{L}_{\phi})}{3\sinh^3(\oper{L}_{\phi})}
+\frac{3-4\oper{L}_{\phi}}{\sinh^3(\oper{L}_{\phi})}-(3+4\oper{L}_{\phi})\frac{\cosh(\oper{L}_{\phi})}{\sinh^4(\oper{L}_{\phi})}\,, \\
\partial_x^2 F_U^1(0,\oper{L}_{\phi}) & = \frac{3}{2}\frac{\sinh(2\oper{L}_{\phi})}{\sinh^5(\oper{L}_{\phi})}
-4\oper{L}_{\phi}\frac{\cosh(2\oper{L}_{\phi})}{\sinh^5(\oper{L}_{\phi})}-8\frac{\oper{L}_{\phi}}{\sinh^5(\oper{L}_{\phi})}\,. \\
&\vdots \nonumber 
\end{align}

Let us proceed to the mixed term $\oper{T}_{mix}$ defined as
\begin{equation}
\oper{T}_{mix} := \lineint y\, \oper{T}_{mix}(\hat y) = \oper{T}_{mix}^0 + \oper{T}_{mix}^1+c_1O\Bigl((c_1/c_2)^2\Bigr)\,,
\end{equation} 
where
\begin{align}
\oper{T}_{mix}^0  = - \frac{c_1}{2m}\sum_{n=2}^{\infty}\frac{(-1)^n}{n!}\lineint x\, &
\biggl[\Bigl(\frac{1}{\cosh x}\Bigr)^{(n)}\frac{1}{\cosh x }+\Bigl(\tanh x\Bigr)^{(n)}\tanh x\biggr] \nonumber \\
&\times  \Tr\biggl[U^{\dagger}\oper{D}_{\mu}U\oper{L}_{\phi}^{n-1}
\Bigl(\oper{D}^{\mu}\phi\Bigr)\biggr]\,, \\
\oper{T}_{mix}^1  = - \frac{c_1^2}{8c_2m}\sum_{n=2}^{\infty}\frac{(-1)^n}{n!}\lineint x\, &
\Tr\biggl\{U^{\dagger}\oper{D}_{\mu}U\oper{L}_{\phi}^{n-1}
\Bigl(\oper{D}^{\mu}\phi\Bigr)\Bigl[\Bigl(\frac{1}{\cosh(\hat x)}\Bigr)^{(n)}\frac{1}{\cosh(\hat x)}
\nonumber \\& +\Bigl(\tanh(\hat x)\Bigr)^{(n)}\tanh(\hat x)
\Bigr]\biggr\} 
 \Tr\Bigl[\frac{\mathbf{1}_N}{\cosh^2(\hat x)}\Bigr]\,.
\end{align}
We find the result in the form:
\begin{equation}
\oper{T}_{mix}^0 =  \frac{c_1}{2m}\Tr\Bigl[U^{\dagger}\oper{D}_{\mu}UF_{mix}^0(\oper{L}_{\phi})(\oper{D}^{\mu}\phi)\Bigr]\,,
\end{equation}
where
\begin{align}
F_{mix}^0(\oper{L}_{\phi}) & = \lineint y\, \frac{1}{\oper{L}_{\phi}}\Bigl[1-\frac{1}{\cosh(y)\cosh(y-\oper{L}_{\phi})}-\tanh(y)\tanh(y-\oper{L}_{\phi})\Bigr] \nonumber \\ 
& =\frac{\cosh(\oper{L}_{\phi})-1}{\oper{L}_{\phi}\sinh(\oper{L}_{\phi})}
\ln\biggl(\frac{1+\tanh(\oper{L}_{\phi})}{1-\tanh(\oper{L}_{\phi})}\biggr)\,.
\end{align}
And
\begin{equation}
\oper{T}_{mix}^1 =  \frac{c_1^2}{8c_2m}\Tr\Bigl[U^{\dagger}\oper{D}_{\mu}UF_{mix}^1(\partial_x,\oper{L}_{\phi})(\oper{D}^{\mu}\phi)\Exp{x\phi}\Bigr]
\Tr\Bigl[\Exp{-x\phi}\Bigr]\bigg|_{x=0}\,,
\end{equation}
with
\begin{equation}
F_{mix}^1(x,\oper{L}_{\phi}) = \lineint y\, \frac{\cosh(\oper{L}_\phi)-1}{\cosh(y)\cosh(y-\oper{L}_{\phi})}\frac{1/\oper{L}_{\phi}}{\cosh^2(y-x)}\,.
\end{equation}
Again, the result of this integral can be obtain in the closed form. First few terms of the Taylor expansion reads
\begin{align}
F_{mix}^1(0,\oper{L}_{\phi}) & = \frac{2}{\oper{L}_{\phi}}\frac{\sinh(2\oper{L}_{\phi})-2\oper{L}_{\phi}}{\sinh(2\oper{L}_{\phi})+2\sinh(\oper{L}_{\phi})}\,, \\
\partial_x F_{mix}^{1}(0,\oper{L}_{\phi}) & = \frac{4}{3\oper{L}_{\phi}}\frac{\sinh(3\oper{L}_{\phi})+9\sinh(\oper{L}_{\phi})-12\oper{L}_{\phi}\cosh(\oper{L}_{\phi})}
{\cosh(3\oper{L}_{\phi})+2\cosh(2\oper{L}_{\phi})-\cosh(\oper{L}_{\phi})-2}\,, \\
\partial_x^2 F_{mix}^{1}(0,\oper{L}_{\phi}) & = -\frac{16}{\oper{L}_{\phi}}\frac{2\oper{L}_{\phi}\cosh(2\oper{L}_{\phi})-3\sinh(2\oper{L}_{\phi})+4\oper{L}_{\phi}}
{\sinh(4\oper{L}_{\phi})+2\sinh(3\oper{L}_{\phi})-2\sinh(2\oper{L}_{\phi})-6\sinh(\oper{L}_{\phi})}\,. 
\end{align}

Finally, let us calculate kinetic term for $\phi$ field:
\begin{equation}
\oper{T}_{\phi} := \lineint y\, \oper{T}_{\phi}(\hat y) = \oper{T}_{\phi}^{0}+\oper{T}_{\phi}^1+c_1O\Bigl((c_1/c_2)^2\Bigr)\,,
\end{equation}
where\footnote{We included the pure kinetic term for $\phi$ into the summation as $n=2$ term in contrast to \refer{ch6:tphi}.}
\begin{align}
\oper{T}_{\phi}^{0} = - \frac{c_1}{2m}\sum_{n=2}^{\infty}\frac{1}{n!}\lineint x\, &
\Bigl[\Bigl(\frac{1}{\cosh x}\Bigr)^{(n)}\frac{1}{\cosh x}+\Bigl(\tanh x\Bigr)^{(n)}\tanh x\Bigr]
\nonumber \\
&\times \Tr\Bigl[\oper{L}_{\phi}^{n-2}
\Bigl(\oper{D}_{\mu}\phi\Bigr)\oper{D}^{\mu}\phi
\Bigr]\,, \\
\oper{T}_{\phi}^{1} = - \frac{c_1^2}{8c_2m}\sum_{n=2}^{\infty}\frac{1}{n!}\lineint x\,&
\Tr\biggl\{\oper{L}_{\phi}^{n-2}
\Bigl(\oper{D}_{\mu}\phi\Bigr)\oper{D}^{\mu}\phi\Bigl[\Bigl(\frac{1}{\cosh(\hat x)}\Bigr)^{(n)}\frac{1}{\cosh(\hat x)}\nonumber \\
&+\Bigl(\tanh(\hat x)\Bigr)^{(n)}\tanh(\hat x)
\Bigr]\biggr\}\Tr\Bigl[\frac{\mathbf{1}_N}{\cosh^2(\hat x)}\Bigr]\,.
\end{align}
It is easy to check that the result is
\begin{equation}
\oper{T}_{\phi}^0 =  \frac{c_1}{2m}\Tr\Bigl[\oper{D}_{\mu}\phi F_{\phi}^0(\oper{L}_{\phi})(\oper{D}^{\mu}\phi)\Bigr]\,,
\end{equation}
where
\begin{align}
F_{\phi}^0(\oper{L}_{\phi}) & = \lineint y\, \frac{1}{\oper{L}_{\phi}^2}\Bigl[1-\frac{1}{\cosh(y)\cosh(y+\oper{L}_{\phi})}-\tanh(y)\tanh(y+\oper{L}_{\phi})\Bigr] \nonumber \\ 
& =\frac{\cosh(\oper{L}_{\phi})-1}{\oper{L}_{\phi}^2\sinh(\oper{L}_{\phi})}
\ln\biggl(\frac{1+\tanh(\oper{L}_{\phi})}{1-\tanh(\oper{L}_{\phi})}\biggr)
\end{align}
and
\begin{equation}
\oper{T}_{\phi}^1 =  \frac{c_1^2}{8c_2m}\Tr\Bigl[\oper{D}_{\mu}\phi F_{\phi}^1(\partial_x,\oper{L}_{\phi})(\oper{D}^{\mu}\phi)\Exp{x\phi}\Bigr]
\Tr\Bigl[\Exp{-x\phi}\Bigr]\bigg|_{x=0}\,,
\end{equation}
with
\begin{equation}
F_{\phi}^1(x,\oper{L}_{\phi}) = \lineint y\, \frac{\cosh(\oper{L}_\phi)-1}{\cosh(y)\cosh(y+\oper{L}_{\phi})}\frac{1/\oper{L}_{\phi}^2}{\cosh^2(y-x)}\,,
\end{equation}
where the first few coefficients in the Taylor expansion are given as
\begin{align}
F_{\phi}^1(0,\oper{L}_{\phi}) & = \frac{2}{\oper{L}_{\phi}^2}\frac{\sinh(2\oper{L}_{\phi})-2\oper{L}_{\phi}}{\sinh(2\oper{L}_{\phi})+2\sinh(\oper{L}_{\phi})}\,, \\
\partial_x F_{\phi}^{1}(0,\oper{L}_{\phi}) & = -\frac{4}{3\oper{L}_{\phi}^2}\frac{\sinh(3\oper{L}_{\phi})+9\sinh(\oper{L}_{\phi})-12\oper{L}_{\phi}\cosh(\oper{L}_{\phi})}
{\cosh(3\oper{L}_{\phi})+2\cosh(2\oper{L}_{\phi})-\cosh(\oper{L}_{\phi})-2}\,, \\
\partial_x^2 F_{\phi}^{1}(0,\oper{L}_{\phi}) & = \frac{16}{\oper{L}_{\phi}^2}\frac{2\oper{L}_{\phi}\cosh(2\oper{L}_{\phi})-3\sinh(2\oper{L}_{\phi})+4\oper{L}_{\phi}}
{\sinh(4\oper{L}_{\phi})+2\sinh(3\oper{L}_{\phi})-2\sinh(2\oper{L}_{\phi})-6\sinh(\oper{L}_{\phi})}\,.
\end{align}

There are still two terms in \refer{ch6:labor} which remain to be evaluated. These are
\begin{align}
\oper{T}_{\phi}^{\prime} & := -\frac{c_1^2}{16c_2}\lineint y\,\Exp{\eta}\Bigl(1-\frac{\sigma}{m}\Bigr)\Tr\Bigl[\frac{1}{\cosh^2(\hat y)}\oper{D}_{\mu}\phi\Bigr]
\Tr\Bigl[\frac{1}{\cosh^2(\hat y)}\oper{D}^{\mu}\phi\Bigr]\,, \\
\oper{T}_{U}^{\prime} & := \frac{c_1^2}{16c_2}\lineint y\,\Exp{\eta}\Bigl(1-\frac{\sigma}{m}\Bigr)\Tr\Bigl[\frac{1}{\cosh^2(\hat y)}\oper{D}_{\mu}U^{\dagger} U\Bigr]
\Tr\Bigl[\frac{1}{\cosh^2(\hat y)}\oper{D}^{\mu}U^{\dagger}U\Bigr]\,.
\end{align}
In our approximation, these terms are already of the order $c_1O(c_1/c_2)$, therefore we can set $\Exp{-\eta} \approx 1$ and $\sigma \approx 0$.
Thus
\begin{align}
\oper{T}_{\phi}^{\prime} & = -\frac{c_1^2}{16c_2m}F(\partial_x)\Tr\Bigl[\Exp{x\phi}\oper{D}_{\mu}\phi\Bigr]
\Tr\Bigl[\Exp{-x\phi}\oper{D}^{\mu}\phi\Bigr]\bigg|_{x=0} + c_1O\Bigl((c_1/c_2)^2\Bigr)\,, \\
\oper{T}_{U}^{\prime} & = \frac{c_1^2}{16c_2m}F(\partial_x)\Tr\Bigl[\Exp{x\phi}\oper{D}_{\mu}U^{\dagger}U\Bigr]
\Tr\Bigl[\Exp{-x\phi}\oper{D}^{\mu}U^{\dagger}U\Bigr]\bigg|_{x=0}+c_1O\Bigl((c_1/c_2)^2\Bigr)\,,
\end{align}
where
\begin{equation}
F(x) := \lineint y\, \frac{1}{\cosh^2(y)\cosh^2(y-x)} = 16\frac{x\cosh(x)-\sinh(x)}{\sinh(3x)-3\sinh(x)}\,.
\end{equation}

\subsection{General case IV: conclusion}

To conclude this section, let us summarize our findings.
The low-energy effective Lagrangian up to the second order in $c_1/c_2$ is given as 
\begin{equation}\label{eq:efflagr}
\oper{L}_{\rm eff} = \oper{L}_{\rm eff}^{(0)}
+\oper{T}_{\phi}^{(1)}+\oper{T}_{U}^{(1)}
+\oper{T}_{mix}^{(1)}+\oper{T}_{\phi}^{\prime}
+\oper{T}_{U}^{\prime}
+c_1O\Bigl((c_1/c_2)^2\Bigr)\,,
\end{equation}
where 
\begin{align}
 \oper{L}_{{\rm eff}}^{(0)} & = \frac{c_1}{2m}
\Tr\biggl[\oper{D}_{\mu} \phi\,
 \frac{\cosh(\oper{L}_{ \phi})-1}{\oper{L}_{ \phi}^2
\sinh(\oper{L}_{ \phi})}
\ln\biggl(\frac{1+\tanh(\oper{L}_{ \phi})}
{1-\tanh(\oper{L}_{ \phi})}\biggr)(\oper{D}^{\mu}\phi) 
\nonumber \\
& +U^{\dagger}\oper{D}_{\mu}U\,
\frac{\cosh(\oper{L}_{\phi})-1}{\oper{L}_{ \phi}
\sinh(\oper{L}_{ \phi})}
\ln\biggl(\frac{1+\tanh(\oper{L}_{ \phi})}
{1-\tanh(\oper{L}_{ \phi})}\biggr)(\oper{D}^{\mu} \phi) 
\nonumber \\\label{eq:efflagr0}
&+\frac{1}{2}\oper{D}_{\mu}U^{\dagger}U\frac{1}
{\tanh(\oper{L}_{\phi})}\ln\biggl(
\frac{1+\tanh(\oper{L}_{\phi})}{1-\tanh(\oper{L}_{ \phi})}
\biggr)
(U^{\dagger}\oper{D}^{\mu}U)\biggr]\,.
\end{align}
Interestingly, this is the same effective Lagrangian we 
obtained previously \refer{eq:result}. 
The rest of terms of the first order in $c_1/c_2$ are given by 
\begin{align}
\oper{T}_{U}^{(1)} & = \frac{c_1^2}{16c_2m}
\Tr\Bigl[\oper{D}_{\mu}U^{\dagger}UF_U(\partial_x,
\oper{L}_{\phi})(U^{\dagger}\oper{D}^{\mu}U)
\Exp{x\phi}\Bigr]\Tr\Bigl[\Exp{-x\phi}\Bigr]\bigg|_{x=0}\,, \\
\oper{T}_{mix}^{(1)} & 
=  \frac{c_1^2}{8c_2m}\Tr\Bigl[U^{\dagger}
\oper{D}_{\mu}UF_{mix}(\partial_x,\oper{L}_{\phi})
(\oper{D}^{\mu}\phi)\Exp{x\phi}\Bigr]
\Tr\Bigl[\Exp{-x\phi}\Bigr]\bigg|_{x=0}\,, \\
\oper{T}_{\phi}^{(1)} & =  \frac{c_1^2}{8c_2m}
\Tr\Bigl[\oper{D}_{\mu}\phi F_{\phi}
(\partial_x,\oper{L}_{\phi})(\oper{D}^{\mu}\phi)
\Exp{x\phi}\Bigr]
\Tr\Bigl[\Exp{-x\phi}\Bigr]\bigg|_{x=0}\,, \\
\oper{T}_{\phi}^{\prime} & = -\frac{c_1^2}{16c_2m}
F(\partial_x)\Tr\Bigl[\Exp{x\phi}\oper{D}_{\mu}\phi\Bigr]
\Tr\Bigl[\Exp{-x\phi}\oper{D}^{\mu}\phi\Bigr]\bigg|_{x=0}\,, 
\\
\oper{T}_{U}^{\prime} & = \frac{c_1^2}{16c_2m}
F(\partial_x)\Tr\Bigl[\Exp{x\phi}\oper{D}_{\mu}U^{\dagger}
U\Bigr]
\Tr\Bigl[\Exp{-x\phi}\oper{D}^{\mu}U^{\dagger}U\Bigr]
\bigg|_{x=0},
\end{align}
and 
\begin{align}
F_U(x,\oper{L}_{\phi}) & = \lineint y\, 
\frac{\cosh(\oper{L}_{\phi})}{\cosh^2(y-x)\cosh(y)
\cosh(y-\oper{L}_{\phi})}\,, \\
F_{mix}(x,\oper{L}_{\phi})  & = \lineint y\, 
\frac{\cosh(\oper{L}_\phi)-1}
{\oper{L}_{\phi}\cosh(y)\cosh(y-\oper{L}_{\phi})
\cosh^2(y-x)}\,, \\
F_{\phi}(x,\oper{L}_{\phi}) & = \lineint y\, 
\frac{\cosh(\oper{L}_\phi)-1}
{\oper{L}_{\phi}^2\cosh(y)\cosh(y+\oper{L}_{\phi})
\cosh^2(y-x)}\,, \\
F(x) & = \lineint y\, \frac{1}{\cosh^2(y)\cosh^2(y-x)}\,.
\end{align}
All the above integrals can be obtained in closed 
forms.

The formula \refer{eq:efflagr} illustrates the 
complexity of the interactions between moduli fields 
$\phi$ and $U$ in the general case.
Let us offer some explanation for this complexity. 
The structure, which is new in \refer{eq:efflagr} compared 
to \refer{eq:efflagr0}, is of the general form
\begin{equation}
F(\partial_x)\Tr[\Exp{x\phi}M(y)]\Tr[\Exp{-x\phi}N(y)]\Big|_{x=0}\,,
\end{equation}
where $M(y)$ and $N(y)$ are some matrix-valued functions, containing either derivatives of $\phi$ or derivatives of $U$.
If we assume that $\phi = mP^{-1}\mathrm{diag}(y_1,\ldots ,y_N)P$ is diagonalizable, we can rewrite the above as 
\begin{equation}
\sum_{i,j=1}^NF(m(y_i-y_j))(PM(y)P^{-1})_{ii}(PN(y)P^{-1})_{jj}\,.
\end{equation}
This form suggests that interaction (here represented by function $F$) depends on the relative size of fluctuation of each pairs of walls.
Indeed, notice that if two walls have the same position  $y_i = y_j$ $i\not = j$ (meaning that the expectation values of the fluctuations are the same), the above form reduces to
\begin{equation}
F(0)\Tr(M)\Tr(N)\,,
\end{equation}
which is in a sense trivial, since we already encountered this kind of terms in \refer{eq:result}.
Thus, the new kind of complexity in our result \refer{eq:efflagr}
can be understood as a manifestation of the fact, that the interaction does not depend only on various moments of the fluctuation as in 
\refer{eq:efflagr0}, but also on their relative size.

\section{More general models of stable 
position-dependent coupling}

In this section, we wish to show that there are more 
models with the stable position-dependent coupling. 
We will illustrate the point by extending the model 
to include more fields and more gauge symmetry. 

The position-dependent gauge coupling comes from the 
cubic coupling between a singlet scalar field and 
field strengths of non-Abelian gauge fields in $4+1$ 
dimensions 
\begin{eqnarray}
{\mathcal L}_{\rm cubic} = 
a(\sigma_i) \Tr\left[\tilde G_{MN}\tilde G^{MN}\right], 
\end{eqnarray}
where the function $a(\sigma_i)$ of singlet scalar fields 
$\sigma_i$ should be linear, if it is to be embeddable 
into a supersymmetric gauge theory in $4+1$ dimensions 
\begin{eqnarray}
a(\sigma_i) = \sum_i \gamma_i \sigma_i,
\quad \gamma_i \in \mathbb{R}, 
\label{eq:coefficient_singlets}
\end{eqnarray}
where $\gamma_i$ are constant coefficients.

Usually each domain wall has one complex moduli: 
a position and a phase. 
For example, both the massive $\mathbb{C}P^2$ model 
and the massive 
$\mathbb{C}P^1 \times \mathbb{C}P^1$ model 
have two free domain walls, corresponding to the 
two complex moduli. 
Although, two free domain walls can produce the desired 
profile of position-dependent gauge coupling by an 
appropriate choice of parameters in 
Eq.~(\ref{eq:coefficient_singlets}), they provide the 
undesired modulus for the width of the profile. 
To avoid this problem, we are led to consider models 
with a single complex moduli. 
The simplest one of such models is the three-flavor model 
in Ref.~\cite{Otha} (or see Ch.~\ref{ch:4}), where three scalar fields are 
constrained by the two Abelian gauge symmetry 
$U(1)\times U(1)$.  
The generalized three-flavor model of this chapter is an extension of this model 
to non-Abelian gauge group: 
$U(1)\times U(1) \to U(N)\times U(1)$. 
The next simplest possibility is to consider four scalars 
constrained by three Abelian gauge symmetry 
$U(1)\times U(1)\times U(1)$, which we call extended models.
We give quantum numbers of fields of a typical extended 
model in Tab.~\ref{table:04}. 
In the limit of strong gauge couplings, the gauge theory 
becomes a nonlinear sigma model whose target space is 
given by an intersection of three conditions as 
\begin{eqnarray}
\left(\mathbb{C}^2 \times \mathbb{C}P^1\right)
\cap
\left(\mathbb{H}^2 \times \mathbb{C}P^1\right)
\cap
\left(\mathbb{C}^2 \times \mathbb{C}P^1\right)
 \simeq \mathbb{C}P^1\,.
\end{eqnarray}
Here $\mathbb{H}^2$ denotes hyperbolic complex plane.

\begin{table}
\begin{center}
\begin{tabular}{c|ccc|c}
\hline
 & $U(1)_1$ & $U(1)_2$ & $U(1)_3$ & mass\\ \hline
$H_{1}$ & 1 & 0 & 0& $m_1$\\ 
$H_{2}$ & 1 & $-1$ & 0& $m_2$\\ 
$H_{3}$ &  0 & 1 & 1& $m_3$\\ 
$H_{4}$ &  0 & 0 & 1& $m_4$\\ 
$\sigma_1$ & 0 & 0 & 0& 0\\
$\sigma_2$ & 0 & 0 & 0& 0 \\
$\sigma_3$ & 0 & 0 & 0& 0 \\
\hline
\end{tabular}
\end{center}
\caption{Quantum numbers of the $U(1)_1\times U(1)_2
\times U(1)_3$ extended model.}
\label{table:04}
\end{table}

The vacuum condition is given by 
\begin{eqnarray}
|H_1|^2 + |H_2|^2 =  c_1,\quad
- |H_2|^2 + |H_3|^2 = c_2,\quad
|H_3|^2 + |H_4|^2 = c_3,\\
H_1(\sigma_1 - m_1) = 0,\quad
H_2(\sigma_1 - \sigma_2 - m_2) = 0,\qquad\qquad\\
H_3(\sigma_2 + \sigma_3 - m_3) = 0,\quad
H_4(\sigma_3 - m_4) = 0,\qquad\qquad
\end{eqnarray}
where $c_i$ is the Fayet-Iliopoulos parameter of $U(1)_i$ and $m_a$ is 
the mass for $H_i$. 
All possible solutions to these equations are shown 
in Tab.~\ref{table:05}.
There are four solutions but only two of them are 
valid solutions for any choice of real parameters 
of $c_i$. 
When we choose $c_1>0$, $c_2>0$ and $c_3 > c_1 + c_2$, 
we are left with the vacua $\left<1\right>$ and 
$\left<2\right>$ in Tab.~\ref{table:05}.
\begin{table}
\begin{center}
\begin{tabular}{c|ccccccc}
 & $|H_1|$ & $|H_2|$ & $|H_3|$ & $|H_4|$ & $\sigma_1$ 
& $\sigma_2$ & $\sigma_3$\\
\hline
$\left<1\right>$ & $0$ & $\sqrt{c_1}$ & $\sqrt{c_{1+2}}$ 
& $\sqrt{c_{3-1-2}}$ & $m_{2+3-4}$ & $m_{3-4}$ & $m_4$\\
$\left<2\right>$ & $\sqrt{c_1}$ & $0$ & $\sqrt{c_2}$ 
& $\sqrt{c_{3-2}}$ & $m_1$ & $m_{3-4}$ & $m_4$\\
$\left<3\right>$ & $\sqrt{c_{1+2}}$ & $\sqrt{-c_2}$ 
& $0$ & $\sqrt{c_3}$ & $m_1$ & $m_{1-2}$ & $m_4$\\
$\left<4\right>$ & $\sqrt{c_{1+2-3}}$ & $\sqrt{c_{3-2}}$ 
& $\sqrt{c_3}$ & $0$ & $m_1$ & $m_{1-2}$ & $m_{2-1+3}$
\end{tabular}
\caption[VEVs of candidate vacua]{VEVs of candidate vacua: 
We use abbreviations like 
$c_{3-1-2} \equiv c_3-c_1-c_2$.}
\label{table:05}
\end{center}
\end{table}

The moduli matrix formalism
is powerful enough to give generic solutions of the 
BPS equations of this nonlinear sigma model. 
Especially, we are 
interested in the kink profiles of $\sigma_1$, $\sigma_2$ 
and $\sigma_3$. They 
can be expressed as derivatives of real functions 
$\eta_i$ 
\begin{eqnarray}
\sigma_i = \frac{1}{2}\partial_y \eta_i,\quad (i=1,2,3),
\end{eqnarray}
where the real functions $\eta_i$ are determined by 
the following algebraic conditions 
\begin{eqnarray}
\Exp{-\eta_1+2m_1y}+\Exp{-\eta_1+\eta_2+2m_2y-2a}=c_1,\\
-\Exp{-\eta_1+\eta_2+2m_2y-2a}+\Exp{-\eta_2-\eta_3+2m_3y}=c_2,\\
\Exp{-\eta_2-\eta_3+2m_3y}+\Exp{-\eta_3+2m_4y}=c_3,
\end{eqnarray} 
with a real constant $a$. 
The parameter $a$ is (the real part of) the unique modulus 
of the solution, corresponding to the position of the domain 
wall.
As we expected, we find only a single modulus. 
The width of the domain wall is not a modulus, but 
fixed by the theory.  

\begin{figure}
\begin{center}
\begin{minipage}[b]{0.48\linewidth}
\centering
\includegraphics[width=\textwidth]{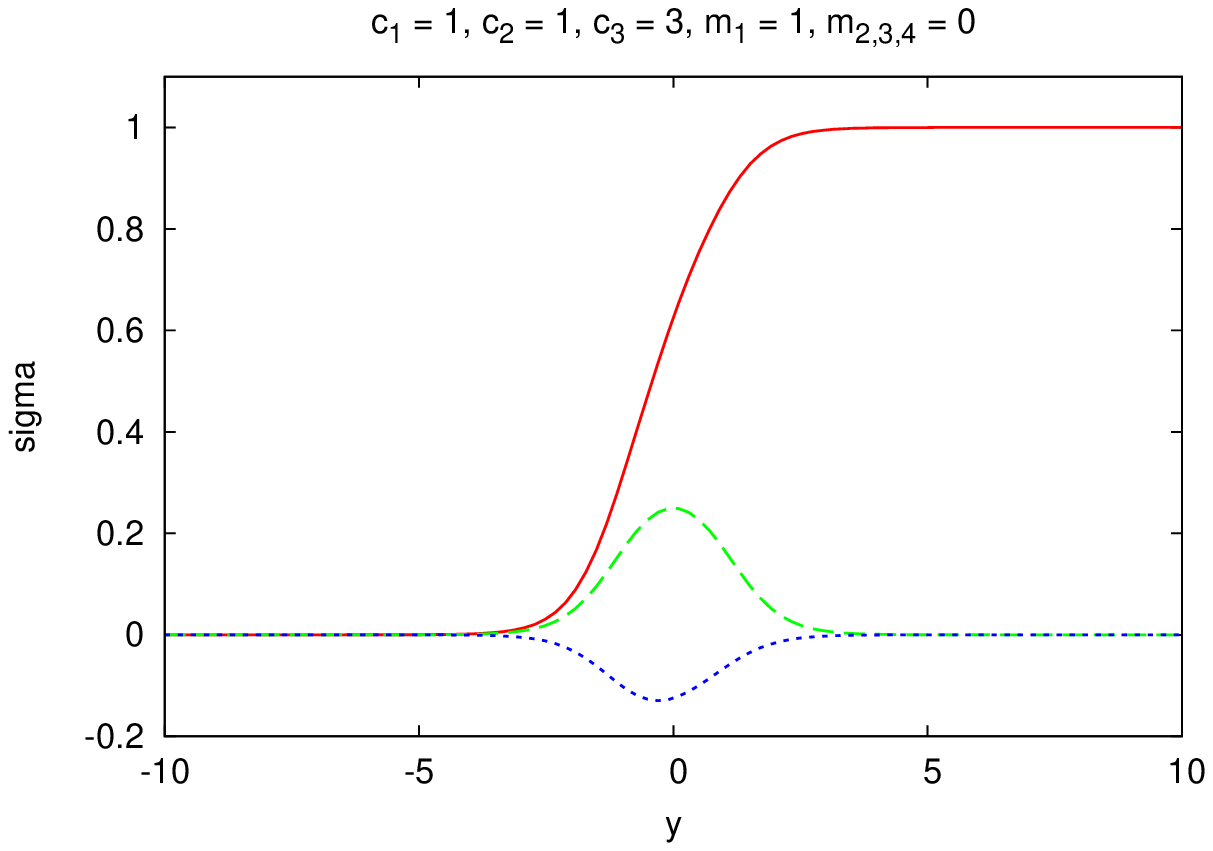}
\end{minipage}
\begin{minipage}[b]{0.48\linewidth}
\centering
\includegraphics[width=\textwidth]{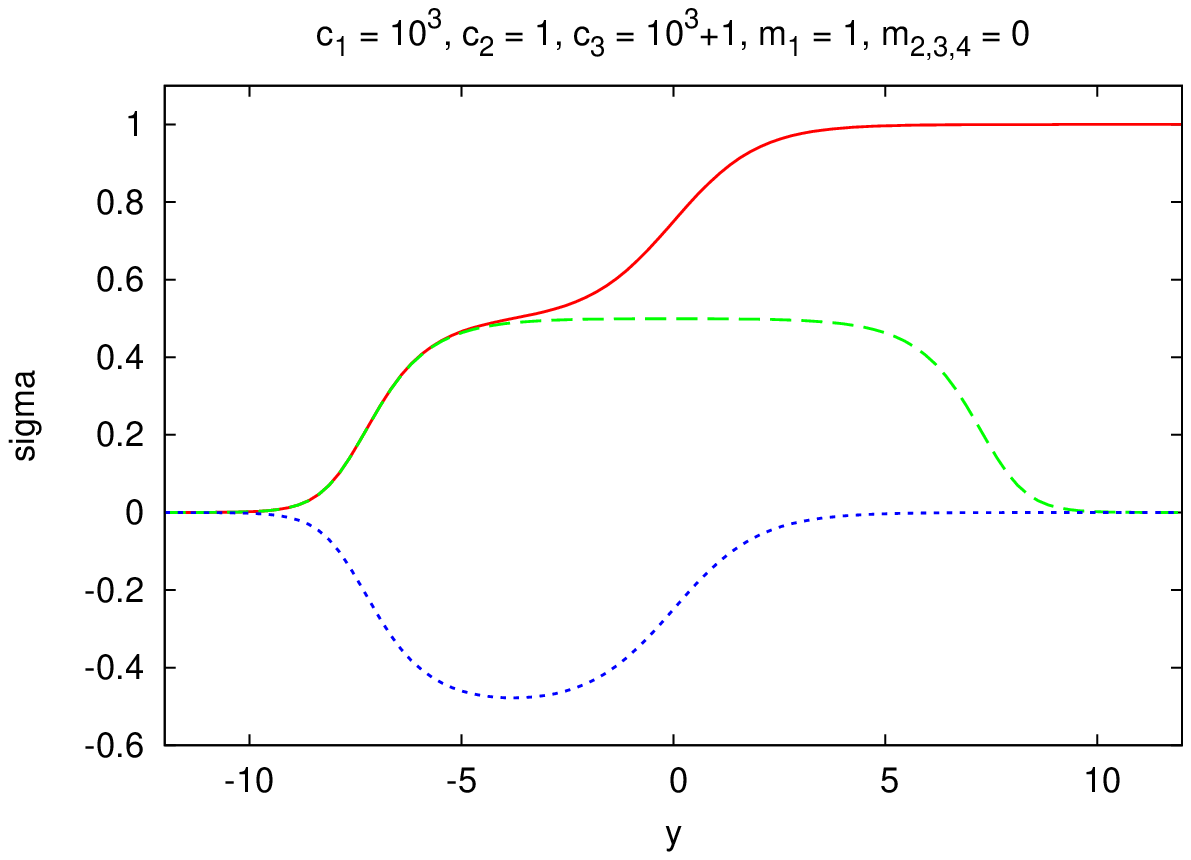}
\end{minipage}
\caption{The kink profiles of $\sigma_1$ (red solid line
), 
$\sigma_2$ (green dashed line
) and $\sigma_3$ (blue dotted line
) for 
two different sets of model parameters. }
\label{fig:sigma_extended}
\end{center}
\end{figure}

Since we want a configuration that $\sigma_i \to 0$ at 
both spacial infinities, we can choose $m_3=m_4=0$. 
Then $\sigma_2 = \sigma_3 \to 0$ at $y=\pm \infty$. 
Several numerical solutions are displayed in 
Fig.~\ref{fig:sigma_extended}.
From Fig.~\ref{fig:sigma_extended}, we clearly see 
that $\sigma_2 \ge 0$ and $\sigma_3 \le 0$ for all the 
values of $y$.
As a position-dependent gauge coupling, we can choose a 
two-parameter family
\begin{eqnarray}
{\mathcal L}_{\rm cubic} = - \left(\gamma_2 \sigma_2 
- \gamma_3\sigma_3\right)\Tr\left[\tilde G_{MN}
\tilde G^{MN}\right],
\end{eqnarray}
where $\gamma_{2,3}$ can be any non-negative real numbers. 
This class of models can be easily made to localize 
non-Abelian gauge fields and minimally interacting 
matter fields by extending two of the $U(1)$ factor 
groups to (possibly different) $U(N)$ gauge groups 
with $N$ scalar fields in the fundamental 
representations, similarly to our model in this chapter. 
A new interesting feature of the extended model 
(and its non-Abelian extensions) is that two possible 
profile of singlet fields $\sigma_1, \sigma_2$ can 
provide a different profile for different non-Abelian 
gauge groups such as $SU(3)$, $SU(2)$ and $U(1)$ 
with their associated matter fields.



\chapter{Conclusion and discussion}\label{ch:7} 




In this thesis we have investigated the possibility that the brane-world scenario might be realized in the field-theoretical framework by a topological soliton.
As a model of the brane we used a simplest soliton possible: a BPS domain wall. A natural consequence of this choice was that the laboratory, where our
models were constructed, was (4+1)-dimensional field theories embeddable into $N=1$ SUSY gauge theories with eight supercharges.
Within this class of theories we showed that localization of non-Abelian gauge fields, together with minimally interacting matter fields, is quite natural, once it is realized that the gauge symmetry, which is to be localized on the domain wall, should not be broken in the bulk. 

One of the interesting features of our models of Ch.~\ref{ch:5} and Ch.~\ref{ch:6} is the geometric realization of the Higgs mechanism. Since we used as a background a coincident domain wall, representing the most symmetric point in the moduli space of solutions, we managed to localize a large
gauge group. It is quite possible that quantum or other corrections will result in departing from this coincident point and some gauge fields becomes massive, eating corresponding NG bosons in the process.
Thus, we see an analogy to the usual Higgs mechanism based on purely geometrical basis. 
Amongst the possible future investigations, it would be worth-while to study non-coincident solution to further clarify this 
geometrical Higgs mechanism. 

We have also noticed that a part of moduli fields in the low-energy effective Lagrangian resembles the pion in QCD. 
Similar attempts have been quite successful using 
D-branes \cite{Sakai}. 
We believe that our methods can provide more insight 
in various aspects of low-energy hadron physics. 

Moreover, at the end of the chapter \ref{ch:6} we explored possible generalizations of our stabilization 
mechanism by including more fields and more gauge  
symmetries
and found a class of more generic models with an added 
flexibility  
for
model building with different localization 
profile for different gauge groups.

To build realistic models of brane-world with our 
scenario of localized gauge fields and matter
fields, 
we should address several questions. 
Perhaps the most important question is to obtain 
(massless) matter fields in representations like the 
fundamental rather than the adjoint of localized gauge 
fields. 
One immediate possibility is to use the localization 
mechanism of fermions in a kink background. 
It has been found that zero modes of such fermions are 
localized in such a way to give automatically chiral 
fermions \cite{Rubakov}. 

Secondly, we should devise a way to give small masses 
to our matter fields in order to do phenomenology. 
Since some of our matter fields are the NG 
modes of a broken global symmetry, we need to consider 
an explicit breaking of such global symmetry. 

Thirdly, another question is to study possibility of 
supersymmetric model of gauge field localization. 
We need to settle the issue of possible new moduli 
in that case, which we discussed in Ref.~\cite{Us1}. 
Moreover, we should examine the mechanism of supersymmetry 
breaking. For example, we can consider a model similar to the one we introduced in Ch.~\ref{ch:5}, but where different halves of 
supercharges are preserved at each sector (BPS and 
anti-BPS walls). In such a model the 
SUSY is completely broken in the system as a whole. 
It has been proposed that the coexistence of BPS and anti-BPS 
walls gives the supersymmetry breaking in a controlled 
manner \cite{Maru}. 
In our present case, BPS and anti-BPS sectors interact 
only weakly. 
If we choose flavor gauge fields for each sector separately, 
we have only higher derivative interactions induced by 
massive modes. 
If we choose the diagonal subgroup of 
each sector as flavor gauge group, we have a more 
interesting possibility of the massless gauge field, propagating 
as a messenger field between two sectors. 

Another interesting possibility for a model building is 
to localize gauge fields of different gauge groups with 
different profiles. 
This situation is often proposed in recent brane-world 
phenomenology, for instance in Ref.~\cite{Hosotani}. 

Finally, let us examine similarities and differences 
of our domain walls compared to D-branes. 
The most interesting similarity of our domain wall 
with D-branes is the realization of geometrical 
Higgs mechanism, where massless gauge fields in the 
coincident wall become massive as walls separate. 
On the other hand, 
there are differences as well. 
D-branes in string theory are defined by the Dirichlet 
condition for fundamental string attached to it,  
but no such condition is visible in our domain walls. 
Domain walls similar to ours have been constructed in Ref.~\cite{Shifman2}, where a
probe magnetic charge is placed in the bulk Higgs phase. The magnetic flux from the probe
magnetic monopole is carried by a vortex which can end on the domain wall. They observed
that this wall-string junction configuration resembles an open string ending on D-branes in
string theory. However, this phenomenon is in theories with one dimension less, namely the
world volume of domain walls has only $(2 + 1)$-dimensions (fundamental theory is in $(3 + 1)$-dimensions). 
In our model based on a theory in $(4 + 1)$-dimensions, monopole (codimension
three) is a string-like soliton and might possibly be a candidate of something similar to the
fundamental string. It is worth pursuing a possibility of composite solitons consisting of
domain walls and other solitons, such as monopoles, in order to clarify more similarities of
solitons with D-branes. 
Another interesting issue of the effective action on 
D-branes is the non-Abelian generalization of the 
Dirac-Born-Infeld (DBI) action. There have been studies to 
obtain first few corrections to Yang-Mills action in 
string theories. 
Since we are at present interested in up to 
quadratic terms in derivatives, we have found the ordinary 
quadratic action of Yang-Mills fields together with 
the action of moduli fields interacting minimally with 
the Yang-Mills 
fields in addition to the nonlinear interactions among 
themselves. Our result is trivially consistent with the 
quadratic approximation of DBI action, but does not give 
us informations on non-Abelian generalization of DBI 
action. 
In order to shed light on that issue, 
we need to compute higher 
derivative corrections to our effective Lagrangian.

We plan to consider all these issues in future studies.






\begin{multicols}{2} 
\begin{tiny} 

\bibliographystyle{h-physrev}
\renewcommand{\bibname}{References} 

\bibliography{references} 

\begin{thebibliography}{10}

\bibitem{Kaluza}
T.~{Kaluza},
\newblock Sitzungsber. Preuss. Akad. Wiss. Phys. Math. Klasse , 966 (1921).

\bibitem{Klein}
O.~{Klein},
\newblock Z.F. Physik {\bf 37}, 895 (1926).

\bibitem{Rubakov}
V.~A. {Rubakov} and M.~E. {Shaposhnikov},
\newblock Physics Letters B {\bf 125}, 136 (1983).

\bibitem{Rubakov2}
V.~A. {Rubakov} and M.~E. {Shaposhnikov},
\newblock Physics Letters B {\bf 125}, 139 (1983).

\bibitem{Dvali}
G.~{Dvali} and M.~{Shifman},
\newblock Physics Letters B {\bf 475}, 295 (2000), arXiv:hep-ph/0001072.

\bibitem{RS}
L.~{Randall} and R.~{Sundrum},
\newblock Physical Review Letters {\bf 83}, 3370 (1999), arXiv:hep-ph/9905221.

\bibitem{RS2}
L.~{Randall} and R.~{Sundrum},
\newblock Physical Review Letters {\bf 83}, 4690 (1999), arXiv:hep-th/9906064.

\bibitem{Polchinski}
J.~{Polchinski},
\newblock Physical Review Letters {\bf 75}, 4724 (1995), arXiv:hep-th/9510017.

\bibitem{Einstein}
A.~{Einstein} and P.~G. {Bergmann},
\newblock Ann. Math. {\bf 39}, 810 (1931).

\bibitem{Mandel}
H.~{Mandel},
\newblock Z.F. Physik {\bf 39}, 136 (1926).

\bibitem{Pauli}
W.~{Pauli},
\newblock Ann. d. Physik {\bf 18}, 305 (1933).

\bibitem{Straumann}
N.~{Straumann},
\newblock ArXiv General Relativity and Quantum Cosmology e-prints  (2000),
  arXiv:gr-qc/0012054.

\bibitem{Scherk}
J.~{Scherk} and J.~H. {Schwarz},
\newblock Physics Letters B {\bf 57}, 463 (1975).

\bibitem{Candelas}
P.~{Candelas}, G.~T. {Horowitz}, A.~{Strominger}, and E.~{Witten},
\newblock Nuclear Physics B {\bf 258}, 46 (1985).

\bibitem{Cremades}
D.~{Cremades}, L.~E. {Ibanez}, and F.~{Marchesano},
\newblock ArXiv High Energy Physics - Phenomenology e-prints  (2002),
  arXiv:hep-ph/0212048.

\bibitem{Cremades2}
D.~{Cremades}, L.~E. {Ibanez}, and F.~{Marchesano},
\newblock ArXiv High Energy Physics - Phenomenology e-prints  (2002),
  arXiv:hep-ph/0212064.

\bibitem{Akama}
K.~{Akama},
\newblock ArXiv High Energy Physics  (2000), arXiv:hep-th/0001113.

\bibitem{Poli}
N.~Poli {\em et~al.},
\newblock Phys. Rev. Lett. {\bf 106}, 038501 (2011).

\bibitem{ADD}
N.~{Arkani-Hamed}, S.~{Dimopoulos}, and G.~{Dvali},
\newblock Physics Letters B {\bf 429}, 263 (1998), arXiv:hep-ph/9803315.

\bibitem{ADD2}
N.~{Arkani-Hamed}, S.~{Dimopoulos}, and G.~{Dvali},
\newblock Physical Review D {\bf 59}, 086004 (1999), arXiv:hep-ph/9807344.

\bibitem{AADD}
I.~{Antoniadis}, N.~{Arkani-Hamed}, S.~{Dimopoulos}, and G.~{Dvali},
\newblock Physics Letters B {\bf 436}, 257 (1998), arXiv:hep-ph/9804398.

\bibitem{Shifman}
M.~{Shifman},
\newblock International Journal of Modern Physics A {\bf 25}, 199 (2010),
  arXiv:hep-ph/0907.3074.

\bibitem{Krauss}
L.~M. {Krauss} and F.~{Wilczek},
\newblock Physical Review Letters {\bf 62}, 1221 (1989).

\bibitem{Nath}
P.~{Nath} and P.~{Fileviez P{\'e}rez},
\newblock Physics Reports {\bf 441}, 191 (2007), arXiv:hep-ph/0601023.

\bibitem{Arkani-Hamed}
N.~{Arkani-Hamed} and M.~{Schmaltz},
\newblock Physical Review D {\bf 61}, 033005 (2000), arXiv:hep-ph/9903417.

\bibitem{ADDM}
N.~{Arkani-Hamed}, S.~{Dimopoulos}, G.~{Dvali}, and J.~{March-Russell},
\newblock Physical Review D {\bf 65}, 024032 (2002), arXiv:hep-ph/9811448.

\bibitem{Maldacena}
J.~M. Maldacena,
\newblock Adv.Theor.Math.Phys. {\bf 2}, 231 (1998), hep-th/9711200.

\bibitem{Witten:1998qj}
E.~Witten,
\newblock Adv.Theor.Math.Phys. {\bf 2}, 253 (1998), hep-th/9802150.

\bibitem{Gubser:1998bc}
S.~Gubser, I.~R. Klebanov, and A.~M. Polyakov,
\newblock Phys.Lett. {\bf B428}, 105 (1998), hep-th/9802109.

\bibitem{ArkaniHamed:2000ds}
N.~Arkani-Hamed, M.~Porrati, and L.~Randall,
\newblock JHEP {\bf 0108}, 017 (2001), hep-th/0012148.

\bibitem{Otha}
K.~{Ohta} and N.~{Sakai},
\newblock Progress of Theoretical Physics {\bf 124}, 71 (2010),
  arXiv:hep-th/1004.4078.

\bibitem{Manton}
N.~{Manton} and P.~{Sutcliffe},
\newblock {\em {Topological Solitons}} (, 2007).

\bibitem{Goodman}
R.~H. {Goodman} and R.~{Haberman},
\newblock SIAM Journal on Applied Dynamical Systems {\bf 4}, 1195 (2005).

\bibitem{Low}
I.~{Low} and A.~V. {Manohar},
\newblock Physical Review Letters {\bf 88}, 101602 (2002),
  arXiv:hep-th/0110285.

\bibitem{Shifmanch2}
M.~{Shifman}, A.~{Vainshtein}, and M.~{Voloshin},
\newblock Physical Review D {\bf 59}, 045016 (1999), arXiv:hep-th/9810068.

\bibitem{Coleman}
S.~Coleman and J.~Mandula,
\newblock Phys. Rev. {\bf 159}, 1251 (1967).

\bibitem{Golfand}
Y.~Golfand and E.~Likhtman,
\newblock JETP Lett. {\bf 13}, 323 (1971).

\bibitem{Haag}
R.~Haag, J.~T. Łopuszański, and M.~Sohnius,
\newblock Nuclear Physics B {\bf 88}, 257  (1975).

\bibitem{Witten}
E.~Witten and D.~Olive,
\newblock Physics Letters B {\bf 78}, 97  (1978).

\bibitem{Wess}
J.~Wess and J.~Bagger,
\newblock {\em Supersymmetry and supergravity}Princeton series in physics
  (PRINCETON University Press, 1992).

\bibitem{Seiberg}
N.~{Seiberg},
\newblock Physics Letters B {\bf 388}, 753 (1996), arXiv:hep-th/9608111.

\bibitem{Hassan}
L.~{Alvarez-Gaum{\'e}} and S.~F. {Hassan},
\newblock Fortschritte der Physik {\bf 45}, 159 (1997), arXiv:hep-th/9701069.

\bibitem{Bilal}
A.~{Bilal},
\newblock ArXiv High Energy Physics - Theory e-prints  (1996),
  arXiv:hep-th/9601007.

\bibitem{Cortes}
V.~{Cortes}, C.~{Mayer}, T.~{Mohaupt}, and F.~{Saueressig},
\newblock Journal of High Energy Physics {\bf 3}, 28 (2004),
  arXiv:hep-th/0312001.

\bibitem{Zumino}
B.~Zumino,
\newblock Phys.Lett. {\bf B87}, 203 (1979).

\bibitem{Fayet}
P.~Fayet and J.~Iliopoulos,
\newblock Phys.Lett. {\bf B51}, 461 (1974).

\bibitem{Galperin}
e.~a. Galperin, A.~S.,
\newblock  (Cambridge University Press, 2001).

\bibitem{Rocek}
A.~Karlhede, U.~{Lindstr{\"o}m}, and M.~Ro\v{c}ek,
\newblock Physics Letters B {\bf 147}, 297  (1984).

\bibitem{Unge}
F.~{Gonzalez-Rey}, M.~{Ro{\v c}ek}, S.~{Wiles}, U.~{Lindstr{\"o}m}, and R.~{von
  Unge},
\newblock Nuclear Physics B {\bf 516}, 426 (1998), arXiv:hep-th/9710250.

\bibitem{Kuzenko}
S.~M. {Kuzenko},
\newblock Journal of Physics A Mathematical General {\bf 43}, 3001 (2010),
  arXiv:hep-th/1004.0880.

\bibitem{ArkaniHamed}
N.~{Arkani-Hamed}, T.~{Gregoire}, and J.~{Wacker},
\newblock Journal of High Energy Physics {\bf 3}, 55 (2002),
  arXiv:hep-th/0101233.

\bibitem{Eto2}
M.~{Eto}, T.~{Fujimori}, M.~{Nitta}, K.~{Ohashi}, and N.~{Sakai},
\newblock Physical Review D {\bf 77}, 125008 (2008), arXiv:hep-th/0802.3135.

\bibitem{Arai}
M.~{Arai}, M.~{Nitta}, and N.~{Sakai},
\newblock Progress of Theoretical Physics {\bf 113}, 657 (2005),
  arXiv:hep-th/0307274.

\bibitem{Isozumi}
Y.~{Isozumi}, M.~{Nitta}, K.~{Ohashi}, and N.~{Sakai},
\newblock Physical Review D {\bf 70}, 125014 (2004), arXiv:hep-th/0405194.

\bibitem{Dvali2}
G.~{Dvali} and M.~{Shifman},
\newblock Physics Letters B {\bf 396}, 64 (1997), arXiv:hep-th/9612128.

\bibitem{Eto}
M.~{Eto}, Y.~{Isozumi}, M.~{Nitta}, K.~{Ohashi}, and N.~{Sakai},
\newblock Journal of Physics A Mathematical General {\bf 39}, 315 (2006),
  arXiv:hep-th/0602170.

\bibitem{Us1}
M.~{Arai}, F.~{Blaschke}, M.~{Eto}, and N.~{Sakai},
\newblock Progress of Theoretical and Experimental Physics {\bf 2013}, 010003
  (2013), arXiv:hep-th/1208.6219.

\bibitem{George}
R.~{Davies}, D.~P. {George}, and R.~R. {Volkas},
\newblock Physical Review D {\bf 77}, 124038 (2008), arXiv:hep-ph/0705.1584.

\bibitem{Susskind}
J.~Kogut and L.~Susskind,
\newblock Phys. Rev. D {\bf 9}, 3501 (1974).

\bibitem{Fukuda}
R.~{Fukuda},
\newblock ArXiv e-prints  (2008), arXiv:hep-th/0805.3864.

\bibitem{Us2}
M.~{Arai}, F.~{Blaschke}, M.~{Eto}, and N.~{Sakai},
\newblock ArXiv e-prints  (2013), arXiv:hep-th/1303.5212.

\bibitem{Sakai}
T.~{Sakai} and S.~{Sugimoto},
\newblock Progress of Theoretical Physics {\bf 113}, 843 (2005),
  arXiv:hep-th/0412141.

\bibitem{Maru}
N.~{Maru}, N.~{Sakai}, Y.~{Sakamura}, and R.~{Sugisaka},
\newblock Physics Letters B {\bf 496}, 98 (2000), arXiv:hep-th/0009023.

\bibitem{Hosotani}
Y.~{Hosotani} and Y.~{Kobayashi},
\newblock Physics Letters B {\bf 674}, 192 (2009), arXiv:hep-ph/0812.4782.

\bibitem{Shifman2}
M.~{Shifman} and A.~{Yung},
\newblock Phys. Rev. D {\bf 67}, 125007 (2003), arXiv:hep-th/0212293.

\end{thebibliography}

\end{tiny}
\end{multicols}









\end{document}